\documentclass[a4paper,fleqn,usenatbib]{mnras}

\usepackage{newtxtext,newtxmath}
\usepackage[T1]{fontenc}
\usepackage{ae,aecompl}

\usepackage{hyperref}
\usepackage{graphicx}
\usepackage{multirow}
\usepackage{mathtools}
\usepackage{makecell}
\usepackage{rotating}
\usepackage[toc,page]{appendix}
\usepackage{pdflscape}

\newcommand{\msun}{\mbox{M}_{\odot}}
\newcommand{\mstar}{$\mbox{M}_*$}

\newcommand{\hii}{H~\textsc{ii}\ }
\newcommand{\ii}{~\textsc{ii}}
\newcommand{\iii}{~\textsc{iii}}
\newcommand{\iv}{~\textsc{iv}}

\newcommand{\ttwo}{T$_2$}
\newcommand{\te}{T$_3$}

\title[The MOSDEF Survey: Direct-Method Metallicities at $z\sim1.5-3.5$]{The MOSDEF Survey: Direct-Method Metallicities and ISM Conditions at $\lowercase{z}\sim1.5-3.5$}

\author[R. L. Sanders et al.]{
Ryan L. Sanders,$^{1}$\thanks{E-mail: rlsand@ucdavis.edu (RLS)}
Alice E. Shapley,$^{2}$
Naveen A. Reddy,$^{3,4}$
Mariska Kriek,$^{5}$
\newauthor Brian Siana,$^{3}$
Alison L. Coil,$^{6}$
Bahram Mobasher,$^{3}$
Irene Shivaei,$^{7,8}$
\newauthor William R. Freeman,$^{3}$
Mojegan Azadi,$^{9}$
Sedona H. Price,$^{10}$
Gene Leung,$^{6}$
\newauthor Tara Fetherolf,$^{3}$
Laura de Groot,$^{11}$
Tom Zick,$^{5}$
Francesca M. Fornasini,$^{9}$
\newauthor and Guillermo Barro$^{12}$
\\
$^{1}$Department of Physics, University of California, Davis, One Shields Ave, Davis, CA 95616, USA\\
$^{2}$Department of Physics \& Astronomy, University of California, Los Angeles, 430 Portola Plaza, Los Angeles, CA 90095, USA\\
$^{3}$Department of Physics \& Astronomy, University of California, Riverside, 900 University Avenue, Riverside, CA 92521, USA\\
$^{4}$Alfred P. Sloan Fellow\\
$^{5}$Astronomy Department, University of California, Berkeley, CA 94720, USA\\
$^{6}$Center for Astrophysics and Space Sciences, University of California, San Diego, 9500 Gilman Dr., La Jolla, CA 92093-0424, USA\\
$^{7}$Department of Astronomy/Steward Observatory, 933 North Cherry Ave, Rm N204, Tucson, AZ, 85721-0065, USA\\
$^{8}$Hubble Fellow\\
$^{9}$Harvard-Smithsonian Center for Astrophysics, 60 Garden Street, Cambridge, MA 02138, USA\\
$^{10}$Max-Planck-Institut f{\"u}r extraterrestrische Physik, Postfach 1312, Garching, 85741, Germany\\
$^{11}$Department of Physics, The College of Wooster, 1189 Beall Avenue, Wooster, OH 44691, USA\\
$^{12}$Department of Phyics, University of the Pacific, 3601 Pacific Ave, Stockton, CA 95211, USA
}

\date{Accepted XXX. Received YYY; in original form ZZZ}

\pubyear{2019}

\begin{document}
\label{firstpage}
\pagerange{\pageref{firstpage}--\pageref{lastpage}}
\maketitle

% Abstract of the paper
\begin{abstract}
We present detections of [O\iii]$\lambda$4363 and direct-method metallicities for star-forming
 galaxies at $z=1.7-3.6$.
We combine new measurements from the MOSFIRE Deep Evolution Field (MOSDEF) survey
 with literature sources to construct a sample of 18 galaxies with
 direct-method metallicities at $z>1$, spanning $7.5<1$2+log(O/H$)<8.2$ and log(\mstar/$\msun)=7-10$.
We find that strong-line calibrations based on local analogs of high-redshift galaxies
 reliably reproduce the metallicity of the $z>1$ sample on average.
We construct the first mass-metallicity relation at $z>1$ based purely on direct-method O/H, finding
 a slope that is consistent with strong-line results.
Direct-method O/H evolves by $\lesssim0.1$ dex at fixed \mstar\ and SFR from $z\sim0-2.2$.
We employ photoionization models to constrain the ionization parameter and ionizing spectrum in the high-redshift sample.
Stellar models with super-solar O/Fe and binary evolution of massive stars are required to reproduce
 the observed strong-line ratios.
We find that the $z>1$ sample falls on the $z\sim0$ relation between ionization parameter and O/H, suggesting
 no evolution of this relation from $z\sim0$ to $z\sim2$.
These results suggest that the offset of the strong-line ratios of this sample from local excitation sequences is driven primarily by
 a harder ionizing spectrum at fixed nebular metallicity compared to what is typical at $z\sim0$, naturally
 explained by super-solar O/Fe at high redshift caused by rapid formation timescales.
Given the extreme nature of our $z>1$ sample, the implications for representative $z\sim2$ galaxy samples
 at $\sim10^{10}~\msun$ are unclear, but similarities to $z>6$ galaxies suggest that these conclusions can be
 extended to galaxies in the epoch of reionization.
\end{abstract}

% Select between one and six entries from the list of approved keywords.
% Don't make up new ones.

%%%%%%%%%%%%%%%%%%%%%%%%%%%%%%%%%%%%%%%%%%%%%%%%%%

%%%%%%%%%%%%%%%%% BODY OF PAPER %%%%%%%%%%%%%%%%%%

\section{Introduction}\label{sec:intro}

A full understanding of galaxy formation and evolution requires knowledge of the chemical enrichment of
 galaxies over cosmic time.
In particular, the gas-phase oxygen abundance (metallicity) of the interstellar medium (ISM) is a sensitive probe
 of the cycle of baryons into and out of galaxies.
The scaling of metallicity with global galaxy properties such as stellar mass (\mstar) and star-formation rate (SFR)
 can reveal the role of gas accretion, star formation, feedback, and outflows in shaping galaxy
 growth \citep[e.g.,][]{fin08,pee11,dav12,lil13}.
The shape and normalization of metallicity scaling relations also provide valuable tests of baryonic physics
 prescriptions adopted in numerical simulations \citep[e.g.,][]{ma16,dav17,dav19,der17,tor18,tor19}.
It is thus of critical importance to obtain accurate metallicity measurements of galaxies at low and high redshifts.

A robust determination of oxygen abundance can be obtained based on atomic physics via the ``direct" method.
In this approach, the electron temperature
 of the ionized gas is determined from the ratio of a weak auroral emission line (e.g., [O\iii]$\lambda$4363 in the rest-optical
 or O\iii]$\lambda\lambda$1661,1666 in the rest-ultraviolet) to a strong emission line from the same ionic species
 (e.g., [O\iii]$\lambda$5007).  In combination with hydrogen recombination lines and strong oxygen emission
 features ([O\ii]$\lambda\lambda$3726,3729, [O\iii]$\lambda$5007), the total oxygen abundance can be derived from the
 electron temperature \citep[e.g.,][]{izo06,ost06,pil12,and13,ber15,cro15,cro16,pil16}.

Since auroral lines are $\sim100\times$ fainter than the strong lines and not typically detected in the spectra
 of most galaxies in large spectroscopic surveys, empirical calibrations have been constructed relating strong-line
 ratios to direct-method metallicity using local \hii regions and $z\sim0$ galaxies with auroral-line detections
 \citep[e.g.,][]{pet04,pil05,mar13,pil16,cur17}.
Theoretical strong-line calibrations have also been developed based on photoionization models that
 reproduce the properties of local \hii regions \citep[e.g.,][]{kew02,tre04,kob04,dop16}.
Strong-line metallicity calibrations have led to a detailed understanding of galaxy metallicity scaling relations at $z\sim0$
 \citep[e.g.,][]{tre04,kew08,man10,lar10,mai19}.

Large samples of galaxies ($N>1000$) at $z>1$ with rest-optical strong-line ratio measurements are now available
 thanks to large spectroscopic surveys including the MOSFIRE Deep Evolution Field (MOSDEF) survey \citep[MOSDEF;][]{kri15},
 Keck Baryonic Structure Survey \citep[KBSS;][]{ste14}, FMOS-COSMOS \citep{kas19a},
 and 3D-HST \citep{bra12a}.
Initial work to establish the metallicities of $z>1$ galaxies across a wide range of \mstar\ and SFR has taken place
 \citep[e.g.,][]{erb06,cul14,ste14,tro14,zah14b,san15,san18,kas17,mai19}.
However, efforts to utilize these large data sets to characterize the metallicity evolution of galaxies
 as a function of redshift have been hindered by systematic uncertainties associated with the use of
 $z\sim0$ strong-line metallicity calibrations at high redshifts.

The strong-line ratio properties of $z>1$ star-forming galaxies suggest evolution in the physical conditions of the ionized gas
 in \hii regions \citep[e.g.,][]{kew13,kew13t,mas14,ste14,sha15,san16a,kas17,str17,str18}.
In particular, there is evidence that the ionization parameter, hardness of the ionizing spectrum, ISM density/pressure,
 and/or N/O ratio evolve with redshift at fixed O/H.
Much focus has been given to the position of $z\sim2$ galaxies in the [O\iii]$\lambda$5007/H$\beta$ vs.\ [N\ii]$\lambda$6584/H$\alpha$
 \citep[BPT;][]{bpt81} diagram, which is systematically offset from the $z\sim0$ star-forming sequence \citep[e.g.][]{sha05,erb06,hai09,ste14,ste16,sha15,san16a,str17,str18}.
As a result of these evolving \hii region physical conditions, metallicity calibrations
 produced from $z\sim0$ empirical datasets or from photoionization models assuming
 $z\sim0$ ISM conditions will yield systematically biased metallicities when applied to high-redshift
 samples.  The precise nature of this bias is not currently known given the lack of a consensus on the nature of the
 ISM conditions at high redshifts.

The most straightforward approach to resolving the problem of measuring metallicities at high redshifts is to obtain
 determinations of metallicity via the electron temperature method, independent of $z\sim0$ strong-line calibrations,
 to characterize the bias of $z\sim0$ calibrations and understand which low-redshift samples most reliably reproduce
 high-redshift metallicities.
Obtaining electron temperature measurements at $z>1$ has been observationally difficult due to the extreme faintness
 of temperature-sensitive auroral emission lines, but efforts over the past decade have produced a sample of
 $>10$ auroral-line measurements at $z\sim1-3$
 \citep{vil04,yua09,bra12b,chr12a,chr12b,sta13,sta14,bay14,jam14,san16b,koj17,ber18,pat18,gbu19}.
  Many of these $z>1$ auroral-line detections have taken advantage of strong gravitational lensing to boost line fluxes,
 or otherwise target particularly bright line emitters in which auroral lines can be detected with current facilities.

In this paper, we present three new detections of [O\iii]$\lambda$4363 and derive temperature-based metallicities for galaxies
 at $z=1.7-3.6$ from the MOSDEF survey \citep{kri15}, increasing the number of galaxies at $z>1$ with [O\iii]$\lambda$4363 detections
 by $\sim50\%$.
We combine these measurements with targets from the literature to construct the largest sample of $z>1$ galaxies
 with direct-method metallicities to-date ($N=18$).
We utilize this $z>1$ auroral-line sample to investigate the evolution of strong-line calibrations with redshift,
 construct metallicity scaling relations based purely on the direct method at $z>1$ for the first time,
 and place constraints on the ionization parameters and ionizing spectra present in these galaxies
 via photoionization modeling.

This paper is organized as follows.
In Section~\ref{sec:data}, we describe the observations, $z>1$ auroral-line sample, and derived galaxy properties.
Using temperature-based metallicities, we characterize the evolution of strong-line metallicity calibrations and
 metallicity scaling relations, including the mass-metallicity and fundamental metallicity relations, in
 Section~\ref{sec:empirical}.
In Section~\ref{sec:models}, we leverage the unique combination of direct-method metallicities and photoionization
 models to constrain the ionization state of the ionized ISM in star-forming galaxies at $z>1$.
However, our sample of high-redshift auroral-line emitters is not representative of typical $\sim L^*$ galaxies at $z\sim1-3$.
In Section~\ref{sec:discussion},
 we discuss the implications of our results for representative spectroscopic samples at $z\sim2$, extreme emission-line
 galaxies at $z\sim1-2$, and typical galaxies in the epoch of reionization at $z>6$.
Finally, we summarize our results and conclusions in Section~\ref{sec:summary}.

We adopt the following abbreviations for strong emission-line ratios:
\begin{equation}
\text{O}3 = [\text{O}\iii]\lambda5007/\text{H}\beta
\end{equation}
\begin{equation}
\text{O}2 = [\text{O}\ii]\lambda\lambda3726,3729/\text{H}\beta
\end{equation}
\begin{equation}
\text{Ne}3 = [\text{Ne}\iii]\lambda3869/\text{H}\beta
\end{equation}
\begin{equation}
\text{R}23 = ([\text{O}\iii]\lambda\lambda4959,5007+[\text{O}\ii]\lambda\lambda3726,3729)/\text{H}\beta
\end{equation}
\begin{equation}
\text{O}32 = [\text{O}\iii]\lambda5007/[\text{O}\ii]\lambda\lambda3726,3729
\end{equation}
\begin{equation}
\text{Ne}3\text{O}2 = [\text{Ne}\iii]\lambda3869/[\text{O}\ii]\lambda\lambda3726,3729
\end{equation}
Emission-line wavelengths are given in air, and all magnitudes are on the AB scale \citep{oke83}.
We assume a $\Lambda$CDM cosmology with H$_0$=70~km~s$^{-1}$~Mpc$^{-1}$, $\Omega_{\text{m}}$=0.3, and
 $\Omega_{\Lambda}$=0.7.

\section{Data, Measurements, \& Derived Quantities}\label{sec:data}

\subsection{The MOSDEF Survey}

The MOSDEF survey was a four-year program that utilized the MOSFIRE
 spectrograph \citep{mcl12} on the 10~m Keck~I telescope to obtain near-infrared (rest-frame optical) spectra
 of $\sim1500$ galaxies at $1.37\le z\le3.80$ \citep{kri15}.
Targets were selected in the five CANDELS fields \citep[AEGIS, COSMOS, GOODS-N, GOODS-S, and UDS;][]{gro11,koe11}
 down to limiting magnitudes of $H=24.0$, 24.5, and 25.0 at $1.37\le z\le1.70$, $2.09\le z\le2.61$,
 and $2.95\le z\le3.80$, respectively, as determined from \textit{HST}/WFC3 F160W imaging.
Strong rest-optical emission lines fall in near-infrared windows of atmospheric transmission
 in these three redshift ranges.  Target selection was performed using the photometric and spectroscopic catalogs of the 3D-HST survey
 \citep{bra12a,ske14,mom16} based on observed $H$-band magnitudes and redshifts (spectroscopic when available, otherwise
 photometric).  Completed in 2016, the full survey targeted $\sim1500$ galaxies and yielded $\sim1300$ robust redshifts:
 $\sim300$ at $z\sim1.5$, $\sim700$ at $z\sim2.3$, and $\sim300$ at $z\sim3.4$.
Exposure times were typically 1~h per filter for masks targeting galaxies in the lowest redshift bin at $z\sim1.5$, and
 2~h per filter for $z\sim2.3$ and $z\sim3.4$ masks.  A custom IDL pipeline was used to reduce the data and produce
 two-dimensional science and error spectra for each slit on a mask that has been flat-fielded, sky-subtracted,
 cleaned of cosmic rays, rectified, and wavelength- and flux-calibrated.  One-dimensional science and error spectra were optimally
 extracted \citep{hor86} using \texttt{BMEP}\footnote{
Source code and installation instructions available at https://github.com/billfreeman44/bmep
} \citep{fre19} and corrected for slit losses based on the \textit{HST}/WFC3 F160W image for each object
 convolved with the average seeing for each filter and mask combination \citep{red15}.
Full details of the survey design, execution, and data reduction can be found in \citet{kri15}.

\subsection{Measurements \& Derived Quantities}\label{sec:measurements}

We measured emission-line fluxes by fitting Gaussian profiles to the one-dimensional science spectra,
 and flux uncertainties are estimated using a Monte Carlo method by perturbing the science spectrum according
 to the error spectrum and remeasuring the flux 1,000 times.  The uncertainty in the line flux was taken to be the
 68th percentile half-width of the resulting distribution.  Redshifts were measured from the highest signal-to-noise (S/N)
 ratio line, almost always H$\alpha$ or [O\iii]$\lambda$5007.

Stellar masses (\mstar) were estimated by fitting flexible stellar population synthesis models \citep{con09} to rest-frame UV through
 near-infrared photometry using the spectral energy distribution (SED)
 fitting code \texttt{FAST} \citep{kri09}.  We assumed solar metallicity, constant star-formation
 histories, the \citet{cal00} attenuation curve, and a \citet{cha03} initial mass function (IMF).
Redshift is fixed to the spectroscopic redshift determined from strong optical emission lines.
Photometry was corrected for emission line contamination based on the measured line fluxes prior to fitting.

The best-fit SED model was used to correct for the effects of stellar absorption on hydrogen Balmer emission line fluxes.
Emission-line equivalent widths were measured using the observed line flux relative to the continuum level inferred from
 the best-fit SED model at the line center \citep{red18}.
We calculated nebular reddening using the available hydrogen Balmer emission lines, assuming a \citet{car89} extinction curve
and intrinsic ratios of H$\alpha$/H$\beta$=2.86, H$\gamma$/H$\beta$=0.468, and H$\delta$/H$\beta$=0.259 \citep{ost06}.
Star-formation rates (SFRs) were calculated from the dust-corrected luminosity of the highest S/N hydrogen Balmer line
 (always H$\alpha$ or H$\beta$) using the solar metallicity calibration of \citet{hao11} converted to a \citet{cha03} IMF.

For density and [O\iii]$\lambda$4363 temperature calculations, we adopted the effective collision strengths of \citet{tay07}
 for O\ii\ and \citet{sto14} for O\iii, with transition probabilities for both species taken from the NIST MCHF database
 \citep{fis14}.  For temperature calculations using O\iii]$\lambda\lambda$1661,1666, the \citet{sto14} collision
 strengths cannot be used since they include only five levels, and we instead employed the six-level O\iii\ collision strengths
 from \citet{agg99}.
Calculations of electron temperature from [O\iii]$\lambda$4363 using either set of O\iii\ collision strengths fall within
 $\sim3\%$ of each other.
When both components of the [O\ii]$\lambda\lambda$3726,3729 doublet are detected with S/N$\ge$3, electron density
 was calculated using a five-level atom approximation \citep{san16a} and solved iteratively with the electron temperature.
Otherwise, an electron density of $n_e$=250~cm$^{-3}$ was assumed,\footnote{When $n_e<3,000$~cm$^{-3}$, changes in density affect
 inferred O$^{2+}$ electron temperatures by at most 1\%.}
 a value typical of star-forming galaxies at $z\sim2$ \citep[e.g.,][]{san16a,ste16,str17}.
The electron temperature of the O$^{2+}$ zone (\te) was determined using the IRAF routine \texttt{nebular.temden} \citep{sha94}.
The temperature of the O$^+$ zone (\ttwo) was estimated assuming the \te-\ttwo\ relation of \citet{cam86}:
 $\mbox{T}_2=0.7\mbox{T}_3 + 3,000~\mbox{K}$.
The equations of \citet{izo06} were used to calculate ionic abundances.  The total oxygen abundance was taken to
 be $\mbox{O/H}=\mbox{O}^+/\mbox{H} + \mbox{O}^{2+}/\mbox{H}$, assuming neutral and O$^{3+}$ contributions are negligible
 within the ionized nebulae.
Assuming other \te-\ttwo\ relations \citep[e.g.,][]{pil09} changes the inferred \ttwo\ by $\lesssim10$\%, corresponding
 to changes in $\mbox{O/H}\lesssim0.05$~dex for the vast majority of our sample that are high-excitation and dominated
 by O$^{2+}$ over O$^+$.  Thus, assuming other \te-\ttwo\ relations does not affect our results.

Uncertainties on properties calculated from emission line strengths were estimated using a Monte Carlo method
 in which the measured line fluxes were perturbed according to their uncertainties and all such properties
 (reddening, equivalent widths, SFR, line ratios, density, temperature, and abundances) were remeasured 1000 times.
The uncertainty on each property was taken to be the 16th and 84th percentile of the resulting distribution.
In this way, uncertainties on line ratios, SFRs, temperatures, and abundances include errors on line fluxes and
 nebular reddening.
Uncertainties on stellar masses were estimated by perturbing the photometry according to the errors and refitting
 the SED 500 times.

For objects taken from the literature, all properties excluding stellar masses and equivalent widths were calculated
 from the observed emission-line fluxes using the same method described above, such that abundances and line ratios
 can be fairly compared.  Any exceptions to this methodology are noted in Appendices~\ref{app:o4363} and~\ref{app:o1663}.
Stellar masses and equivalent widths were taken from literature sources, with all masses converted to a \citet{cha03} IMF.

\subsection{Sample Selection}

\subsubsection{MOSDEF [O\iii]$\lambda$4363 emitters}\label{sec:MOSDEFo4363}

\begin{figure*}
 \includegraphics[width=\textwidth]{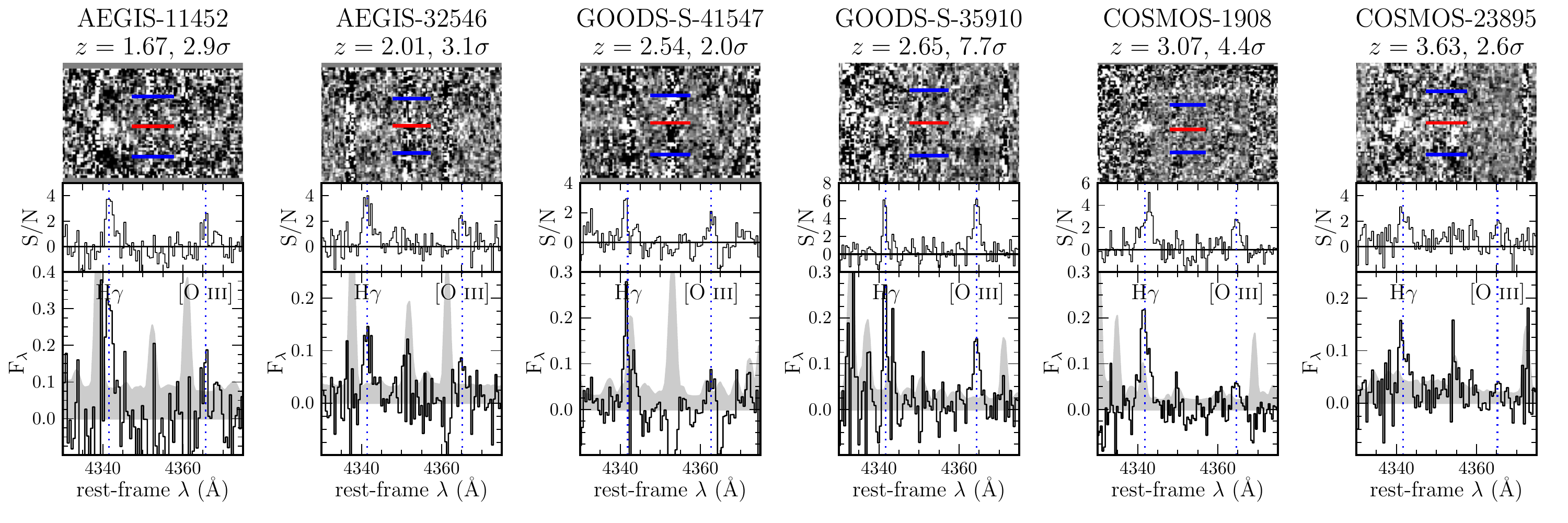}
 \centering
 \caption{
Spectra of MOSDEF targets with [O\iii]$\lambda$4363 detections.
The redshift and integrated signal-to-noise ratio (S/N) of [O\iii]$\lambda$4363 are given below the object ID.
 \textsc{Top:} Two-dimensional science spectrum.  Red and blue lines show the locations of the positive and negative traces, respectively, of the target based on the y-position of strong lines.  A positive and negatives can be identified for each [O\iii]$\lambda$4363 detection.
 \textsc{Middle:} S/N spectrum.  Positive peaks in S/N can be seen for both H$\gamma$ and [O\iii]$\lambda$4363.
 \textsc{Bottom:} One-dimensional science (black line) and error (gray shaded region) spectra.  The central wavelengths of H$\gamma$ and [O\iii]$\lambda$4363 are shown by dotted blue lines.
}\label{fig1}
\end{figure*}

We visually inspected the spectra of all MOSDEF targets with robust redshifts and wavelength coverage
 of 4364.436~\AA\ (the vacuum wavelength of [O\iii]$\lambda$4363) and searched for detections of the line.
  We found that several galaxies with formal
 S/N$>$2 for [O\iii]$\lambda$4363 in fact have spurious detections due to noise associated with sky lines.
After culling out such false-positives, this search yielded 6 detections of [O\iii]$\lambda$4363,
 the spectra of which are presented in Figure~\ref{fig1}.  The 6 sources are star-forming galaxies spanning
 $z=1.67-3.63$ with [O\iii]$\lambda$4363 significance of 2.0-7.7$\sigma$.
While two of the galaxies (GOODS-S-41547 and COSMOS-23895) have low-S/N detections of [O\iii]$\lambda$4363 (2.0$\sigma$
 and 2.6$\sigma$, respectively), these line identifications were made in conjunction with a precise determination of
 the systemic redshift and line width from multiple strong lines.
Additionally, we required identification of both postive and negative traces for [O\iii]$\lambda$4363
 at the correct $y$-pixel locations (Fig.~\ref{fig1}).
None of these galaxies host
 an active galactic nucleus (AGN) according to their X-ray and rest-frame infrared properties \citep{coi15,aza17}.
All six targets are low-mass (log(\mstar/$\msun)\lesssim9.5$) and highly star-forming (log(sSFR/Gyr$^{-1})\gtrsim0.75$).

We cannot determine direct-method metallicities for two of the six
 targets because of wavelength coverage: AEGIS-32546 at $z=2.01$ lacks coverage of H$\alpha$, H$\beta$, and [O\ii];
 GOODS-S-35910 at $z=2.65$ lacks coverage of [O\iii]$\lambda\lambda$4959,5007 and [O\ii].  The other four galaxies have
 detections of [O\iii]$\lambda$4363, [O\iii]$\lambda\lambda$4959,5007, [O\ii], H$\beta$, and at least one other
 hyrogen Balmer line, comprising the MOSDEF [O\iii]$\lambda$4363 emitter sample for which robust determinations
 of nebular reddening, electron temperature, and nebular oxygen abundance can be obtained.
The emission-line fluxes of these four galaxies are given in Table~\ref{tab:fluxes}, and their properties are
 presented in Table~\ref{tab:props}.  The [O\iii]$\lambda$4363 detection for COSMOS-1908 was previously published
 in \citet{san16b}.  We note that the emission-line fluxes and properties of COSMOS-1908 presented in this work
 differ slightly due to updates to the data reduction pipeline and SED fitting procedure.

\begin{table*}
 \centering
 \caption{Observed emission-line fluxes of MOSDEF [O\iii]$\lambda$4363 emitters in units of 10$^{-17}$ erg s$^{-1}$ cm$^{-2}$.
 }\label{tab:fluxes}
 \begin{tabular}{ l l l l l l l l l l l l }
   \hline\hline
   Object & [O\ii]$\lambda$3726 & [O\ii]$\lambda$3729 & [Ne\iii]$\lambda$3869 & H$\delta$ & H$\gamma$ & [O\iii]$\lambda$4363 & H$\beta$ & [O\iii]$\lambda$5007 & H$\alpha$ \\
   \hline
   \makecell{AEGIS\\11452} & 2.57$\pm$0.50 & 4.32$\pm$0.66 & 1.92$\pm$0.30 & 1.93$\pm$0.40 & 3.29$\pm$0.64 & 0.82$\pm$0.28 & 7.54$\pm$0.35 & 36.1$\pm$0.29 & 19.7$\pm$0.39 \\ 
   \makecell{GOODS-S\\41547} & 13.3$\pm$1.41 & 15.6$\pm$0.75 & --- & --- & 5.56$\pm$1.79 & 1.40$\pm$0.70 & 11.1$\pm$1.33 & 57.5$\pm$0.53 & 31.7$\pm$0.83 \\
   \makecell{COSMOS\\1908} & 1.06$\pm$0.16 & 1.22$\pm$0.14 & 1.74$\pm$0.19 & 1.22$\pm$0.36 & 2.29$\pm$0.28 & 0.53$\pm$0.12 & 4.82$\pm$0.25 & 33.8$\pm$0.65 & --- \\ 
   \makecell{COSMOS\\23895} & 2.17$\pm$0.35 & 2.18$\pm$0.22 & --- & --- & 1.57$\pm$0.30 & 0.26$\pm$0.10 & 1.96$\pm$0.21 & 13.1$\pm$0.28 & --- \\ 
   \hline
   \makecell{composite$^a$} & 0.10$\pm$0.01 & 0.17$\pm$0.01 & --- & --- & 0.08$\pm$0.007 & 0.02$\pm$0.003 & 0.17$\pm$0.008 & 1.00$\pm$0.007 & --- \\ 
   \hline
 \end{tabular}
 \begin{flushleft}$^a$ Line fluxes of the composite spectrum of the four MOSDEF [O\iii]$\lambda$4363 emitters are reddening-corrected and normalized to [O\iii]$\lambda$5007.
 \end{flushleft}
\end{table*}

\begin{figure*}
 \includegraphics[width=\textwidth]{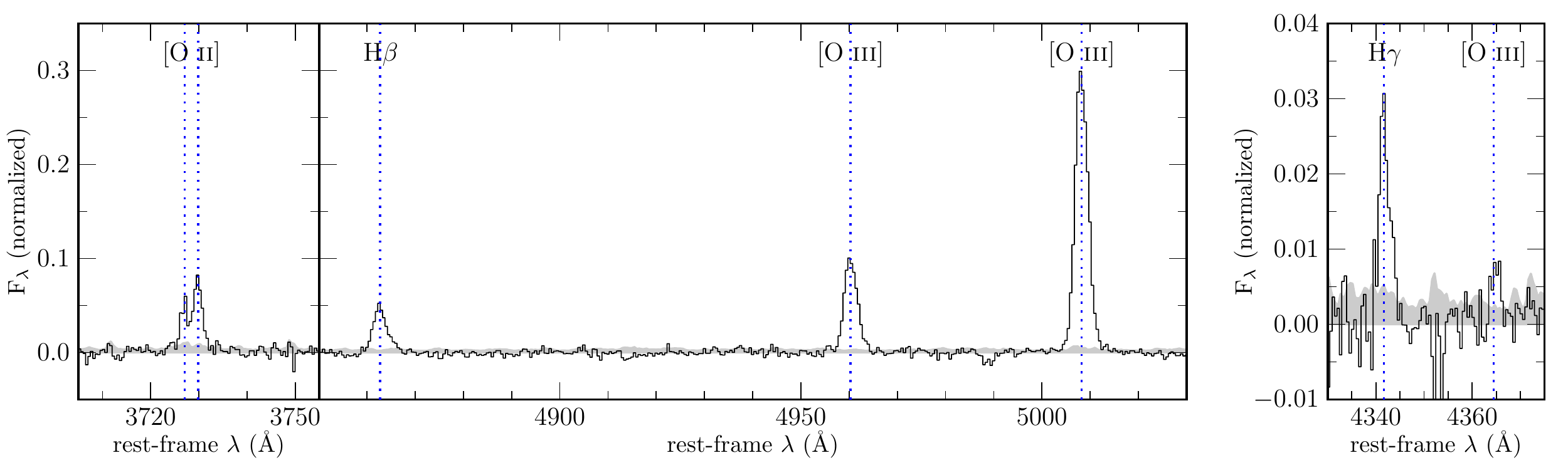}
 \centering
 \caption{
Composite spectrum of four MOSDEF targets with detections of [O\iii]$\lambda$4363 and wavelength coverage of [O\ii]$\lambda\lambda$3726,3729, H$\beta$, and [O\iii]$\lambda\lambda$4959,5007.  The gray shaded region denotes the composite error spectrum.  The left and middle panels show the strong lines, while the right panel displays a zoom-in on H$\gamma$ and [O\iii]$\lambda$4363.  Central wavelengths of emission features are identified by blue dotted lines.  The significance of [O\iii]$\lambda$4363 is 6.6$\sigma$ in the composite spectrum.
}\label{fig2}
\end{figure*}

We produced a composite spectrum of the four MOSDEF [O\iii]$\lambda$4363 emitters following the median stacking methodology
 described in \citet{san18}, with the exception that we normalized the spectra by [O\iii]$\lambda$5007
 flux instead of H$\alpha$ flux to prevent the composite electron temperature measurement from being dominated
 by the brightest [O\iii]$\lambda$5007 emitter in the sample.  The composite spectrum is displayed in
 Figure~\ref{fig2}.  The significance of [O\iii]$\lambda$4363 in the stacked spectrum is 6.6$\sigma$.
The \mstar\ and SFR of the composite were taken to be the median \mstar\ and SFR of the individual galaxies.
Emission-line equivalent widths were derived for the composite by dividing the composite line
 luminosity by the median luminosity density at line center of the set of best-fit SED models of galaxies in the composite.
We have confirmed that this methodology reproduces the average equivalent widths of samples of individual MOSDEF 
 galaxies for which the lines of interest are detected.  The emission line luminosities and properties derived
 from the MOSDEF [O\iii]$\lambda$4363 emitter composite spectrum are given in Tables~\ref{tab:fluxes}
 and~\ref{tab:props}, respectively.

\begin{landscape}
\begin{table}
 \centering
 \caption{Derived properties of the $z>1$ auroral-line sample.
 }\label{tab:props}
 \begin{tabular}{ l l l l l l l l l l l l l }
   \hline\hline
   Object & $z$ & $\log{(\frac{\mbox{M}_*}{\mbox{M}_\odot})}$ & SFR & log($\frac{\text{sSFR}}{\text{Gyr}^{-1}}$) & E(B$-$V)$_{\text{gas}}$ & $n_e$ & T$_3$ & 12+log(O/H) & S/N$_{\text{aur}}$$^{i}$  & EW$_0$([O\iii]$\lambda$5007) & EW$_0$(H$\beta$) \\
          &     &          & {\scriptsize ($\msun\mbox{~yr}^{-1}$)} &  &  & {\scriptsize (cm$^{-3}$)} & {\scriptsize (K)}  &  & {\scriptsize (\AA)} & {\scriptsize (\AA)} \\
   \hline
   \multicolumn{11}{c}{MOSDEF [O\iii]$\lambda$4363} \\
   \hline
   \makecell{AEGIS\\11452} & 1.6715 & 9.48$^{+0.08}_{-0.36}$ & 17$^{+1}_{-1}$ & 0.75$^{+0.38}_{-0.08}$ & 0.00$^{+0.00}_{-0.00}$ & $<$381$^{a}$              & 16200$^{+3100}_{-2600}$ & 7.72$^{+0.19}_{-0.16}$ & 2.9 & 348$\pm$3 & 105$\pm$5 \\
   \makecell{GOODS-S\\41547} & 2.5451 & 9.3$^{+0.13}_{-0.06}$ & 77$^{+30}_{-18}$ & 1.58$^{+0.14}_{-0.16}$ & 0.27$^{+0.13}_{-0.11}$ & 278$^{+194}_{-176}$ & 16800$^{+4700}_{-4200}$ & 7.84$^{+0.33}_{-0.24}$ & 2.0 & 448$\pm$4 & 110$\pm$13 \\
   \makecell{COSMOS\\1908} & 3.0767 & 8.93$^{+0.1}_{-0.1}$ & 53$^{+65}_{-2}$ & 1.79$^{+0.37}_{-0.06}$ & 0.00$^{+0.23}_{-0.00}$ & 280$^{+373}_{-240}$ & 13700$^{+1800}_{-1200}$ & 8.02$^{+0.10}_{-0.13}$ & 4.4 & 1845$\pm$36 & 343$\pm$18 \\
   \makecell{COSMOS\\23895} & 3.6372 & 9.43$^{+0.13}_{-0.07}$ & 32$^{+4}_{-3}$ & 1.07$^{+0.08}_{-0.14}$ & 0.00$^{+0.00}_{0.00}$ & 531$^{+482}_{-346}$ & 15300$^{+2800}_{-3000}$ & 7.99$^{+0.26}_{-0.17}$ & 2.6 & 491$\pm$11 & 98$\pm$11 \\
   \makecell{composite} & 2.81$^{b}$ & 9.37$^{+0.05}_{-0.14}$$^{b}$ & 42$^{+2}_{-2}$$^{b}$ & 1.25$^{+0.09}_{-0.11}$$^{b}$ & --- & $<$543$^{a}$ & 15400$^{+2400}_{-500}$ & 7.90$^{+0.09}_{-0.09}$ & 6.6 & 770$\pm$231 & 120$\pm$36 \\
   \hline
   \multicolumn{11}{c}{Literature [O\iii]$\lambda$4363} \\
   \hline
   \makecell{S13$^{c}$} & 1.425 & 8.33$^{+0.1}_{-0.14}$ & 7$^{+1}_{-1}$ & 1.48$^{+0.18}_{-0.09}$ & 0.03$^{+0.08}_{-0.03}$ & 201$^{+132}_{-118}$ & 15400$^{+1100}_{-1000}$ & 7.95$^{+0.07}_{-0.07}$ & 7.3 & $>$320 & --- \\
   \makecell{J14$^{c,d}$} & 1.433 & --- & 5$^{+2}_{-1}$ & --- & 0.08$^{+0.09}_{-0.07}$ & 324$^{+112}_{-120}$ & 11700$^{+2200}_{-2500}$ & 8.12$^{+0.34}_{-0.21}$ & 1.7 & --- & --- \\
   \makecell{C12a$^{c,e}$} & 1.8339 & 7.7$^{+0.1}_{-0.1}$ & 1.4$^{+3}_{-0.9}$ & 1.44$^{+0.54}_{-0.44}$ & 0.31$^{+0.32}_{-0.34}$ & 171$^{+204}_{-145}$ & 20900$^{+5900}_{-4000}$ & 7.46$^{+0.23}_{-0.22}$ & 3.4 & 482$\pm$124 & 100$\pm$26 \\
   \hline
   \multicolumn{11}{c}{Literature O\iii]$\lambda$1663} \\
   \hline
   \makecell{S14a$^{c}$} & 1.7024 & 7.78$^{+0.1}_{-0.1}$ & 0.2$^{+0.1}_{-0.1}$ & 0.52$^{+0.10}_{-0.10}$ & 0.03$^{+0.03}_{-0.03}$ & ---             & 13300$^{+700}_{-1400}$ & 8.04$^{+0.10}_{-0.08}$ & 3.5 & 1033$\pm$100 & 172$\pm$20 \\
   \makecell{B18$^{c,f}$} & 1.8443 & 8.26$^{+0.1}_{-0.1}$ & 12$^{+4}_{-1}$ & 1.82$^{+0.18}_{-0.07}$ & 0.00$^{+0.08}_{-0.00}$ & ---                 & 15400$^{+1600}_{-100}$ & 7.52$^{+0.08}_{-0.04}$ & 103 & 1575$\pm$75 & 520$\pm$40 \\
   \makecell{C12b$^{c}$} & 1.9634 & 10.57$^{+0.08}_{-0.08}$ & 10$^{+2}_{-2}$ & -0.5$^{+0.12}_{-0.10}$ & 0.23$^{+0.01}_{-0.01}$ & 82$^{+10}_{-10}$ & 12400$^{+400}_{-400}$ & 8.14$^{+0.04}_{-0.03}$ & 7.5 & --- & --- \\
   \makecell{S14b$^{c}$} & 2.0596 & 7.41$^{+0.1}_{-0.1}$ & 1.6$^{+0.3}_{-0.1}$ & 1.80$^{+0.12}_{-0.09}$ & 0.00$^{+0.01}_{-0.00}$ & ---           & 22800$^{+3500}_{-2700}$ & 7.30$^{+0.09}_{-0.14}$ & 3.5 & --- & --- \\
   \makecell{K17} & 2.159 & 9.24$^{+0.15}_{-0.15}$ & 14$^{+19}_{-1}$ & 0.90$^{+0.38}_{-0.11}$ & 0.00$^{+0.36}_{-0.00}$ & ---               & 12300$^{+6100}_{-300}$ & 8.26$^{+0.02}_{-0.30}$ & 6.5 & --- & --- \\
%   \makecell{E16c} & 2.1742 & 9.73$^{+0.1}_{-0.1}$ & 30$^{+1}_{-1}$ & 0.75$^{+0.08}_{-0.10}$ & 0.00$^{+0.00}_{-0.00}$ & ---                  & 13800$^{+700}_{-800}$ & 7.98$^{+0.08}_{-0.09}$ & 5.2 & 1051$\pm$34 & 170$\pm$10 \\
%   \makecell{E16a} & 2.1889 & 9.72$^{+0.1}_{-0.1}$ & 67$^{+21}_{-14}$ & 1.10$^{+0.15}_{-0.12}$ & 0.37$^{+0.11}_{-0.09}$ & ---                & 17500$^{+3500}_{-2100}$ & 7.90$^{+0.12}_{-0.13}$ & 12.1 & 698$\pm$68 & 80$\pm$15 \\
%   \makecell{E16b} & 2.3054 & 9.45$^{+0.1}_{-0.1}$ & 59$^{+8}_{-6}$ & 1.32$^{+0.11}_{-0.11}$ & 0.34$^{+0.05}_{-0.04}$ & ---                  & 19900$^{+2600}_{-2700}$ & 7.73$^{+0.12}_{-0.10}$ & 3.5 & 759$\pm$89 & 102$\pm$15 \\
   \makecell{Ste14c} & 2.1742 & 8.97$^{+0.1}_{-0.1}$ & 30$^{+3}_{-1}$ & 1.51$^{+0.12}_{-0.08}$ & 0.00$^{+0.04}_{-0.00}$ & 380$^{+161}_{-142}$   & 12700$^{+800}_{-400}$   & 8.15$^{+0.04}_{-0.06}$ & 5.5 & 1051$\pm$34 & 170$\pm$10 \\
   \makecell{Ste14a} & 2.1889 & 9.02$^{+0.1}_{-0.1}$ & 44$^{+9}_{-7}$ & 1.62$^{+0.12}_{-0.13}$ & 0.19$^{+0.07}_{-0.06}$ & 2899$^{+1385}_{-834}$ & 14600$^{+1400}_{-1100}$ & 8.03$^{+0.06}_{-0.06}$ & 7.5 & 698$\pm$68  & 80$\pm$15 \\
   \makecell{Ste14b} & 2.3054 & 8.87$^{+0.1}_{-0.1}$ & 27$^{+4}_{-1}$ & 1.55$^{+0.12}_{-0.08}$ & 0.00$^{+0.05}_{-0.00}$ & 781$^{+220}_{-172}$   & 12800$^{+800}_{-300}$   & 8.10$^{+0.02}_{-0.06}$ & 7.7 & 759$\pm$89  & 102$\pm$15 \\
   \makecell{V04$^{c}$} & 3.357 & 7.67$^{+0.1}_{-0.1}$$^{g}$ & 48$^{+13}_{-9}$ & 3.01$^{+0.16}_{-0.14}$$^{e}$ & 0.00$^{+0.00}_{-0.00}$ & ---                   & 17600$^{+400}_{-400}$ & 7.76$^{+0.04}_{-0.03}$ & 14.3 & --- & --- \\
   \makecell{C12c$^{c}$} & 3.5073 & 9.16$^{+0.21}_{-0.21}$ & 4$^{+8}_{-1}$ & 0.39$^{+0.60}_{-0.17}$ & 0.00$^{+0.23}_{-0.00}$ & 415$^{+301}_{-206}$ & 16200$^{+6200}_{-200}$ & 7.76$^{+0.01}_{-0.30}$ & 12.7 & --- & --- \\
   \makecell{B14$^{c}$} & 3.6252 & 9.5$^{+0.35}_{-0.35}$ & 27$^{+14}_{-8}$ & 0.93$^{+0.40}_{-0.38}$ & 0.32$^{+0.06}_{-0.06}$ & 560$^{+84}_{-73}$ & 13700$^{+1000}_{-1000}$ & 8.10$^{+0.08}_{-0.08}$ & 6.9 & --- & --- \\
   \makecell{KBSS-LM1$^{h}$} & 2.40$^{b}$ & 9.8$^{+0.3}_{-0.3}$$^{b}$ & 28$^{+1.3}_{-1.2}$ & 0.65$^{+0.3}_{-0.3}$ & 0.24$^{+0.01}_{-0.02}$ & 404$^{+67}_{-54}$ & 12400$^{+400}_{-400}$ & 8.14$^{+0.03}_{-0.03}$ & 8.6 & --- & --- \\
   \hline
 \end{tabular}
 \begin{flushleft}
 $^{a}$ {3$\sigma$ upper limit.}
 $^{b}$ {Median redshift, \mstar, and SFR of the galaxies in the composite.  sSFR is calculated from the median \mstar\ and SFR.}
 $^{c}$ {Gravitationally-lensed.}
 $^{d}$ {Also has a O\iii]$\lambda$1663 detection reported in \citet{jam14}.  We adopt the electron temperature and O/H derived from [O\iii]$\lambda$4363 (see Appendix~\ref{app}).}
 $^{e}$ {Also has a O\iii]$\lambda$1663 detection reported in \citet{chr12b}.  We adopt the electron temperature and O/H derived from [O\iii]$\lambda$4363 that yields smaller uncertainties.}
 $^{f}$ {Also has a [O\iii]$\lambda$4363 detection reported in \citet{bra12b}.  We adopt the electron temperature and O/H derived from O\iii]$\lambda$1663 that yields smaller uncertainties.}
 $^{g}$ {Lower limit on the total stellar mass and upper limit on sSFR.  See Appendix~\ref{app}.}
 $^{h}$ {Composite spectrum of 30 galaxies from the KBSS survey, presented in \citet{ste16}.  SFR and sSFR are calculated using the dust-corrected H$\alpha$ luminosity of the composite.}
 $^{i}$ {The signal-to-noise ratio on the auroral feature [O\iii]$\lambda$4363 or O\iii]$\lambda$1663, as appropriate for each subsample.}
 \end{flushleft}
\end{table}
\end{landscape}

\subsubsection{Literature [O\iii]$\lambda$4363 and O\iii]$\lambda\lambda$1661,1666 emitters}\label{sec:literature}

We have searched the literature for galaxies at $z>1$ with detections of [O\iii]$\lambda$4363,
 and found 4 targets for which detections of the requisite lines to determine both
 dust reddening and total oxygen abundance are additionally available \citep{bra12b,chr12b,sta13,jam14}.\footnote{
\citet{yua09} also reported a detection of [O\iii]$\lambda$4363 in a lensed $z=1.706$ star-forming galaxy.  The claimed detection
 appears to be a misidentification of H$\gamma$, as described in Appendix~\ref{app:y09}.  Accordingly, we do not
 include the source from \citet{yua09} in our sample.}
%These sources are SL2SJ02176-0513 \cite[B18;][]{bra12b}, Abell 1689 arc ID 31.1 \citep[C12a;][]{chr12a,chr12b},
% SDSS J0846+0446/CSWA 141 \citep[S13;][]{sta13}, and CSWA 20 \citep[J14;][]{jam14}.
Based on the work presented in \citet{koj17} and \citet{pat18}, we have additionally compiled a sample of 12 star-forming
 galaxies at $z>1$ from the literature with detections of the O\iii]$\lambda\lambda$1661,1666 doublet (hereafter O\iii]$\lambda$1663)
 with the additional detection of lines necessary to determine both the reddening correction and total oxygen abundance
 \citep{vil04,chr12a,chr12b,bay14,sta14,ste14,erb16,koj17,ber18}.
%The O\iii]$\lambda$1663 literature sample has 12 sources.
The stellar masses and SFRs of literature targets that are gravitationally lensed have been corrected for magnification, and include
 uncertainties on the magnification when available.
There are 3 literature sources that have both [O\iii]$\lambda$4363 and O\iii]$\lambda$1663 detections (J14, C12a, and B18).
The metallicities based on either auroral line are consistent with one another in each case.
We calculate the electron temperature and oxygen abundance using [O\iii]$\lambda$4363 for J14, [O\iii]$\lambda$4363 for C12a,
 and O\iii]$\lambda$1663 for B18 (see Appendix~\ref{app}).
The derived properties of the literature targets are presented in Table~\ref{tab:props}, and detailed descriptions of the calculations
 for each literature source are provided in Appendix~\ref{app}.

The literature $z>1$ auroral sample thus comprises 14 individual galaxies, and the total $z>1$ auroral-line sample size is 18
 when combined with the MOSDEF [O\iii]$\lambda$4363 emitters.
The significance of the auroral-line detections ([O\iii]$\lambda$4363 or O\iii]$\lambda$1663) of the total sample spans
 $1.7-103\sigma$, with a median value of 6.0$\sigma$ and an interquartile spanning $3.5-7.5\sigma$.
Only two objects have S/N$<$2.5 (GOODS-S-41547 and J14), and our results do not significantly change if these
 objects are excluded.
For comparison, we additionally include the KBSS-LM1 composite rest-UV and rest-optical spectrum of 30 galaxies
 at $2.1<z<2.6$ from \citet{ste16}, with a 8.6$\sigma$ O\iii]$\lambda$1663 detection.  The stellar mass of the
 KBSS-LM1 composite was taken to be the median mass of the individual galaxies.  Composites are not included in any fitting
 function or sample median calculations.

\section{Direct-method metallicity calibrations \& scaling relations at $\lowercase{z}>1$}\label{sec:empirical}

In this section, we use the largest temperature-based metallicity sample at $z>1$ to-date to assess
 the applicability of several strong-line metallicity indicators from $z\sim0-3.5$, construct the mass-metallicity
 relation based on the direct method for the first time at $z>1$, and investigate dependence of O/H on SFR.

\subsection{Metallicity calibrations at $z>1$}\label{sec:cals}

Using our sample of 18 individual galaxies at $z>1$ with auroral-line detections and temperature-based
 metallicities, we investigate whether strong-line metallicity calibrations change with redshift.
Recently, \citet{pat18} used a sample of 16 $z>1$ auroral-line emitters to test the reliability
 of various sets of strong-line metallicity calibrations at high redshifts by comparing the derived metallicity
 from each calibration to the temperature-based metallicity.  Our approach differs from that of \citeauthor{pat18}
 in that we do not test specific sets of parameterized calibrations, but instead perform a
 purely empirical comparison of strong-line ratios at fixed direct-method metallicity between various samples
 to understand which calibrating samples best match the high-redshift galaxies.  
This empirical approach was employed in earlier works by \citet{jon15} at $z\sim0.8$ and \citet{san16b} at $z\sim1-3$.
We perform our analysis on individual galaxies at $z>1$ and only include stacks
% (MOSDEF [O\iii]$\lambda$4363 emitter and KBSS-LM1 composites)
 for comparison.

\begin{figure*}
 \includegraphics[width=\textwidth]{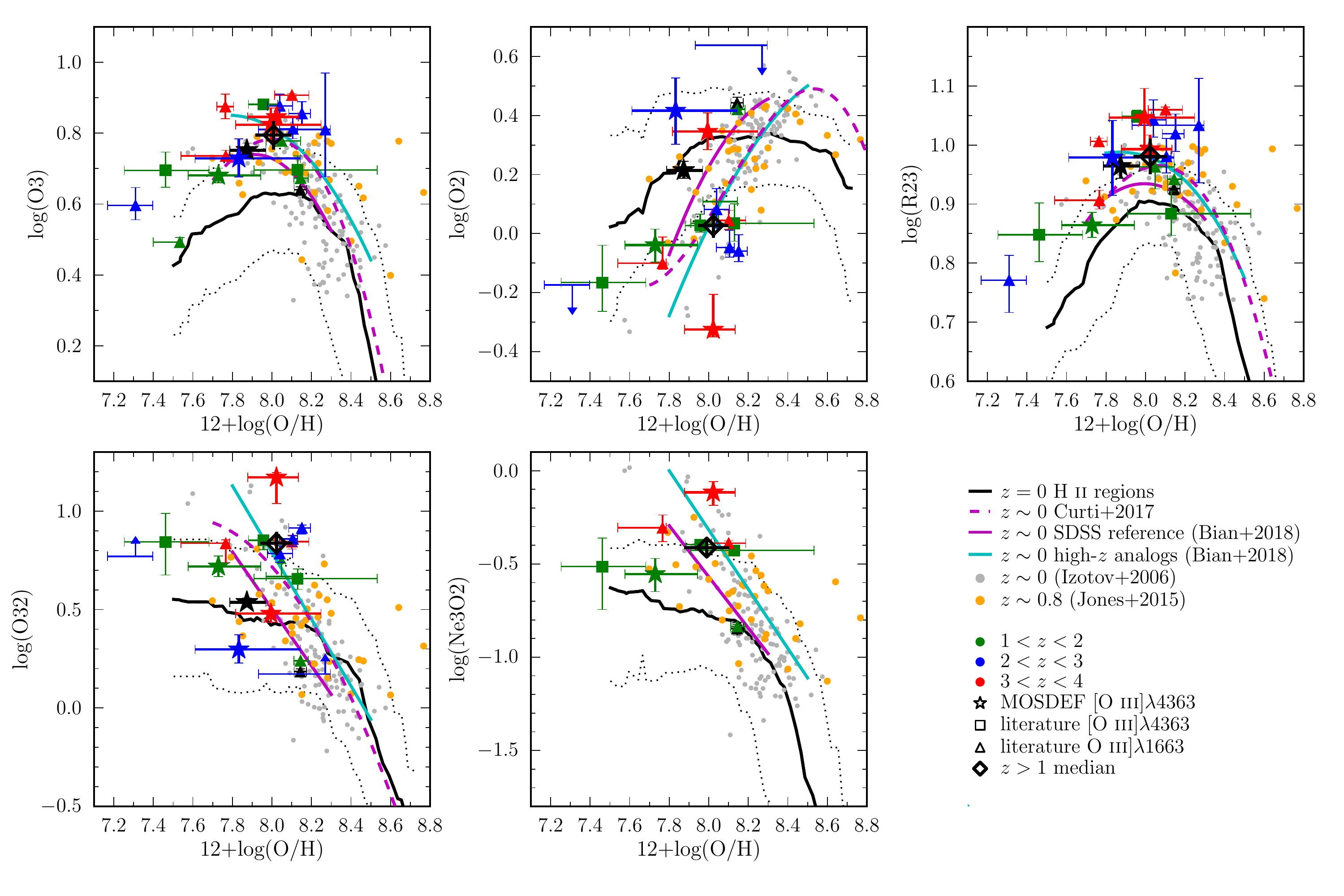}
 \centering
 \caption{
Strong optical emission-line ratios vs.\ direct-method metallicity for $z>1$ galaxies and low-redshift samples.  Points are color-coded by redshift and shapes show the auroral line used for electron temperature calculation ([O\iii]$\lambda$4363 or O\iii]$\lambda$1663).
  MOSDEF targets are given by stars.
  The composite spectra of MOSDEF [O\iii]$\lambda$4363 emitters and \citet[KBSS-LM1]{ste16}
% Steidel et al. (2016; $z\sim2.2$, based on O\iii]$\lambda$1663)
 are shown as a black star and triangle, respectively.
  The median of individual $z>1$ galaxies is given by the black open diamond in each panel.
  The low-redshift comparison samples include galaxies at $z\sim0$ \citep[gray circles;][]{izo06} and $z\sim0.8$ \citep[orange circles;][]{jon15}. 
  The cyan line shows the best-fit relations of local analogs of $z\sim2$ galaxies from \citet{bia18}, while the magenta line displays
 their local reference sample.
  The dashed magenta line shows the $z\sim0$ calibrations of \citet{cur17}.
  The median relation and 68th percentile scatter of individual $z=0$ H\ii\ regions are displayed,
 respectively, by the solid and dotted black lines.
  The median of the $z>1$ galaxies is well-matched to the $z\sim0$ \citet{izo06} and $z\sim0.8$ \citet{jon15} samples, and closely matches the \citet{bia18} high-redshift analog relations for all 5 line ratios.
}\label{fig3}
\end{figure*}

We present strong-line ratios as a function of temperature-based oxygen abundance for the $z>1$ auroral-line
 sample in Figure~\ref{fig3}.
Galaxies are color-coded by redshift, while shapes encode whether [O\iii]$\lambda$4363 or O\iii]$\lambda$1663
 was used to derive the nebular metallicity.
Here, we investigate the behavior of five line ratios: O3, O2, R23, O32, and Ne3O2.
We do not explore the strong-line ratios involving [N\ii] at this time because only 4 $z>1$ galaxies have
 detections of [N\ii]$\lambda$6584, while over half of the sample (10/18) does not have spectral coverage of [N\ii].
It is therefore not possible to draw confident conclusions about nitrogen-based metallicity indicators with
 the current $z>1$ auroral-line sample.
The five line-ratios presented in Figure~\ref{fig3} have the advantage of being accessible from the ground to $z=3.80$,
 and will be observable with \textit{JWST}/NIRCam and NIRSpec spectroscopy to $z\sim9$.

The small sample size ($N=18$), large typical uncertainties in O/H, and small dynamic range in metallicity
 (7.5$\lesssim$12+log(O/H)$\lesssim$8.1; 0.07-0.25~Z$_{\odot}$) of the $z>1$ galaxies precludes the construction
 of new calibrations based on the high-redshift sample alone.  While we cannot reliably constrain the shape
 of high-redshift metallicity calibrations, we can use mean sample properties to look for evolution in the normalization
 of calibrations at 12+log(O/H$)\sim8.0$ (0.2~Z$_{\odot}$).
In each panel of Figure~\ref{fig3}, we calculate the median 12+log(O/H) and line ratio of $z>1$ galaxies with detections of that
 line ratio, displayed as black open diamonds.  Limits and stacks are not included in median calculations.
  We note that there are at most 3 limits in
 any particular panel.  We determine the uncertainty on median values by bootstrap resampling, perturbing the data
 points according to the uncertainties, and remeasuring the median 500 times.  The $1\sigma$ confidence bounds are taken
 from the 68th percentile width of the resulting distribution.  We compare the $z>1$ sample and medians to low- and
 intermediate-redshift samples to investigate potential evolution: individual $z=0$ \hii regions \citep[black line;][]{pil16,san17},
 $z\sim0$ star-forming galaxies from the Sloan Digital Sky Survey \citep[SDSS, gray points;][]{izo06}, and $z\sim0.8$ star-forming galaxies from DEEP2
 \citep[orange points;][]{jon15}.  The oxygen abundances of each sample have been recalculated using our methodology.

Considering the $z>1$ sample alone, we find that O3 and R23 are saturated over the range of metallicity spanned by
 the bulk of the high-redshift galaxies (7.7$<$12+log(O/H)$<$8.1; 0.1$-$0.25~Z$_{\odot}$), which reside
 in the turnover regime between the ``upper" and ``lower" branches of these indices.
O3 and R23 thus cannot be utilized as reliable metallicity indicators for high-redshift galaxies falling in this oxygen
 abundance range, even when used in combination with an additional line-ratio to break the upper/lower branch
 degeneracy.
Earlier works have highlighted similar issues with obtaining metallicities from R23 at $z\sim2$ \citep{ste14,san18}.
The $z>1$ galaxies display a wide dynamic range in O2, but the scatter in line ratio at fixed O/H is largest for O2,
 suggesting that O2 is a poor metallicity indicator at $z>1$ as well.
The large scatter in O2 at fixed O/H may signify significant galaxy-to-galaxy variation in the nebular attenuation curve at $z>1$.

In contrast, O32 and Ne3O2 have a monotonic dependence on direct-method metallicity for all of the samples
 displayed in Figure~\ref{fig3}.  The number of $z>1$ galaxies with [Ne\iii] detections is small and thus
 scatter is difficult to evaluate, but the scatter in O32 at fixed O/H of the $z>1$ sample is comparable to that
 of the lower-redshift samples after accounting for measurement uncertainties (typically much larger at $z>1$).
O32 thus appears to be the best strong-line metallicity indicator for $z>1$ samples with nebular metallicities
 $\lesssim1/3$~Z$_{\odot}$, displaying the highest utility over the range of metallicities in our sample.
Given the tight relation between O32 and Ne3O2 at sub-solar metallicities \citep{lev14,str17},
 Ne3O2 should perform equally well and has the additional advantage of not requiring a reddening-correction.

The $z>1$ auroral-line sample displays significantly higher excitation than local \hii regions at fixed O/H,
 with higher O3, O32, and Ne3O2 and lower O2 on average.
  The $z>1$ sample also has significantly higher R23 values than are typical of the local \hii regions.
While in reasonable agreement at 12+log(O/H$)\sim8.3$, the \hii regions diverge from the $z\sim0$ and $z>1$
 galaxy samples at 12+log(O/H$)<8.0$, flattening in O32 and Ne3O2 and dropping off steeply in O3 and R23.
This divergent behavior may indicate that the local \hii region sample is significantly incomplete in
 the metal-poor regime, lacking high-excitation star-forming regions that must be present in star-forming
 galaxies with high O32 based on integrated spectra.
We caution against utilizing empirical calibrations based on individual local \hii regions at oxygen abundances
 below 0.2~Z$_{\odot}$ (12+log(O/H)=8.0) for either low- or high-redshift galaxies.

\citet[][hereafter BKD18]{bia18} recently constructed metallicity calibrations for use at high redshifts by stacking spectra
 of $z\sim0$ galaxies from SDSS that fall in the same region of the BPT diagram as $z\sim2$ galaxies, and
 deriving temperature-based metallicities from these stacks.
Their approach differs from past studies that have selected analogs based on global galaxy properties
 \citep[e.g., sSFR;][]{bro16,cow16} in that BKD18 have selected high-redshift analogs based
 directly on strong-line ratios that should be more closely linked to the local gas conditions in \hii regions.
Stacks of a local reference sample that traces the average $z\sim0$ BPT sequence were also constructed.
Functional fits to the BKD18 high-redshift analogs and $z\sim0$ reference sample are shown in
 Figure~\ref{fig3} (cyan and magenta lines, respectively).
We find that the median of the $z>1$ sample (black diamonds) closely matches the high-redshift analogs in
 O3, O2, R23, and O32 at fixed O/H, being $\gtrsim$2$\sigma$ inconsistent with the $z\sim0$ reference sample in each case.
For Ne3O2, the $z>1$ median falls midway between the high-redshift analog and $z\sim0$ reference lines, displaying
 1$\sigma$ consistency with both.  [Ne\iii] detections were only reported for 8 out of 18 galaxies in our $z>1$
 sample.  Consequently, the median O/H uncertainty is largest in the Ne3O2 panel.
Collectively, the panels in Figure~\ref{fig3} demonstrate that the high-redshift analog calibrations of
 BKD18 reliably reproduce the properties of the $z>1$ auroral-line sample on average, while the
 $z\sim0$ reference calibrations fail to do so.

We additionally show the $z\sim0$ empirical metallicity calibrations of \citet[][hereafter C17]{cur17} in Figure~\ref{fig3}
 (dashed magenta lines).
The $z>1$ median values display good agreement with the C17 $z\sim0$ O3, O2, R23, and O32 calibrations.
At first glance, this consistency appears to suggest negligible evolution in strong-line calibrations with redshift.
However, this is not the case when the calibration sample of C17 is considered.
C17 stacked SDSS galaxies in bins of strong-line ratio in the O3 vs.\ O2 diagram.
The composites only reach down to 12+log(O/H$)\approx8.1$, and C17 supplemented the stacked sample
 using individual SDSS galaxies with [O\iii]$\lambda$4363 detected at $>$10$\sigma$ to extend the calibrations to
 12+log(O/H)$\approx$7.7.
The C17 fits are dominated by this individual [O\iii]$\lambda$4363-detected sample at 12+log(O/H$)\lesssim8.4$,
 the region of overlap with our $z>1$ sample and the BKD18 calibrations.
The stringent S/N cut on the faint auroral line leads to a sample that is biased significantly high in sSFR
 compared to typical $z\sim0$ galaxies, with 80\% lying $\ge0.7$~dex above the mean $z\sim0$ \mstar-SFR relation and the
 bulk of the sample scattering around the $z\sim2$ \mstar-SFR relation and coincident with the BKD18 high-redshift
 analogs (see Appendix~\ref{app:c17}).
Thus, at 12+log(O/H$)\lesssim8.4$, the C17 calibration sample is largely composed of high-redshift analogs
 that have ISM conditions shifted away from the local norm and towards what is typical at $z>1$.
The coincidence of the C17 calibrations with the $z>1$ median values in Figure~\ref{fig3} is therefore
 \textit{not} indicative of redshift invariant metallicity calibrations, but is instead a reflection of
 the C17 calibrations tracing normal $z\sim0$ ISM conditions at high metallicities (12+log(O/H$)\gtrsim8.4$)
 and extreme ISM conditions typical of $z>1$ at low metallicities (12+log(O/H$)\lesssim8.4$). 

Interestingly, the \citet{izo06} $z\sim0$ SDSS sample does not follow the BKD18 local reference sample,
 but instead lies in between the reference and high-redshift analog lines.
Similarly to the case of the C17
 [O\iii]$\lambda$4363-detected SDSS sample, this trend is due to the
 requirement of a [O\iii]$\lambda$4363 detection and large H$\beta$ flux by \citet{izo06}, which preferentially
 selects high-sSFR and high-excitation galaxies with bright [O\iii]$\lambda$4363.  Such galaxies likely
 have ISM conditions that are shifted away from the local average and towards what is typical at high redshift.
Thus, the \citet{izo06} sample is not representative of $z\sim0$ galaxies, but instead matches the $z>1$ sample suitably
 well within the current uncertainties.  \citet{jon15} have previously shown that their [O\iii]$\lambda$4363
 sample at $z\sim0.8$ matches the \citet{izo06} sample in these line-ratio vs. O/H spaces, and we find that
 the $z\sim0.8$ galaxies also agree with the $z>1$ sample on average.

In summary, the median O3, O2, R23, O32, and Ne3O2 of 18 individual $z>1$ galaxies agree with the values for
 the \citet{izo06} $z\sim0$, \citet{jon15} $z\sim0.8$, and BKD18 high-redshift analog samples at fixed
 temperature-based O/H.  In contrast, the $z>1$ sample displays systematically higher excitation at
 fixed O/H compared to the BKD18 SDSS local reference sample and individual $z\sim0$ \hii regions.
Calibrations based on $z=0$ \hii regions are not reliable at $z>1$ for 12+log(O/H$)\lesssim8.0$ (0.2~Z$_{\odot}$).
Calibrations based on typical $z\sim0$ galaxies tend to underestimate O/H at $z>1$ by $\sim0.1-0.3$ dex on average.
The BKD18 high-redshift analog O32 and Ne3O2 calibrations display the highest utility for application at
 high redshifts, reproducing the metallicities of our $z>1$ auroral-line sample to within $\sim0.1$~dex on average
 while maintaining a monotonic sensitivity to O/H.

We stress that these conclusions only apply over the range of metallicities probed by our $z>1$ auroral-line sample 
(7.5$\lesssim$12+log(O/H)$\lesssim$8.1; 0.07-0.25~Z$_{\odot}$), and only for galaxies with similar properties
 (i.e., low \mstar\ and high sSFR).
The $z>1$ sample presented in Figure~\ref{fig3} is not representative of typical galaxies at $z\sim1-3$ in large
 rest-optical spectroscopic surveys (e.g., MOSDEF, KBSS, 3D-HST, FMOS-COSMOS), which have log(\mstar/$\msun)\sim10.0$
 and lie on the mean \mstar-SFR relation \citep{kri15,ste14,bra12a,kas19a}.
Based on commonly applied $z\sim0$ strong-line methods, the majority of $z\sim2$ galaxies at log(\mstar/$\msun)>9.5$
 are expected to have higher O/H than our $z>1$ auroral-line sample \citep{ste14,san15,san18}, even when allowing for
 a factor of two systematic uncertainty in strong-line metallicities.
We discuss implications for the typical $z\sim1-3$ galaxy population in Section~\ref{sec:typicalcomparison}.
Additionally, our conclusions regarding metallicity calibration evolution are based on the mean properties of the $z>1$ sample,
 and a detailed analysis of the scatter awaits a larger sample.
We thus caution that the stated $<0.1$~dex precision of the BKD18 high-redshift analog calibrations is applicable
 for population averages only, and that the metallicity of an individual galaxy may not be well-determined using this method.

\subsection{Direct-method metallicity scaling relations at $z>1$}

The $z>1$ auroral-line sample spans a wide dynamic range in stellar mass, with log(\mstar/$\msun)\sim7.5-10.0$.
We use this sample to explore the dependence of gas-phase oxygen abundance on stellar mass based entirely on
 direct-method metallicities for the first time at $z>1$, and constrain the evolution of the mass-metallicity
 relation (MZR) from $z\sim0$ to $z\sim2$.
Two galaxies in the $z>1$ auroral-line sample (C12b and C12c) are identified as possible ongoing mergers
 due to morphology and close companions \citep{chr12a,pat16}.
Mergers in the local universe have been shown to deviate from the mean $z\sim0$ MZR, having systematically lower
 metallicities and higher SFRs at fixed \mstar\ \citep{ell08,scu12}.
In contrast, recent theoretical and observational work suggests that the SFR and O/H of close galaxy pairs may not be systematically offset
 from isolated galaxies at $z\sim2$ \citep{fen17,wil19}.
However, it is not clear if the two potential mergers in our sample fall into the merger stage and pair separation over
 which such conclusions are valid.
To avoid introducing systematic biases from these merging systems, we exclude them from the sample used to
 construct metallicity scaling relations with global galaxy properties (SFR and \mstar).
We show these two excluded mergers as open triangles in the figures in this section and discuss their behavior
 in comparison to the rest of the $z>1$ sample where appropriate.
We also exclude J14, which lacks a reliable \mstar\ estimate, and V04, which does not have a total \mstar\ determination available (see Appendix~\ref{app}).

In the sections below, we investigate the shape and evolution of the MZR and its dependence on SFR.
In Section~\ref{sec:mzr}, we present the MZR at $z\sim2.2$ based purely on direct-method metallicities.
We begin with the correlation between \mstar\ and O/H displayed by the derived properties given in Table~\ref{tab:props},
 and then correct for biases in the SFR and redshift distributions of the $z>1$ auroral-line sample to produce a
 direct-method MZR representative of typical star-forming galaxies at $z\sim2$ (equation~\ref{eq:mzrtotcorrz}).
In Section~\ref{sec:fmr}, we investigate the evolution of the relation between \mstar, SFR, and direct-method O/H from $z\sim0-2.2$.
We consider systematic effects on the comparison of galaxy properties of the low- and high-redshift samples in Section~\ref{sec:systematics}.

\subsubsection{The direct-method $z\sim2.2$ mass-metallicity relation}\label{sec:mzr}

\begin{figure*}
 \includegraphics[width=0.495\textwidth]{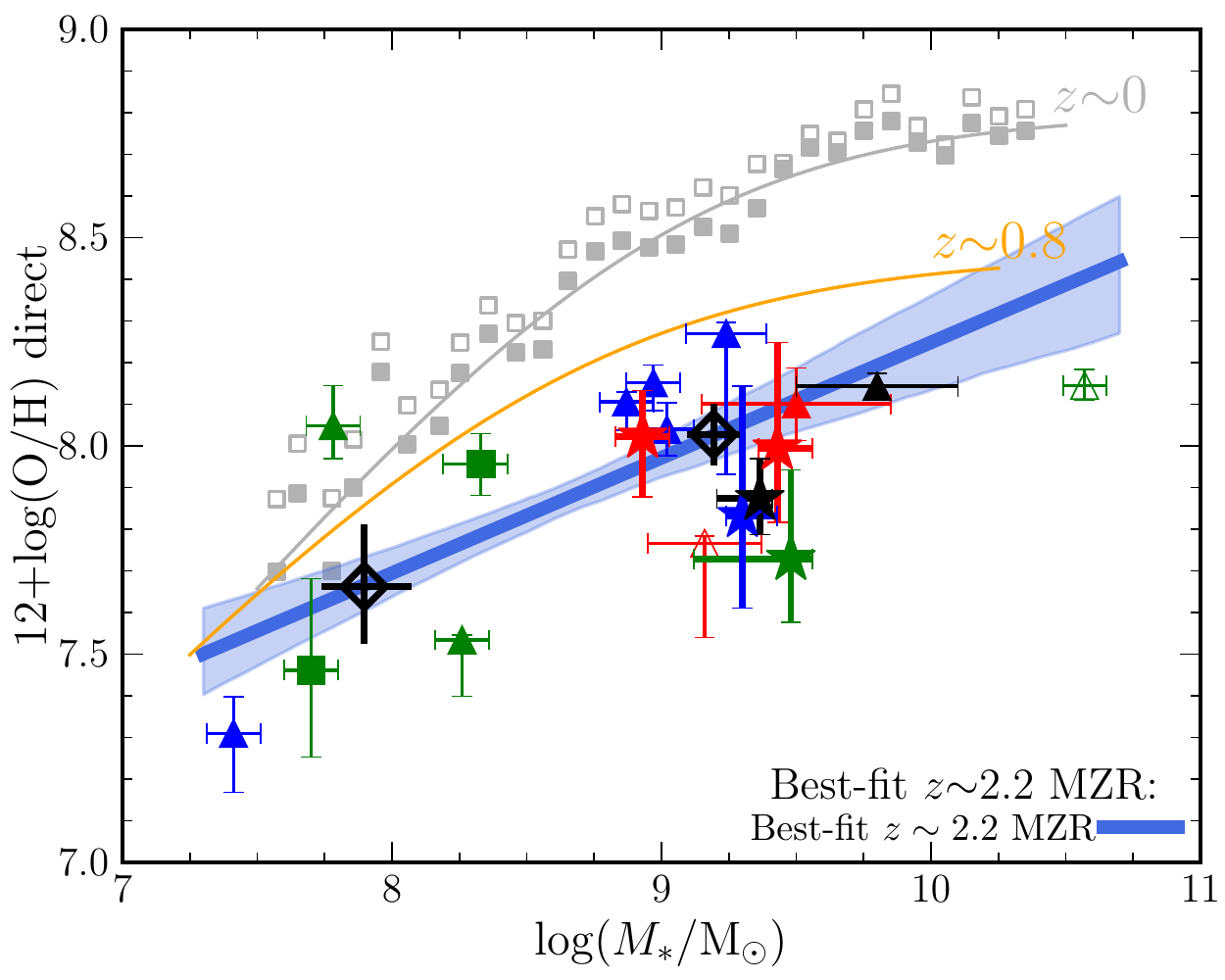}
 \includegraphics[width=0.495\textwidth]{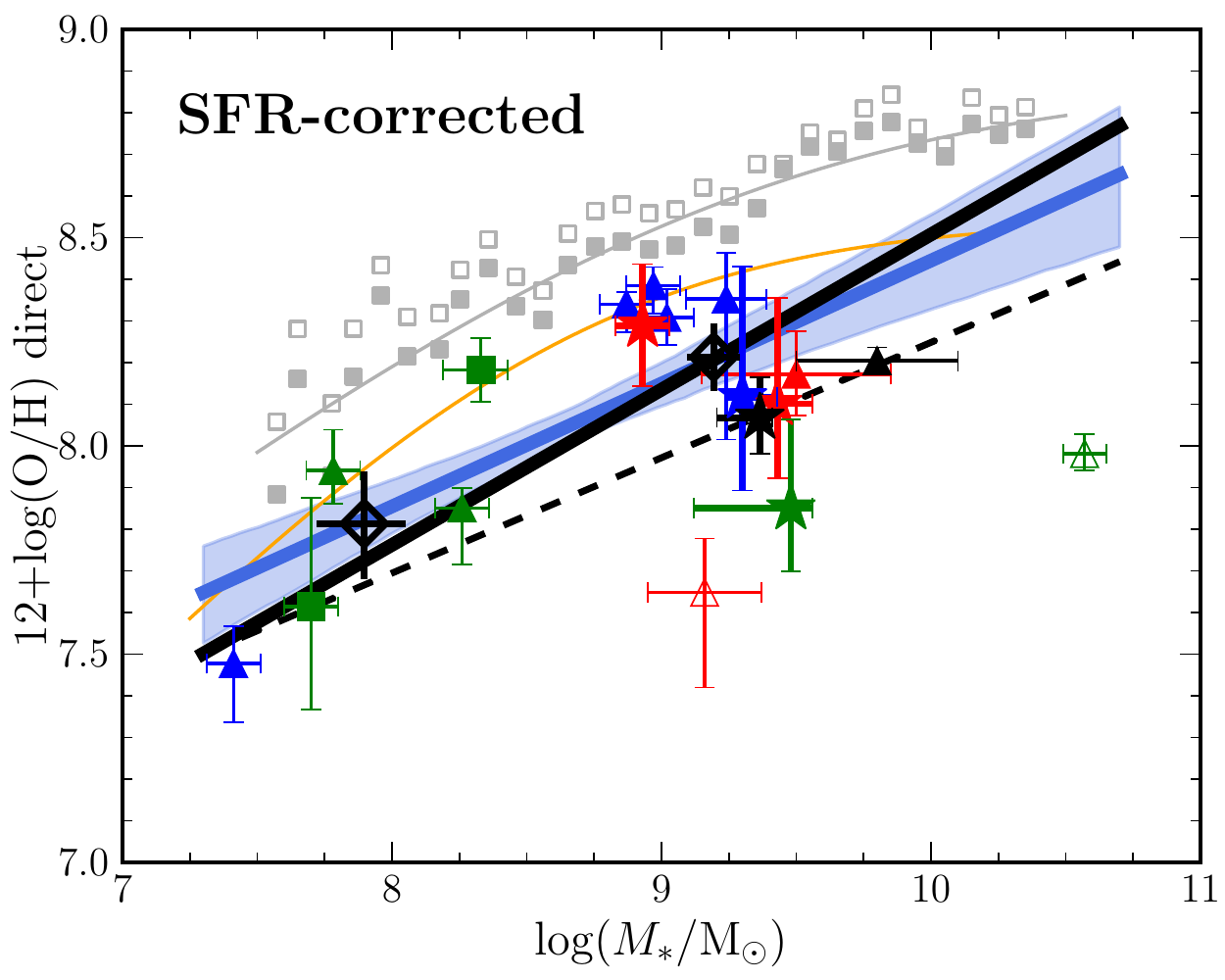}
 \centering
 \caption{
\textsc{Left:} The stellar mass---gas-phase metallicity relation (MZR) at $z\sim2.2$ using electron temperature measurements.
  Points with error bars are as in Figure~\ref{fig3}.
  Hollow triangles denote the two galaxies identified as probable mergers (C12b, C12c), which are excluded when fitting the MZR.
  The solid blue line shows the best-fit linear relation to individual $z>1$ galaxies (equation~\ref{eq:mzrtotuncorr}), with the 1$\sigma$ confidence interval given by the blue-shaded region.
%  The dashed and dotted blue lines denote the best-fit relations to the subset of $z>1$ galaxies with only [O\iii]$\lambda$4363 or O\iii]$\lambda$1663 detections, respectively, and are not significantly different than the best fit to the total sample.
  The mean values in two bins in stellar mass divided at log($M_*$/M$_{\odot})=8.5$ are displayed as black open diamonds.
  The orange line shows the direct-method MZR at $z\sim0.8$ from \citet{ly16}.
  The $z\sim0$ \mstar-binned stacks of \citet{and13} are presented as gray squares, where the filled squares have been corrected for
 DIG emission and flux-weighting effects using the models of \citet{san17}.
  The best-fit $z\sim0$ MZR is given by the gray line.
\textsc{Right:} The MZR based on direct-method metallicities, where O/H has been corrected for biases in SFR of each sample relative
 to the mean \mstar-SFR relation at each redshift.
  Corrected O/H values should represent typical galaxies on the \mstar-SFR relation at each redshift. 
  The blue line in the right panel shows the best-fit $z\sim2.2$ MZR to the SFR-corrected sample (equation~\ref{eq:mzrtotcorr}).
%total, [O\iii]$\lambda$4363, and O\iii]$\lambda$1663 samples.
  The uncorrected best-fit MZR from the left panel is displayed as a black dashed line, lying $\sim0.2$~dex below the SFR-corrected MZR.
  The solid black line displays the best-fit $z\sim2.2$ MZR after additionally correcting for the low-redshift bias of the low-mass bin (equation~\ref{eq:mzrtotcorrz}),
 and is the most robust representation of the MZR at $z\sim2.2$ presented here.
}\label{fig4}
\end{figure*}

With the remaining high-redshift auroral-line sample of 14 galaxies, we construct the direct-method MZR at $z>1$ for the
 first time, shown in the left panel of Figure~\ref{fig4}.
The median redshift of the $z>1$ MZR sample is $z_{\text{med}}=2.17$.
We calculate mean values in two bins of stellar mass divided at
 log(\mstar/$\msun)=8.5$, with 5 galaxies in the low-mass bin and 9 galaxies in the high-mass bin. 
The binned means are shown in Figure~\ref{fig4} as black open diamonds, with
 log($M_*^{\text{mean}}/\msun)=[7.90\pm0.16, 9.19\pm0.10]$
 and 12+log(O/H$)^{\text{mean}}=[7.66\pm0.14, 8.03\pm0.08]$ for the low- and high-mass bins, respectively.
We observe a positive correlation between direct-method O/H and \mstar\ for both the means and individual galaxies,
 though with significant scatter for the latter.
Using an orthogonal distance regression, we fit a linear relation to the 14 individual $z>1$ galaxies
 (excluding the open triangles and stacks), obtaining:
\begin{equation}\label{eq:mzrtotuncorr}
\begin{multlined}
12+\text{log(O/H)} \\ = (0.277\pm0.081)\times \text{log}(M_*/\msun) + (5.48\pm0.70)
\end{multlined}
\end{equation}
with a strong covariance between slope and intercept of $\rho=0.9975$.
The best-fit line is shown in the left panel of Figure~\ref{fig4} as a solid blue line. 
%We also fit lines to the subset of galaxies with [O\iii]$\lambda$4363 detections and the subset with O\iii]$\lambda$1663
% detections, and obtain consistent results (dashed and dotted blue lines).
The $z>1$ mergers do not appear to be strong outliers compared to the full $z>1$ sample.

We compare to two $z<1$ samples with auroral-line detections to investigate metallicity evolution at fixed \mstar.
The \mstar-binned $z\sim0$ SDSS stacks of \citet{and13} are shown as gray squares in the left panel of Figure~\ref{fig4}.
The open squares show the values derived by \citet{and13}, while the filled squares denote the derived metallicities
 after correcting for contamination from diffuse ionized gas (DIG) according to \citet{san17}, shifted $\sim0.05$ dex lower in O/H.
The orange line shows the best-fit MZR to a sample of $z\sim0.8$ galaxies with [O\iii]$\lambda$4363 detections from \citet{ly16}.
The $z\sim0.8$ MZR lies approximately midway between the $z\sim0$ and $z\sim2.2$ MZRs.
Compared to $z\sim0$, we find that O/H at fixed \mstar\ is 0.3~dex lower at $z\sim0.8$ and 0.55~dex lower at $z\sim2.2$
 at log(\mstar/$\msun)\approx9.2$, where the majority of the $z>1$ auroral-line sample lies.
This amount of metallicity evolution is larger than is seen in studies using strong-line methods on representative samples
 of galaxies at similar \mstar, which find $0.3-0.4$~dex evolution between $z\sim0$ and $z\sim2.3$
 \citep{erb06,ste14,san15,san18}.

A major concern when comparing the gas-phase metallicities of different samples is how representative
 the SFRs are over the range of \mstar.  In the local universe, there is a well-established relationship
 between \mstar, SFR, and O/H such that galaxies with higher SFR have lower O/H at fixed \mstar,
 known as the ``fundamental metallicity relation" \citep[FMR; e.g.,][]{man10,lar10,yat12,and13,sal14,cre18}.
A \mstar-SFR-O/H relation has also been shown to exist at $z\sim1-2.5$ using strong-line metallicities
 with a similar strength of SFR dependence as at $z\sim0$ \citep{zah14b,san18}.
Thus, if a sample is biased in SFR at fixed \mstar, then its metallicity will not be representative of typical
 galaxies at the same \mstar.
Samples selected to have detections of [O\iii]$\lambda$4363 are often biased towards higher SFR at fixed \mstar\ due to the faintness
 of the feature, and thus the MZR evolution displayed in the left panel of Figure~\ref{fig4} may be artifically large.

\begin{figure}
 \includegraphics[width=\columnwidth]{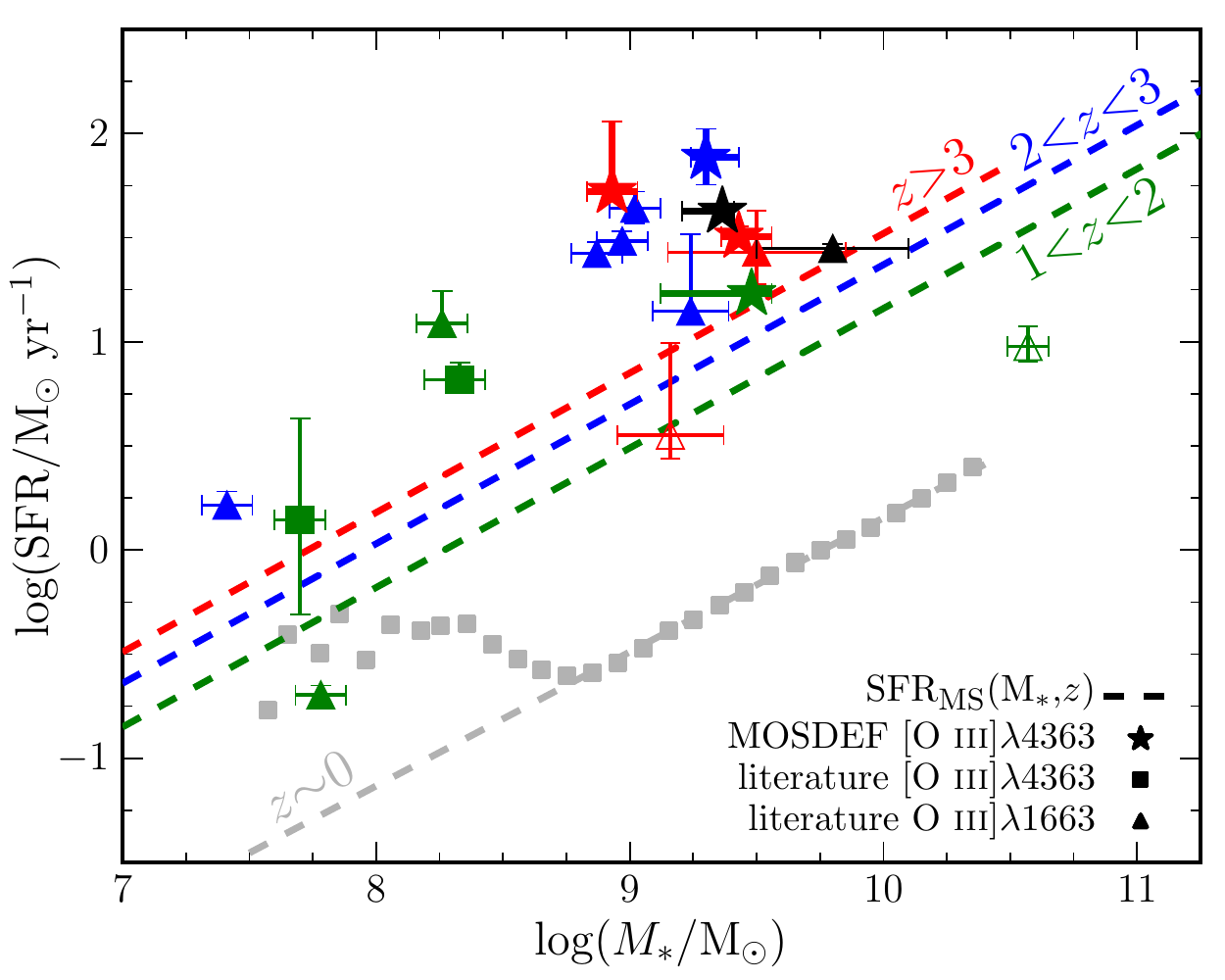}
 \centering
 \caption{
SFR vs.\ \mstar\ for the $z>1$ auroral-line sample.
  Points with error bars are as in Figure~\ref{fig3}.
  The dashed blue line shows the best-fit SFR-\mstar\ relation at $z\sim2.3$ from \citet{san18}.
  The green and red dashed lines show the SFR-\mstar\ relation at $z\sim1.5$ and $z\sim3.4$, respectively, constructed by shifting the normalization of the $z\sim2.3$ relation by $\Delta$log(SFR) as given by the model of \citet{spe14}.
  The majority of the $z>1$ sample is significantly elevated above the mean \mstar-SFR relation in each redshift bin.
  The gray dashed line show the best-fit SFR-\mstar\ relation to the \citet{and13} \mstar-binned stacks at log($M_*$/M$_{\odot})>9.0$.
  The best-fit $z\sim0$ relation displays a nearly identical slope to the $z\sim2.3$ relation despite being fit independently.
  The \citet{and13} stacks below log($M_*$/M$_{\odot})=8.75$ display a large bias in SFR at fixed \mstar.
}\label{fig5}
\end{figure}

In Figure~\ref{fig5}, we place the $z>1$ auroral-line sample on the \mstar-SFR diagram.
We compare to the best-fit \mstar-SFR relation from \citet{san18} for a representative sample of $z\sim2.3$
 star-forming galaxies with log(\mstar/$\msun)=9.0-11.0$ from MOSDEF,
 with SFRs derived from dust-corrected H$\alpha$ luminosities (dashed blue line):
\begin{equation}\label{eq:sfrmstar}
\text{log}(\text{SFR}/\msun~\text{yr}^{-1}) = 0.67\times \text{log}(M_*/\msun) - 5.33
\end{equation}
Given the redshift range of the $z>1$ auroral line sample ($1.4<z<3.7$) and the evolution of the \mstar-SFR
 relation with redshift, we fix the slope of the \mstar-SFR relation to that of equation~\ref{eq:sfrmstar} and shift the normalization to estimate mean \mstar-SFR
 relations at $z\sim1.5$ (green dashed line) and $z\sim3.4$ (red dashed line) with offsets calculated
 according to the best-fit SFR(\mstar,~$z$) relation of \citet{spe14} ($-0.21$~dex for $z\sim1.5$ and $+0.15$~dex for
 $z\sim3.4$).
We find that the $z>1$ galaxies have significantly higher SFR at fixed \mstar\ than is typical for galaxies at the same redshifts,
 with only one galaxy in the $z>1$ MZR sample falling below its corresponding \mstar-SFR relation.

We plot the median \mstar\ and SFR values of the \citet{and13} stacks in Figure~\ref{fig5} (gray squares), and find that their SDSS sample
 is significantly biased in SFR at log(\mstar/$\msun)<8.75$.  We fit a linear relation to the \citet{and13} stacks
 above log($M_*$/$\msun)=9.0$ to determine the mean $z\sim0$ \mstar-SFR relation (gray dashed line),
 recovering a slope that is nearly identical to that of the $z\sim2.3$ relation from \citet{san18}.
The $z\sim0.8$ MZR sample has been shown to be biased in SFR $\sim0.3$~dex higher than representative samples at
 the same redshift \citep{ly16}.

\begin{figure}
 \includegraphics[width=\columnwidth]{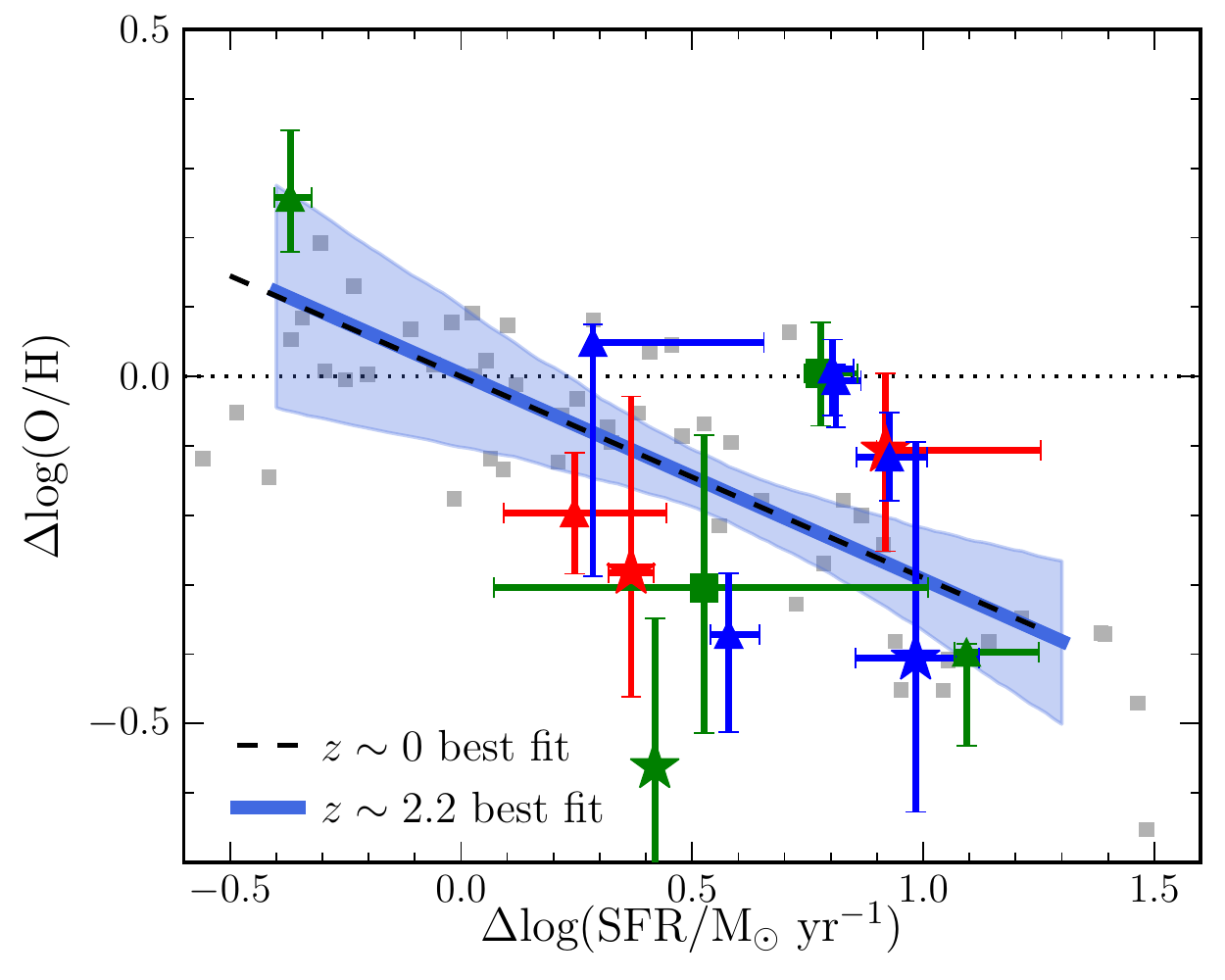}
 \centering
 \caption{
Deviation plot comparing the residuals around the MZR ($\Delta$log(O/H)) to residuals around the mean SFR-\mstar\ relation ($\Delta$log(SFR)).
  The best-fit MZR {\it after SFR correction} (equation~\ref{eq:mzrtotcorr}) and the uncorrected O/H values are used to calculate the O/H residuals at $z>1$.
  Colored points are as in Figure~\ref{fig3}.
  The gray squares show the \mstar-SFR binned $z\sim0$ stacks of \citet{and13}, with a best-fit given by the dashed black line
 (equation~\ref{eq:am13fmr}).
  A weak anticorrelation is present at $z>1$, suggesting a dependence of direct-method O/H on SFR at fixed \mstar\ quantitatively similar to the $z\sim0$ direct-method FMR (blue line, with 1$\sigma$ uncertainty given by the light blue shaded region).
}\label{fig6}
\end{figure}

If the dependence of O/H on SFR at fixed \mstar\ is known, then SFR biases can be corrected for to obtain representative
 MZRs for each sample.
The strength of the SFR dependence of the FMR can be quantified by the slope of the anticorrelation between residuals
 around the \mstar-SFR relation ($\Delta$log(SFR)) and residuals around the MZR ($\Delta$log(O/H))
 \citep[e.g.,][]{sal15,dav17,kas17,san18}.
Using the \citet{and13} $z\sim0$ SDSS stacks binned in \mstar+SFR and corrected for DIG \citep{san17},
 we fit $\Delta$log(O/H) vs.\ $\Delta$log(SFR) using direct-method metallicities
 (gray points and black dashed line in Figure~\ref{fig6}) and find that
\begin{equation}\label{eq:am13fmr}
\Delta\text{log(O/H)}\propto -0.29\times\Delta\text{log(SFR}/\msun~\text{yr}^{-1})
\end{equation}
This dependence of O/H on SFR is approximately twice as strong as is seen when using strong-line metallicity calibrations
 \citep{sal14,san18}, but the FMR is known to have a stronger SFR dependence when direct-method metallicities are employed
 \citep{and13,san17}.
This best-fit relation can be used to correct the metallicity of samples in the left panel of Figure~\ref{fig4} for SFR biases,
 assuming the SFR dependence of direct-method O/H at fixed \mstar\ does not change with redshift.\footnote{At fixed \mstar,
 $z\sim0$ samples display a non-linear SFR dependence of O/H at high masses
 (log(\mstar/M$_{\odot})\gtrsim10.0)$ and low SFRs (i.e., below the mean $z\sim0$ SFR-\mstar\ relation)
 \citep[e.g.,][]{man10,yat12,cre18}.
All of the galaxies to which we apply eq.~\ref{eq:am13fmr} have log(\mstar/M$_{\odot})<10.0$ and lie significantly above the
 local SFR-\mstar\ relation, a regime in which both local and high-redshift studies have demonstrated a linear dependence of
 O/H on SFR at fixed \mstar\ \citep{man10,san18,cre18}.
The linear approximation in eq.~\ref{eq:am13fmr} is thus appropriate to carry out SFR bias correction on these samples.
}

We correct the oxygen abundances of the $z\sim0$ and $z>1$ MZR samples for SFR biases according to equation~\ref{eq:am13fmr}
 and $\Delta$log(SFR) relative to the \mstar-SFR relation at each redshift (Figure~\ref{fig5}).
For the $z\sim0.8$ MZR of \citet{ly16}, we shift their best-fit MZR adopting an average $\Delta$log(SFR)=0.3~dex.
The resulting SFR-corrected MZR at $z\sim0-2.2$ is presented in the right panel of Figure~\ref{fig4}.
Correcting for the SFR biases present in each sample shifts the relations in the the left panel of Figure~\ref{fig4}
 such that they are now representative of galaxy populations falling on the mean \mstar-SFR relation at each redshift.
We fit the SFR-corrected $z\sim2.2$ MZR and obtain
\begin{equation}\label{eq:mzrtotcorr}
\begin{multlined}
12+\text{log(O/H)} \\ = (0.295\pm0.080)\times \text{log}(M_*/\msun) + (5.49\pm0.68)
\end{multlined}
\end{equation}
with a covariance of $\rho=0.998$ between the slope and intercept.
The SFR-corrected $z\sim2.2$ MZR fit is $\sim0.2$ dex higher in metallicity than the uncorrected fit (equation~\ref{eq:mzrtotuncorr}),
 but has a similar slope.
The mean SFR-corrected metallicity of the low- and high-mass bin is
 12+log(O/H$)_{\text{med}}=7.81\pm0.13$ and $8.21\pm0.08$, respectively.

We test whether the SFR dependence described by equation~\ref{eq:am13fmr} was appropriate to apply to the $z>1$
 sample by plotting the high-redshift galaxies in Figure~\ref{fig6}, where $\Delta$log(SFR) is determined relative
 to the \mstar-SFR relation matched in redshift for each galaxy (Fig.~\ref{fig5}) and $\Delta$log(O/H) is determined
 using the uncorrected metallicities (i.e., those displayed in Table~\ref{tab:props})
 relative to the SFR-corrected MZR of equation~\ref{eq:mzrtotcorr}.
We find a weak anticorrelation between $\Delta$log(O/H) and $\Delta$log(SFR), with a significance of $\approx2\sigma$.
We fit a linear relation to the $z>1$ sample, finding a slope of $-0.30\pm0.16$, remarkably consistent with the $z\sim0$ slope.
This relation is tentative evidence for the existence of a \mstar-SFR-O/H relation at $z\sim2.2$ based purely on
 direct-method metallicities, although the current sample has only one galaxy falling below the mean \mstar-SFR relation
 (i.e., with $\Delta$log(SFR$)<0$).
We conclude that equation~\ref{eq:am13fmr} and our method of correcting for SFR biases are applicable at high redshifts.
% TODO: add discussion of M* flattening and appropriateness of this correction

The scatter of the $z>1$ sample around the best-fit MZR is slightly smaller after correcting for SFR.
Indeed, the most significant outliers in the left panel of Figure~\ref{fig4} were also outliers
 in SFR compared to the rest of the sample, and are no longer outliers in the right panel.
The $z>1$ mergers (open triangles) are not strong outliers in the left panel, but fall well below
 the best-fit $z\sim2.2$ MZR in the right panel, suggesting that they do not have the same relation
 between \mstar, SFR, and O/H as the rest of the sample.

The best-fit $z\sim2.2$ MZR in equation~\ref{eq:mzrtotcorr} has a slightly shallower slope than is typically
 seen in strong-line metallicity studies \citep[e.g.,][]{erb06,cul14,mai14,ste14,san15,san18}.
This flattening is due to the redshifts of the galaxies in each
 mass bin.  The median redshift of galaxies in the high-mass bin is 2.31, while the median redshift
 in the low-mass bin is 1.83.  It is thus expected that the high-mass bin will have evolved farther
 from the $z\sim0$ MZR than the low-mass bin, artificially flattening the observed MZR of
 our $z>1$ auroral-line sample.  Consequently, the true $z\sim2.2$ MZR is steeper than
 equation~\ref{eq:mzrtotcorr}.

In the high-mass bin, we find that O/H at log($M_*$/$\msun)\approx9.2$ decreases by 0.2~dex
 from $z\sim0$ to $z\sim0.8$ and by $0.4\pm0.1$~dex from $z\sim0$ to $z\sim2.3$.
At log($M_*$/$\msun)\approx7.9$, O/H is lower by 0.2~dex at $z\sim0.8$ and 0.35$\pm$0.15~dex at $z\sim1.8$
 compared to $z\sim0$.
These values suggest a roughly linear evolution in log(O/H) at fixed \mstar\ with redshift of $d\text{log(O/H)}/dz\approx0.2$
 that does not display a strong dependence on \mstar\ below $\sim$10$^{9.5}~\msun$ out to $z\sim2.3$.
Accordingly, we find that the low-mass bin would have 0.10~dex lower O/H when ``evolved" from $z=1.83$ to $z=2.31$.
We alter the slope and intercept of equation~\ref{eq:mzrtotcorr} such that the line has the same O/H at
 log(\mstar/$\msun)=9.19$ but yields a value 0.10~dex lower at log(\mstar/$\msun)=7.90$, obtaining:
\begin{equation}\label{eq:mzrtotcorrz}
12+\text{log(O/H)} = 0.373\times \text{log}(M_*/\msun) + 4.78
\end{equation}
where the uncertainties in equation~\ref{eq:mzrtotcorr} can be applied to the slope and intercept here.

After correcting for SFR and sample redshift biases, we find that O/H at log(\mstar/$\msun)=7.5-9.5$ decreases by 0.2~dex
 at fixed \mstar\ from $z\sim0$ to $z\sim0.8$, and by $0.4\pm0.1$~dex at fixed \mstar\ from $z\sim0$ to $z\sim2.3$.
This MZR evolution is consistent with what is found using strong-line methods at $z\sim0.5-1$
 \citep{sav05,mai08,zah14a,tro14} and consistent with strong-line studies at $z\sim2$
 that find $\sim0.25-0.4$~dex evolution in O/H at fixed \mstar\ \citep[e.g.,][]{erb06,mai14,ste14,san15,san18}.
%This tension with strong-line results at $z\sim2.3$ is not significant given the current uncertainties.

Our final $z\sim2.3$ slope of 0.373$\pm$0.080 is consistent within the uncertainties
 with previous results at $z\sim2$ based on a number of strong-line metallicity indicators
 \citep[e.g.,][]{erb06,hen13,mai14,san15,san18}.  \citet{san18} found a $z\sim2.3$ MZR slope 
 of 0.26, 0.30, 0.34, and 0.30 when using O32, O3N2, N2, and N2O2 strong-line calibrations, respectively.
\citet{ste14} find a shallow $z\sim2.3$ MZR slope of 0.20 based on the N2 and O3N2 indicators, in contrast
 to other strong-line works.  Our direct-method results disagree with such a shallow MZR slope at the $2.2\sigma$ level,
 favoring a steeper slope but not statistically ruling out a shallower slope.  A larger $z>1$ auroral-line sample is needed to
 tightly constrain the high-redshift MZR slope via the direct-method.

\subsubsection{The direct-method fundamental metallicity relation from $z\sim0-2.2$}\label{sec:fmr}

The $z\sim0$ FMR has been claimed to be redshift independent, such that galaxies at $z=0-2.5$
 all follow the same \mstar-SFR-O/H relation, with high-redshift galaxies having lower metallicities
 in proportion to their higher SFRs at fixed \mstar \citep[e.g.,][]{man10,cre18}.
In \citet{san18}, we showed that $z\sim2.3$ galaxies were offset $\sim0.1$~dex below the $z\sim0$ FMR
 using a direct comparison of strong-line metallicities at fixed \mstar\ and SFR.
Whether the FMR is in fact redshift invariant remains an open question given the uncertainties associated
 with strong-line metallicities at high redshifts.  It is thus of interest to test the universality of the
 FMR using direct-method metallicities.  We cannot perform a direct comparison following \citet{san18} because
 there are not enough local analogs with [O\iii]$\lambda$4363 detections matched in \mstar\ and SFR to our $z>1$ sample.
In lieu of a direct comparison, we rely on parameterized forms of the FMR.

\begin{figure}
 \includegraphics[width=\columnwidth]{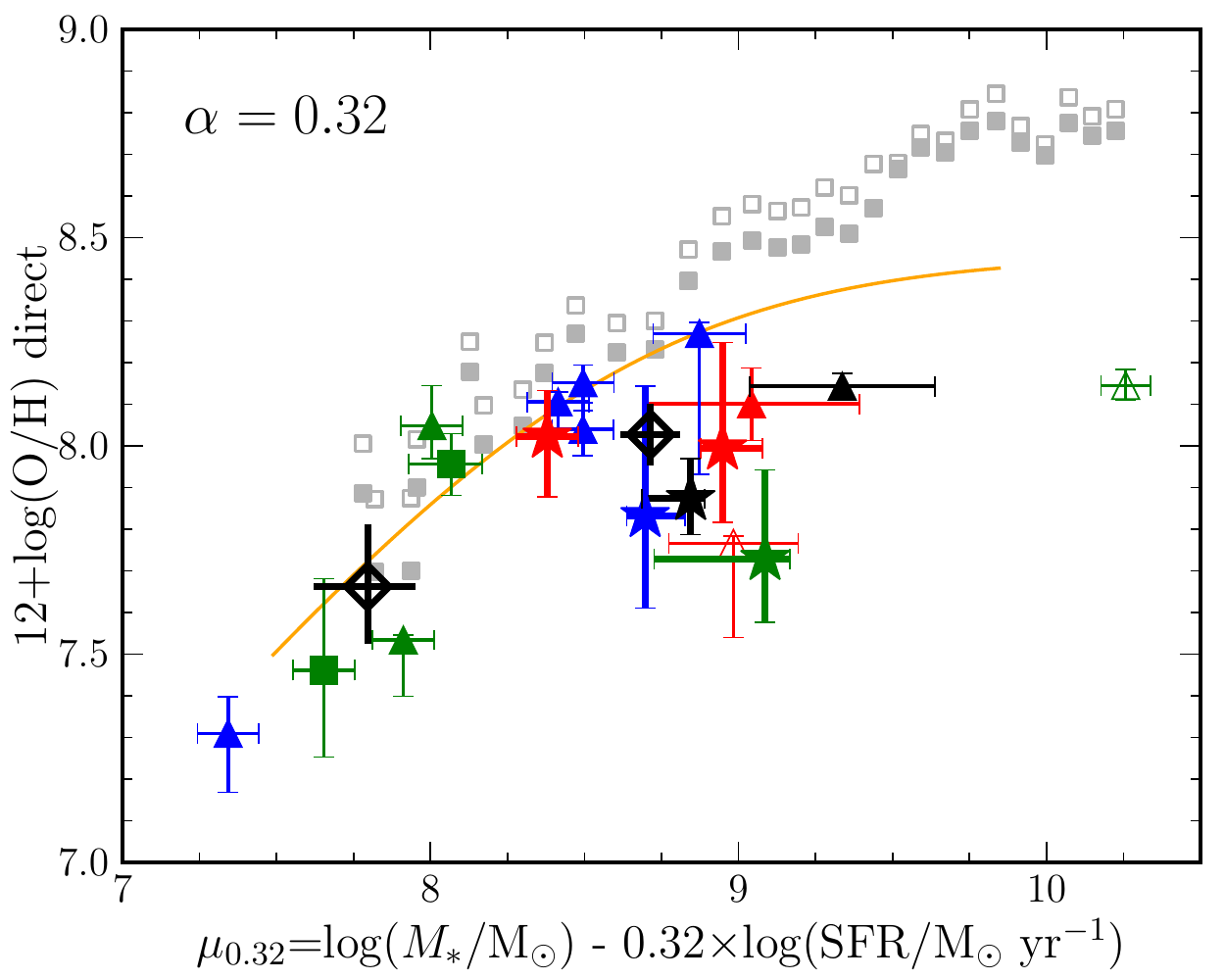}
 \includegraphics[width=\columnwidth]{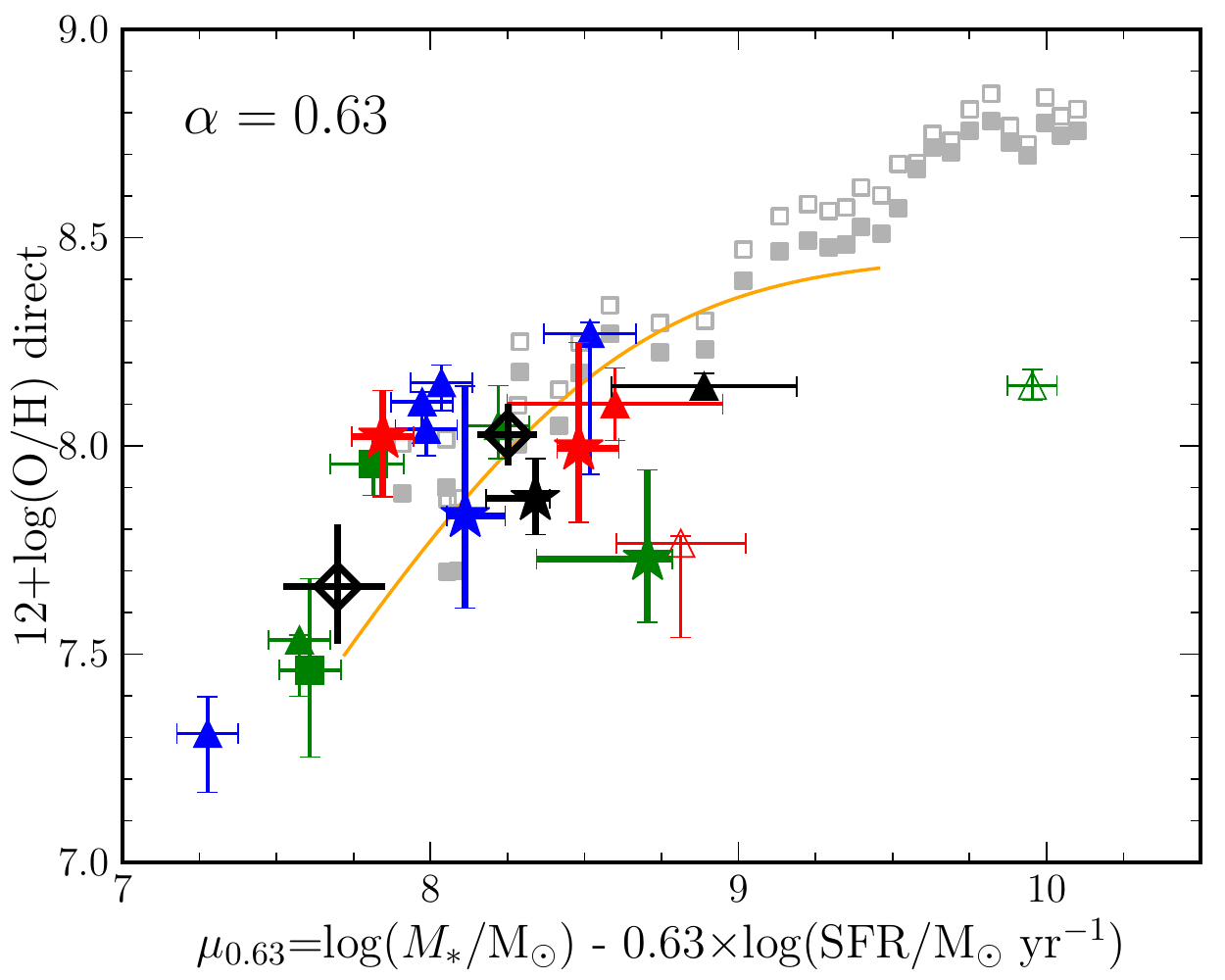}
 \centering
 \caption{
The \citet{man10} planar projection of the FMR.
  Points and lines are as in Figure~\ref{fig4}.
  The $z>1$ sample is significantly offset from the $z\sim0$ FMR when using $\alpha=0.32$ (top panel), the best-fit value from \citet{man10} based on strong-line metallicities.
  When using the best-fit $z\sim0$ value, $\alpha=0.63$ (bottom panel), from \citet{san17} based on direct-method metallicities,
 the $z>1$ galaxies fall within $\sim0.1$~dex of the $z\sim0$ FMR on average.
}\label{fig7}
\end{figure}

\citet{man10} constructed a planar parameterization of the $z\sim0$ \mstar-SFR-O/H relation that is described
 by O/H as a function of $\mu_\alpha$=log(\mstar$)-\alpha\times$log(SFR).
Using strong-line metallicities, \citeauthor{man10}\ found that $\alpha=0.32$ minimized the scatter in O/H at fixed $\mu_\alpha$.
We plot 12+log(O/H) vs.\ $\mu_{0.32}$ for the $z>1$ and low-redshift MZR samples in the top panel of Figure~\ref{fig7}.
There are offsets between the $z\sim0$, $z\sim0.8$, and $z\sim2.2$ samples, such that galaxies at higher redshift fall
 lower in O/H at fixed $\mu_{0.32}$.
However, the direct-method FMR has been found to have stronger SFR dependence than when using strong-line metallicities.
\citet{and13} find a value of $\alpha=0.66$ minimizes the FMR scatter using direct-method metallicity,
 while \citet{san17} find a best-fit value of $\alpha=0.63$ after correcting for DIG contamination.
We adopt $\alpha=0.63$, and present the resulting FMR projection in the bottom panel of Figure~\ref{fig7}.
The $z\sim2.2$ sample is consistent with both the $z\sim0$ and $z\sim0.8$ samples to within $\sim0.1$~dex
 at fixed $\mu_{0.63}$.
This comparison is the first test of the universality of the FMR at high redshifts using temperature-based metallicities.
This result suggests that, on average, O/H does not vary by more than $\sim0.1$~dex at fixed \mstar\ and SFR from
 $z=0$ to $z\sim2.2$, in agreement with results using strong-line metallicities \citep{san18,cre18}.
%The mean O/H of the high-mass $z>1$ bin at $\mu_{0.63}\approx8.5$ is shifted $\sim0.1$~dex below $z\sim0$ galaxies at
% the same $\mu_{0.63}$, hinting at a slight evolution of the \mstar-SFR-O/H relation, as found by \citet{san18} using
% strong-line metallicities.  A larger $z>1$ auroral-line sample is needed to resolve such a small shift with statistical
% significance.

\subsubsection{Systematic effects on O/H, SFR, and \mstar\ comparison}\label{sec:systematics}

A concern when comparing metallicity scaling relations across multiple galaxy samples is
 whether the metallicities, \mstar, and SFRs of each sample can be fairly compared.
The metallicities of all samples in this work have been determined via the direct-method
 based on either [O\iii]$\lambda$4363 or O\iii]$\lambda$1663 and can be directly compared in an
 unbiased manner, even when the samples span a large range in redshift.
The SFRs of all samples have been determined using reddening-corrected Balmer emission lines
 and calibrations within $\approx10\%$ of one another in normalization, such that the comparison of
 SFRs between our low- and high-redshift samples is robust.
Stellar mass determinations require a number of assumptions regarding the stellar population templates,
 attenuation curve, IMF, and star-formation history.
We have shifted all stellar masses and SFRs to a \citet{cha03} IMF.

Even when utilizing the same IMF, differences in SED-fitting assumptions can result in systematic uncertainties
 of $\sim0.3$ dex in \mstar \citep[e.g.,][]{red12}.
Details of the SED fitting were not included in all of the references from which we obtained \mstar\ for the
 $z>1$ aurora-line sample.
However, we have confirmed that over half of the $z>1$ sample was fit with \citet{bru03} stellar population synthesis
 models, a constant or declining star-formation history, a \citet{cal00} attenuation curve, and metallicities
 ranging from $0.2-1.0$~Z$_{\odot}$.
Based on changing the SED-fitting assumptions for the MOSDEF $z\sim2$ star-forming galaxy sample, stellar masses
 under the range of assumptions described above typically vary by $\sim0.1-0.2$~dex compared to \mstar\ derived under our fiducial assumptions
 (Sec.~\ref{sec:measurements}).
Furthermore, this range of assumptions is similar to those used in \mstar\ determination of the $z\sim0$
 and $z\sim0.8$ samples \citep{bri04,and13,ly16}.
Given our best-fit $z\sim2$ MZR slope of $\approx0.37$, a systematic offset of 0.2~dex in \mstar\ between
 the samples at each redshift would introduce a systematic bias of $\sim0.07$~dex in the inferred O/H evolution, smaller than the
 typical uncertainties on O/H in the $z>1$ sample.
We conclude that the metallicity evolution results presented herein are not significantly affected by
 systematic biases associated with the determination of O/H, SFR, or \mstar\ for each sample.

An additional concern is whether sample selection effects are present beyond SFR biases (which we have corrected for).
This is a particular concern at high redshift where the sample size is small and selection requires the detection
 of extraordinarily weak emission lines.
Even at fixed \mstar\ and SFR, the auroral-line detection criterion will select galaxies with higher
 excitation and lower O/H than average such that we are incomplete in metallicity sampling.
Furthermore, since the detection cut is made on a weak emission line, an additional bias will be present where galaxies
 are preferentially detected when positive noise fluctuations occur, leading to an overestimate of the auroral-line fluxes
 among low-S/N detections.
Both of these effects bias the mean metallicity of our sample low.
Given that 10/18 of the $z>1$ auroral-line sample have S/N$_{\text{aur}}>5$ (Tab.~\ref{tab:props}) and a positive noise
 fluctuation is not expected to be present for all low-S/N detections, we expect that the latter bias has a minimal effect
 on the sample average metallicities from which we draw conclusions.
The former bias is likely present to some extent.

We estimate the magnitude of the excitation bias by selecting a sample of MOSDEF star-forming galaxies at $z\sim2.3$
 matched in \mstar\ and
 SFR to a subset of the $z>1$ auroral-line sample.
We can only make such a match in the narrow range of overlap at $9.0\le\text{log}(M_*/\text{M}_{\odot})\le9.5$ and
 $\text{log(SFR}/\msun~\text{yr}^{-1})\ge1.1$, where there are 5 auroral-line sources and 15 MOSDEF $z\sim2.3$ galaxies.
The MOSDEF galaxies should be representative at this redshift, \mstar, and SFR since they are selected according to a
 rest-optical magnitude limit.
The two subsamples are closely matched in \mstar\ and SFR, with mean values of
 $\langle\text{log}(M_*/\text{M}_{\odot})\rangle=[9.33\pm0.08, 9.37\pm0.02]$ and
 $\langle\text{log(SFR}/\msun~\text{yr}^{-1})\rangle=[1.47\pm0.05, 1.34\pm0.04]$ for the auroral-line and MOSDEF targets,
 respectively.
The auroral-line subsample has $\langle$log(O32$)\rangle=0.54\pm0.08$ and $\langle$log(O3$)\rangle=0.80\pm0.03$, systematically
 higher than the \mstar- and SFR-matched MOSDEF subsample with
 $\langle$log(O32$)\rangle=0.42\pm0.02$ and $\langle$log(O3$)\rangle=0.70\pm0.02$.
Based on the BKD18 high-redshift analog O32 calibration,
 the O32 difference of $0.12\pm0.08$~dex corresponds to a bias in O/H of $-0.07\pm0.05$~dex for the auroral-line subsample.
Part of this O/H difference can be accounted to the slightly higher SFR of the auroral-line subsample (see eq.~\ref{eq:am13fmr}).
We conclude that the high-excitation selection biases our sample-averaged metallicities by $<0.1$~dex
 and thus does not significantly affect our conclusions regarding the evolution of metallicity calibrations, the MZR, and the FMR.

\section{Photoionization modeling: stellar metallicity \& ionization parameter}\label{sec:models}

Photoionization models of \hii regions utilizing state-of-the-art stellar models are powerful tools for
 extracting information about the ionization state and chemical abundances of ionized gas from
 observed nebular emission-line strengths \citep[e.g.,][]{kew02,ste14,dop16,str18,kas19b}.
However, degeneracies between the ionization state and nebular metallicity are significant when
 only the strong optical emission-line ratios are available, especially when just a subset of the strong optical nebular emission
 lines are detected, as is the case for nearly all high-redshift galaxies.

Recent observational and theoretical work has suggested that the Fe/H of massive stars may be deficient relative
 to the O/H of the ionized gas on a solar abundance scale in $z>1$ galaxies \citep[i.e., O/Fe$>$O/Fe$_{\odot}$;][]{ste16,mat18,str18,cul19}.
This O/Fe enhancement is thought to be caused by the rapid formation timescales \citep[$\lesssim1$~Gyr; e.g.,][]{red12,red18} of high-redshift galaxies,
 in which chemical enrichment is dominated by Type II SNe that yield super-solar O/Fe ratios \citep{nom06,kob06}.
This enhancement in O/Fe relative to a solar abundance pattern is brought about because the majority of Fe is produced in Type Ia
 SNe, most of which occur on timescales $\sim1$~Gyr after a star-formation event, while O is produced promptly
 in core-collapse Type II SNe on timescales of $\lesssim10$~Myr, the lifetimes of massive stars.

Using the unique direct-method nebular metallicity constraints of our $z>1$ auroral-line sample,
 we fix the input nebular metallicity (i.e., O/H) in grids of photoionization models, allowing us to unambiguously
 constraint the stellar metallicity (i.e., Fe/H) and the ionization parameter, $U$, for this set of
 high-redshift star-forming galaxies.

\subsection{Description of the photoionization models and fitting procedure}\label{sec:fitting}

We utilize the code Cloudy \citep[v17.01;][]{fer17} to produce a grid of photoionization models
 for use in constraining the ionization parameter ($U$) and stellar ionizing spectral shape
 for the $z>1$ auroral-line sample.
We construct simple \hii region models assuming a plane-parallel geometry ionized by a single star cluster,
 where the calculation is stopped at the edge of the fully-ionized H zone.
The input parameters of the models are the ionization parameter $U$, electron density $n_e$,
 nebular metallicity $Z_{\text{neb}}^{\text{in}}$, and stellar metallicity $Z_*$.
In practice, we fix $n_e$ to a constant value and vary $U$, $Z_{\text{neb}}^{\text{in}}$, and $Z_*$
 for the calculation of the model grids, as described below.
Using this set of model grids, we fix $Z_{\text{neb}}^{\text{in}}$ to a value matched to the direct-method metallicity
 for each object and unambiguously fit for $U$ and $Z_*$.

\subsubsection{Photoionization model grids}

The intensity of the radiation field is set by the dimensionless ionization parameter, $U=n_{\text{Lyc}}/n_H$,
 where $n_{\text{Lyc}}$ is the volume density of hydrogen-ionizing photons and $n_H$ is
 the hydrogen gas volume density, which can be approximated by the electron density $n_e$.
We vary log($U$) from $-1.0$ to $-4.0$ in 0.1~dex steps.
We assume a constant density of $n_e=250$~cm$^{-3}$, a typical value for $z\sim2$ star-forming galaxies
 \citep[e.g.,][]{san16a,ste16,str17}.

The gas-phase elemental abundances are assumed to follow a solar abundance pattern, such that the
 abundances of each element is set by the oxygen abundance, O/H.
To avoid confusion with the nebular metallicity measured via the direct-method from observations ($Z_{\text{neb}}$),
 we refer to the model input nebular metallicity as $Z_{\text{neb}}^{\text{in}}$.
While the abundances of elements with secondary production channels (C, N) depend on $Z_{\text{neb}}$,
 we do not utilize any lines of these elements in the fitting procedure described below, and thus
 simply leave them fixed at a solar scale for convenience.
This assumption has a negligible effect on the line ratios of O, Ne, and H across the range of parameters
 explored here.
We adopt a range of values for the nebular metallicity:
 $Z_{\text{neb}}^{\text{in}}/\text{Z}_{\odot}$=[0.05, 0.1, 0.2, 0.3, 0.4, 0.5, 0.6, 0.7, 0.8, 0.9, 1.0, 1.25, 1.5]
 (7.4$\le$12+log(O/H)$^{\text{in}}$$\le$8.9).

Following \citet{ste16} and \citet{str18}, the stellar metallicity, $Z_*$, is allowed to vary freely from $Z_{\text{neb}}^{\text{in}}$.
Only the hydrogen-ionizing photons ($h\nu>13.6$~eV; $\lambda<912$~\AA) produced by the stellar ionizing source affect the
 emission-line production of the ionized nebula, as these photons are responsible for liberating and
 heating the free electron population.
The spectrum of massive stars at $\lambda<912$~\AA\ is primarily governed by the opacity of the photosphere from
 line-blanketing of iron-peak elements.  This opacity, driven by Fe/H, strongly affects the evolution of
 massive stars.  Larger opacity due to higher Fe/H results in stronger stellar winds, driving larger mass loss rates
 and decreasing rotation.  Accordingly, stars with lower Fe/H produce a larger number of ionizing photons,
 have a harder spectral shape due to a hotter effective temperature, and can potentially continue producing
 copious amounts of ionizing photons for a longer time period due to effects such as quasi-homogeneous evolution \citep{eld11,eld12}.
We assume that $Z_*$, which sets the shape of the ionizing stellar spectrum,  effectively traces Fe/H.
Thus, decoupling $Z_*$ and $Z_{\text{neb}}^{\text{in}}$ allows us to investigate non-solar O/Fe ratios and vary the
 hardness of the ionizing spectrum at fixed O/H.

We adopt the ``Binary Population and Spectral Synthesis" v2.2.1 models \citep[BPASS;][]{sta18}, which include the
 effects of binarity on the evolution of massive stars.  We use the binary BPASS models with a high-mass
 IMF slope of $-2.35$ and a high-mass cutoff of 100~$\msun$ as our fiducial set of ionizing spectra.
We vary the BPASS stellar metallicities over the range $Z_*=10^{-5}-0.020$ (0.0007-1.4~Z$_{\odot}$ based on $\text{Z}_{\odot}$=0.014 from \citealt{asp09}).
For comparison, we also utilize single-star models from Starburst99 \citep[SB99;][]{lei14} with the same IMF slope and cutoff,
 with a metallicity range of $0.001-0.040$ (0.07-2.8~Z$_{\odot}$).
The input stellar populations are assumed to have continuous star formation, with an age of 100 Myr.
The ionizing spectra of both BPASS and SB99 models reach equilibrium after 10 Myr of continuous star formation,
 so the nebular emission-line predictions are appropriate for systems with stellar populations older than 10~Myr.

The photoionization models are stopped at the edge of the \hii zone, and thus do not include any DIG component.
We have ensured that the empirical data sets we model can be well represented by H~\textsc{ii}-only emission.
The $z\sim0$ composites of AM13 have been corrected for DIG contamination following \citet{san17}.
For star-forming galaxies at $z>1$, and in particular the extremely high-sSFR $z>1$ auroral-line sample,
 DIG emission is expected to be negligible due to their high SFRs and compact sizes \citep{sha19}.
This expectation assumes no redshift evolution of the relation between SFR surface density and the fractional
 importance of DIG to line emission, empirically established at $z=0$ \citep{oey07}.

No dust grains or depletion of gas-phase elements onto grains are included in the photoionization models.
We thus assume that the effects of dust depletion on the derived gas-phase O/H for the $z>1$ auroral-line sample and
 the $z\sim0$ AM13 stacks is negligible.
\citet{ste16} derived an O depletion factor of 0.09~dex for the KBSS-LM1 composite that has 12+log(O/H$)=8.14$ and
 E(B-V)$_{\text{gas}}=0.24$.
The bulk of the $z>1$ sample has lower metallicity and less dust reddening than KBSS-LM1
 ($\langle$12+log(O/H$)\rangle\approx8.00$ and $\langle$E(B-V)$_{\text{gas}}\rangle\approx0.10$).
At such low metallicities, O depletion has been found to be negligible in local star-forming galaxy samples, and
 O depletion is not observed to much exceed 0.1~dex even at near-solar metallicities \citep{izo06}.
We thus conclude that our assumption of no O depletion is valid for the $z>1$ auroral-line sample and $z\sim0$ galaxies
 at comparably low metallicities, but that depletion may bias metallicities low by up to 0.1~dex for $z\sim0$ galaxies
 at higher metallicities.
Adding a 0.1~dex correction to the derived temperature-based metallicities does not produce qualitatively different results
 from the photoionization model fitting.

\subsubsection{$U$ and $Z_*$ fitting procedure}

\begin{figure}
 \includegraphics[width=\columnwidth]{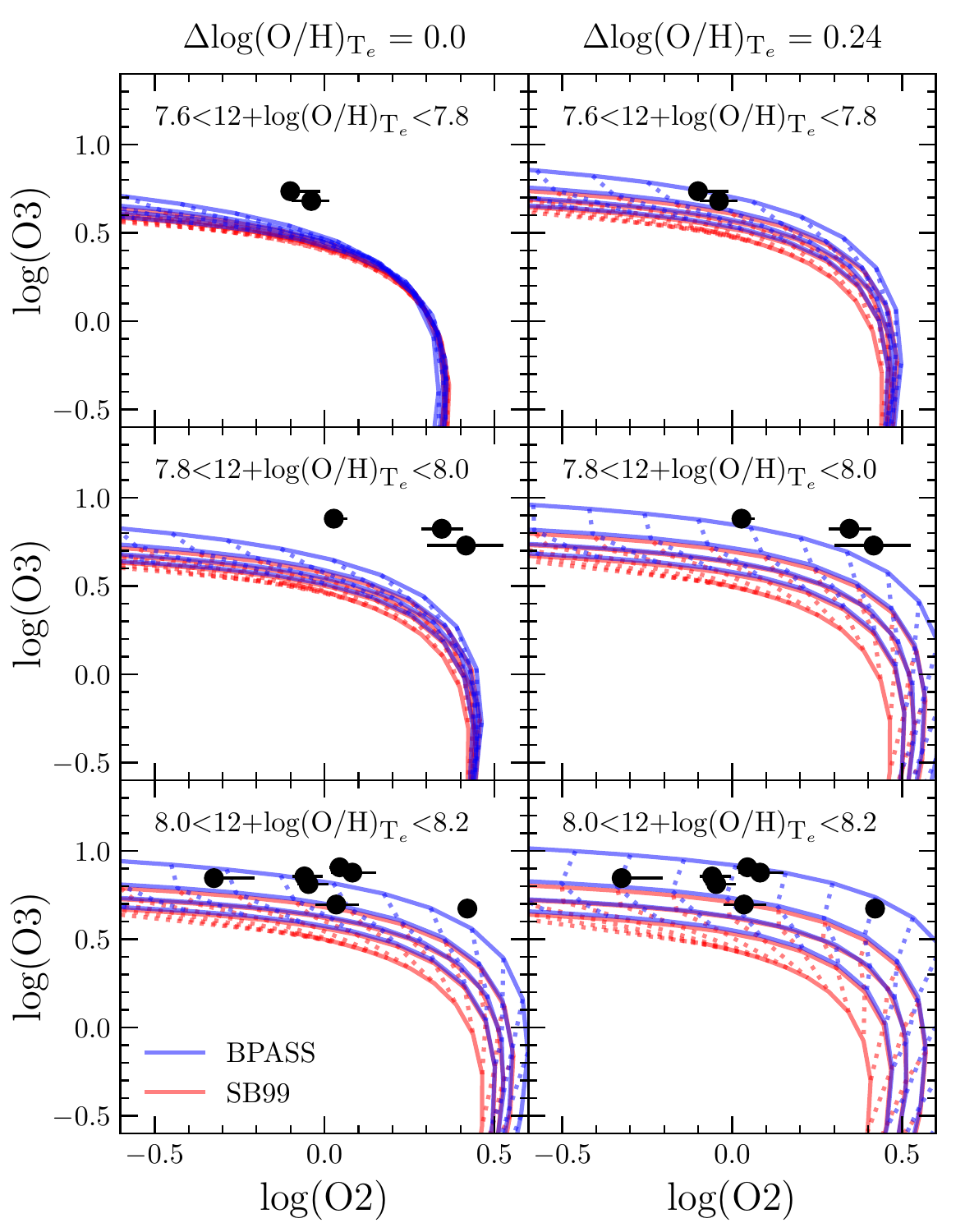}
 \centering
 \caption{
Comparison of photoionization model grids to observed line-ratios (O3 vs. O2) at $z>1$, in three bins of metallicity.
  Grids are shown from the BPASS (blue) and SB99 (red) models at fixed nebular metallicity, with varying stellar metallicity and ionization parameter.  Lines of constant stellar metallicity are solid, while lines of constant ionization parameter are dotted.
  In the left column, the input nebular metallicity to the models is assumed to be the same as the direct-method
 metallicity of the observed galaxies ($\Delta$log(O/H)$_{\text{T}_e}=0.0$), such that the input nebular metallicity
 for the top-, middle-, and bottom-left panels is 12+log(O/H$)=7.7$, 7.9, and 8.1, respectively.
  In the right column, the input nebular metallicity to the models is instead assumed to be larger than the observed direct-method metallicity by 0.24~dex ($\Delta$log(O/H)$_{\text{T}_e}=0.24$), the mean difference between electron-temperature and recombination line metallicities of $z\sim0$ \hii regions.
  The input nebular metallicity of the models in the top-, middle-, and bottom-right panels is thus 12+log(O/H$)=7.94$, 8.14, and 8.34, respectively.
  The SB99 models fail to produce large enough O3 and O2 values to match observations in all panels.
  The BPASS models fail to match observations when $\Delta$log(O/H)$_{\text{T}_e}=0.0$.
  The ability of the BPASS models to reproduce the range of O3 and O2 values observed at fixed O/H significantly improved when assuming $\Delta$log(O/H)$_{\text{T}_e}=0.24$.
  We adopt this as a fiducial assumption when matching models to observations.
}\label{fig8}  
\end{figure}

Utilizing the model grids described above, we produce constraints on the ionization state of the $z>1$ auroral-line sample
 by fixing $Z_{\text{neb}}^{\text{in}}$ to a value appropriate for each galaxy.
Thus, a critical step to utilizing these photoionization models to constrain $U$ and $Z_*$ is matching the
 measured direct-method $Z_{\text{neb}}$ to $Z_{\text{neb}}^{\text{in}}$ of a particular set of models.  
The simplest assumption is $Z_{\text{neb}}^{\text{in}}$=$Z_{\text{neb}}$, that is, the direct-method
 O/H represents the ``true" gas-phase metallicity in an unbiased manner.
In the left column of Figure~\ref{fig8}, we compare the measured O3 and O2 line ratios of $z>1$ galaxies
 to model grids directly matched in metallicity.
We find that both BPASS and SB99 grids significantly underpredict O3 and O2 of the majority of $z>1$
 galaxies when $Z_{\text{neb}}^{\text{in}}$=$Z_{\text{neb}}$, even at the lowest $Z_*$ (hardest ionizing spectrum).

There is evidence that the temperature-based metallicities are systematically biased and thus do not represent
 the ``true" gas-phase metallicity.
There is a longstanding disagreement between temperature-based  metallicities from collisionally-excited auroral lines
 (direct method) and metallicities determined using O recombination lines (RL method), known as the
 abundance discrepancy factor (ADF) problem \citep[e.g.,][]{pei67,est14}.
Direct-method metallicities are systematically lower than RL-method metallicities measured for the same targets.
The value of the ADF appears to be roughly constant as a function of metallicity, with the direct method yielding
$\approx0.24$ dex lower metallicities than the RL method on average \citep[e.g.,][]{est14,bla15}.
The ADF has been proposed to be caused by fluctuations in the temperature field of the ionized nebula, such that
 hotter regions dominate the auroral-line emission and bias temperature measurements high and, consequently,
 direct-method O/H measurements low \citep{pei67,gar07}.
Alternatively, a chemically inhomogeneous medium with metal-rich small-scale overdensities
 provides an explanation of the ADF in which RL metallicities are biased high \citep{tsa05,sta07}.
Both of these phenomena could be present in \hii regions.
There is an ongoing debate whether the direct-method or RL metallicities represent the true gas-phase abundance scale \citep{mai19}.
Attempts have been made to solve this problem by comparing the stellar oxygen abundance of
 young A and B supergiants to the RL and direct method metallicities of the \hii regions they occupy, with conflicting
 results.  The stellar O/H agrees with the RL method in some \hii regions and the direct method in others
 \citep{bre16,tor16,tor17}.  It is not clear which method yields the true $Z_{\text{neb}}$.\footnote{While temperature fluctuations
 may bias the normalization of direct-method metallicities by $\sim0.2-0.3$~dex,
 relative differences in direct-method metallicity are much more robust.
The direct-method metallicity results on strong-line calibrations and scaling relations presented in Sec.~\ref{sec:empirical} would only
 be biased if the magnitude of temperature fluctuations evolves with redshift.  There are currently no constraints on the ADF outside of
 the local universe and no theoretical expectations for an evolving ADF.} 

Regardless of the remaining uncertainties regarding direct vs.\ RL methods, we have found that the models cannot reproduce
 the observations when $Z_{\text{neb}}^{\text{in}}$=$Z_{\text{neb}}$, where the latter is determined using
 the direct method (Fig.~\ref{fig8}, left column).
In the right column of Figure~\ref{fig8}, we show the resulting model grids when we apply the typical ADF of 0.24~dex
 between the RL and direct methods to our measured $z>1$ direct-method O/H values
 (log($Z_{\text{neb}}^{\text{in}}$)=log($Z_{\text{neb}})+0.24$).
The BPASS grids (blue) now span the range of measured O3 and O2 of the majority of the $z>1$ sample.
\citet{ste16} similarly found that applying a 0.24~dex ADF to the direct-method O/H of their stack of $z\sim2.3$
 star-forming galaxies was necessary for model grids to match measured line ratios.
We proceed under the assumption that log($Z_{\text{neb}}^{\text{in}}$)=log($Z_{\text{neb}})+0.24$.
%, and consider the comparison in Figure~\ref{fig8} circumstantial evidence for the existence of temperature fluctuations
% in \hii regions at $z>1$.
We note that the SB99 grids fail to overlap the $z>1$ galaxies even under this assumption because they do not produce
 hard enough ionizing spectra even at the lowest available metallicity, $Z_*=0.07$~Z$_{\odot}$.
Therefore, we conclude that SB99 ionizing spectra cannot reproduce the emission-line properties of our
 $z>1$ sample and only show results using the BPASS binary models for the remainder of the analysis.

\begin{figure*}
 \includegraphics[width=0.95\textwidth]{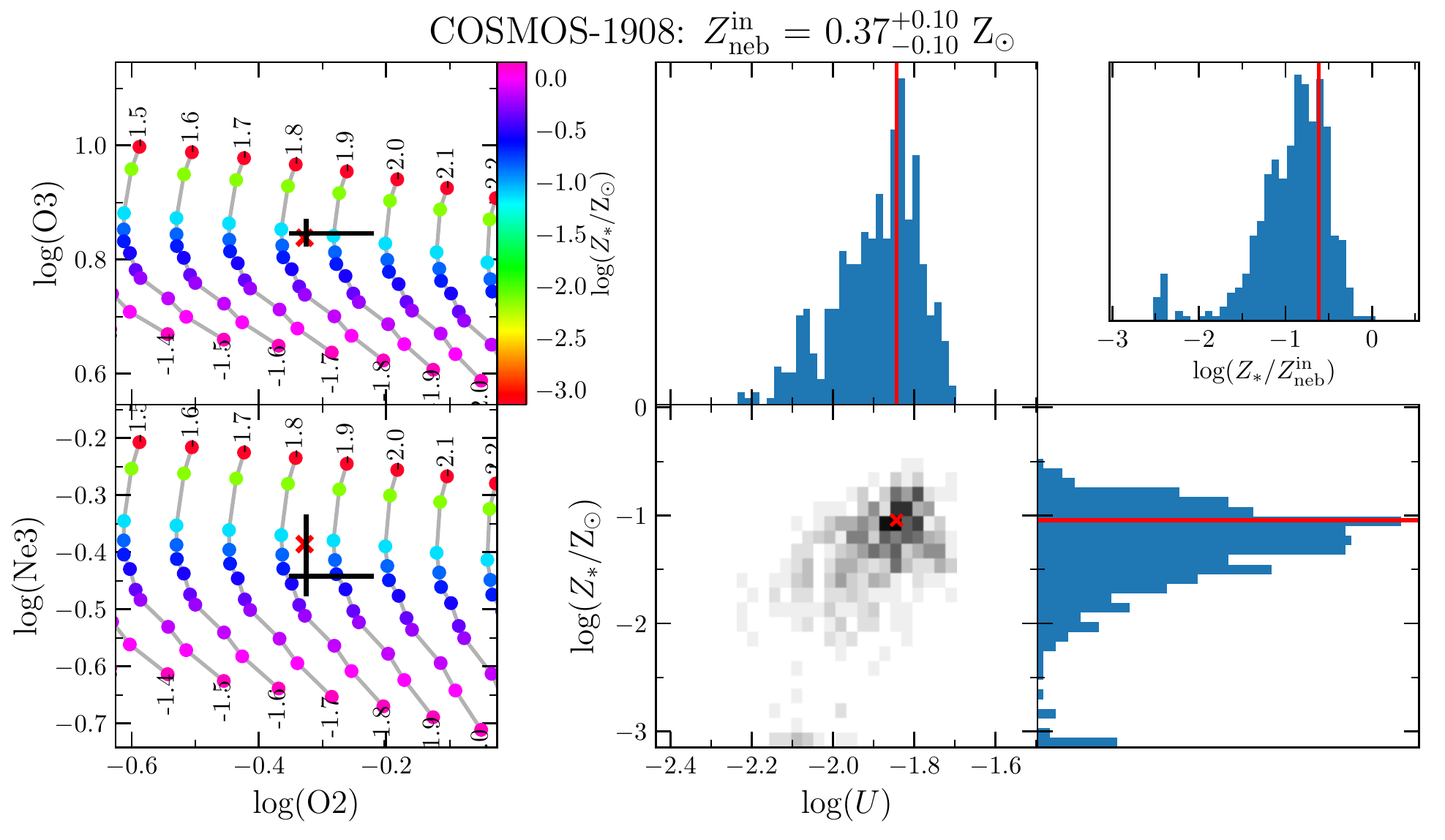}
 \centering
 \caption{
Best-fit solutions for COSMOS-1908 of ionization parameter, $U$, and stellar metallicity, $Z_*$, from the BPASS photoionization models.
  \textsc{Left:} Line-ratio diagrams displaying the observed strong-line ratios (black error bars), model grid (colored points and gray lines), and best-fit solution (red X).
  The displayed model grid has nebular metallicity fixed to the nebular metallicity inferred for COSMOS-1908.
  Model grid-points are color-coded by stellar metallicity.  Gray lines connect points of constant ionization parameter, with the value of log($U$) labeling the lowest and highest stellar metallicity points.
  \textsc{Right:} Distribution of $U$ and $Z_*$ from Monte Carlo simulations.  The best-fit solution is shown by the red lines or red X in each panel.
  The upper-right panel shows the constraint on $Z_*$/Z$_{\text{neb}}^{\text{in}}$, which we take to be equivalent to Fe/O.
}\label{fig9}
\end{figure*}

\begin{figure*}
 \includegraphics[width=0.95\textwidth]{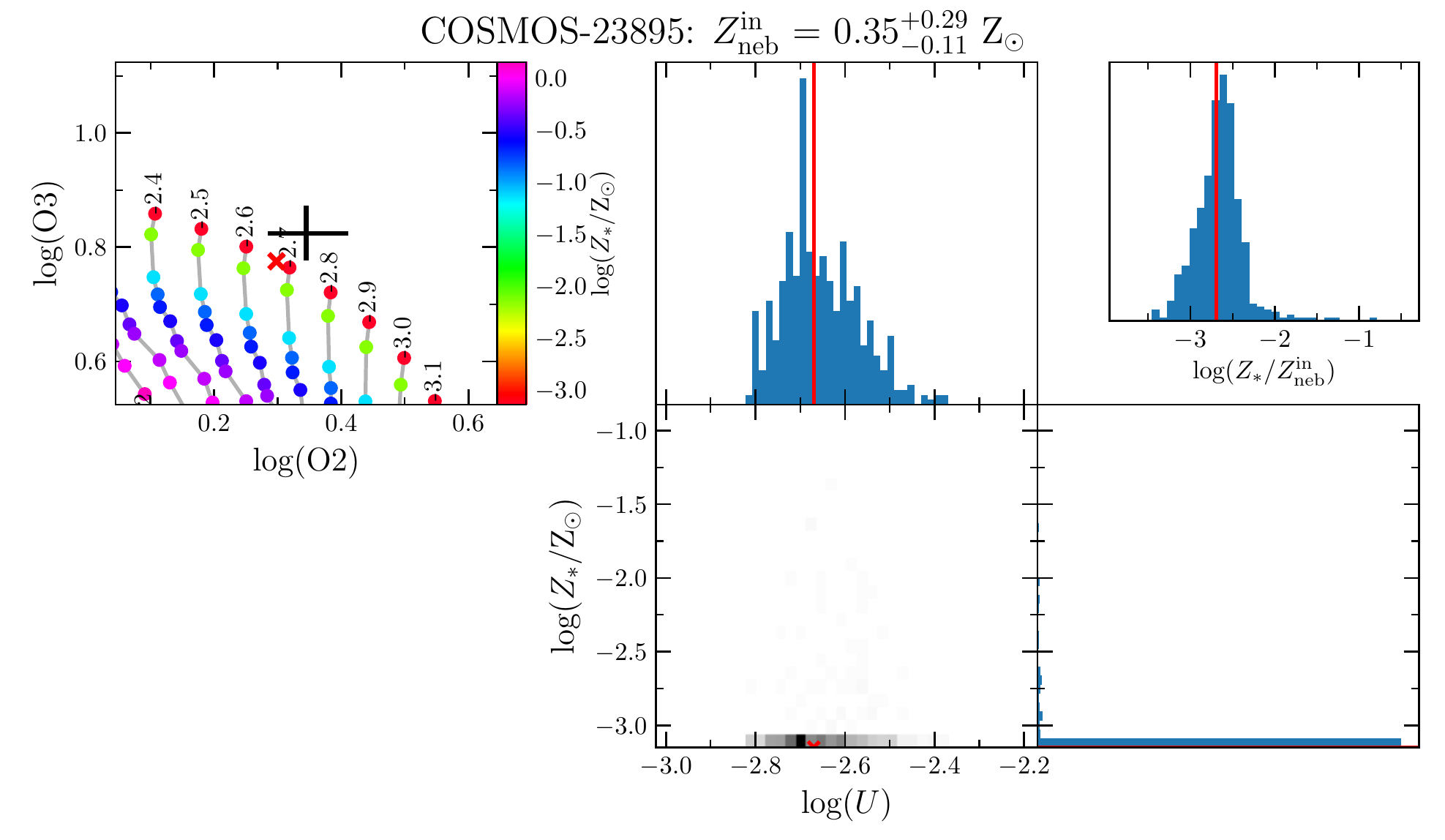}
 \centering
 \caption{
Same as Figure~\ref{fig9} for COSMOS-23895.
The best-fit solution is at the lower bound of the explored parameter space for $Z_*$,
 thus only a 3$\sigma$ lower limit on $Z_*$ is obtained.
}\label{fig10}
\end{figure*}

\begin{figure*}
 \includegraphics[width=0.95\textwidth]{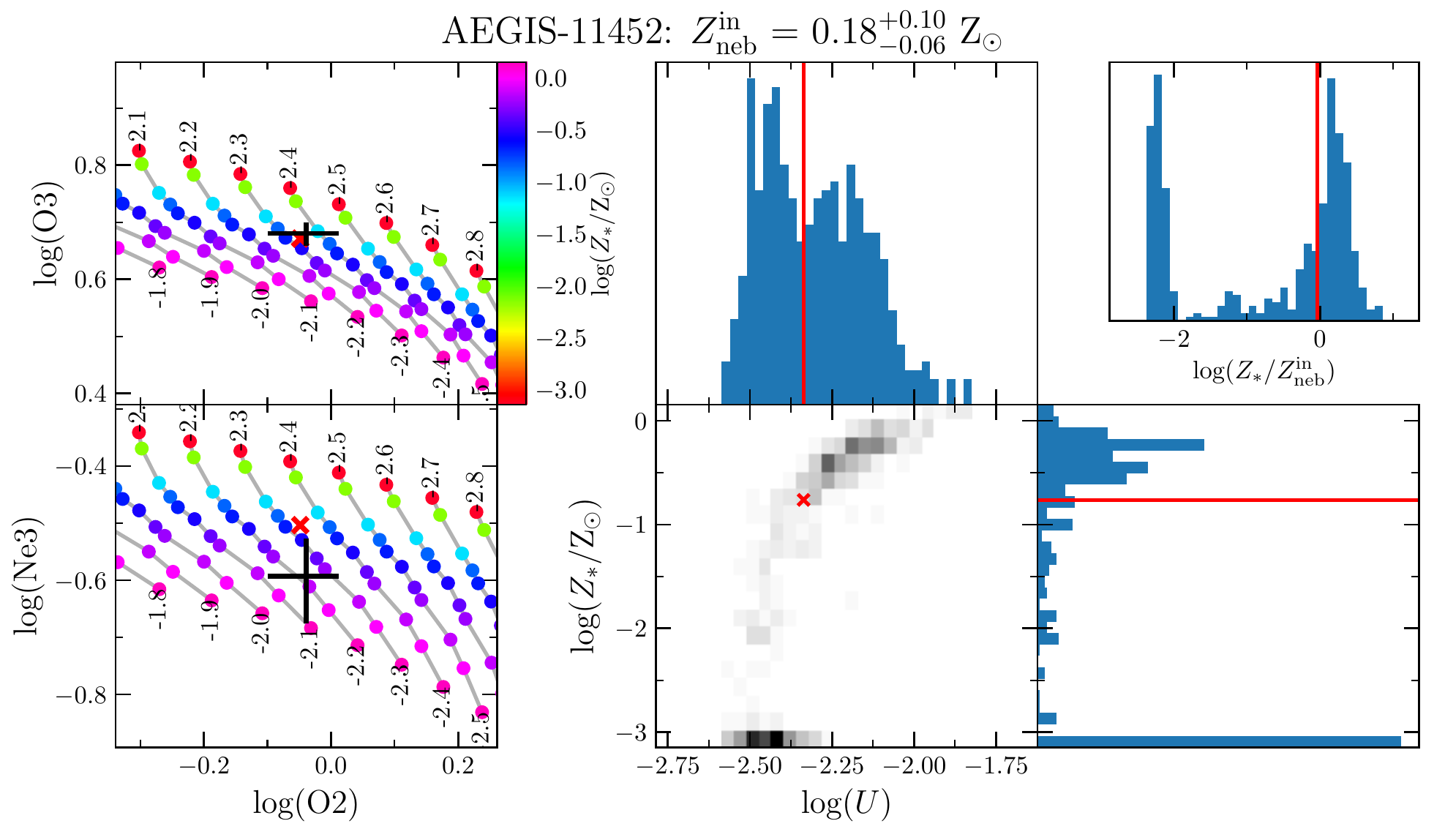}
 \centering
 \caption{
Same as Figure~\ref{fig9} for AEGIS-11452.
}\label{fig11}
\end{figure*}

\begin{figure*}
 \includegraphics[width=0.95\textwidth]{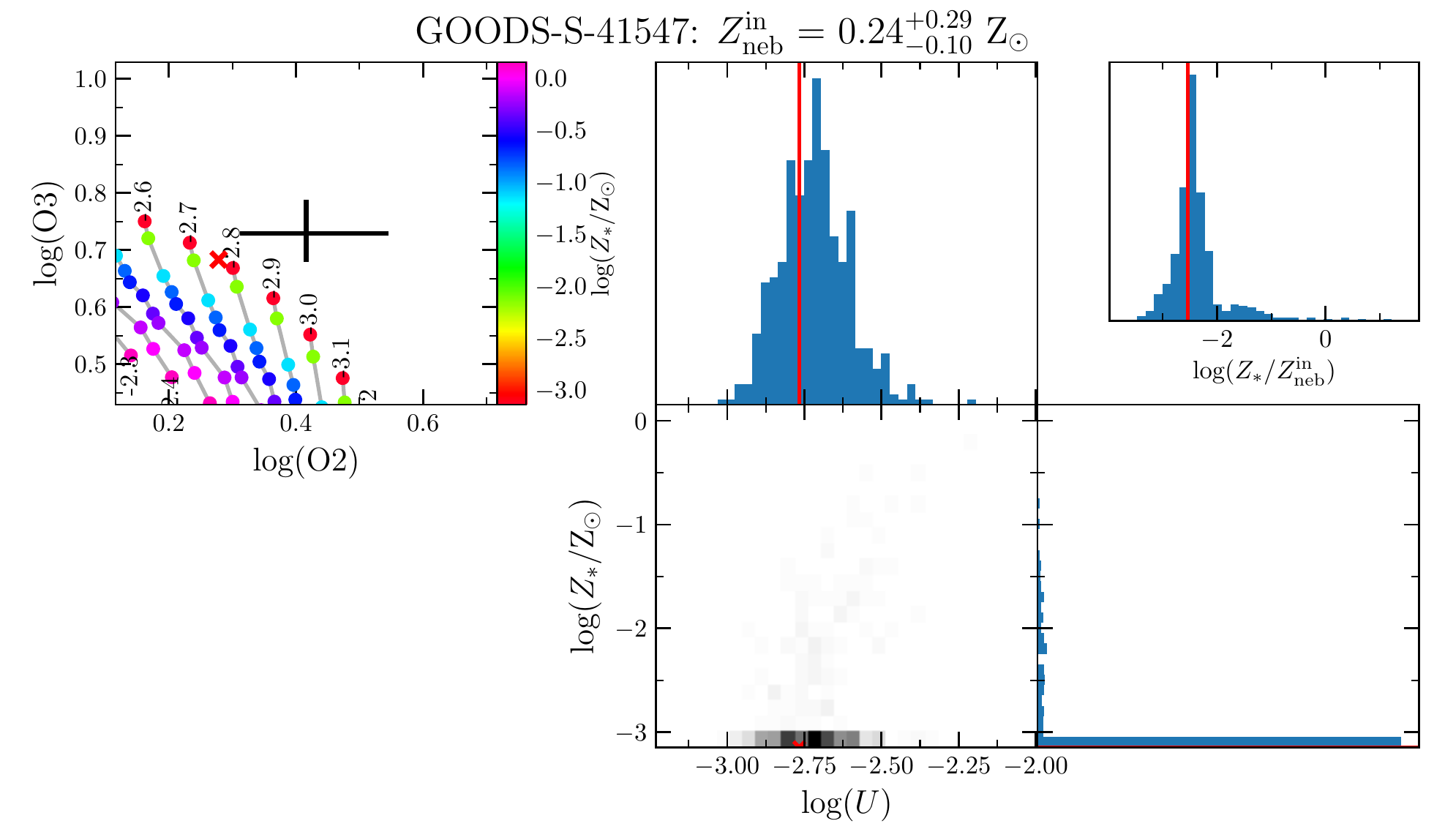}
 \centering
 \caption{
Same as Figure~\ref{fig9} for GOODS-S-41547.
The best-fit solution is at the lower bound of the explored parameter space for $Z_*$,
 thus only a 3$\sigma$ lower limit on $Z_*$ is obtained.
}\label{fig12}
\end{figure*}

We fit for the ionization parameter and stellar metallicity of each galaxy using the following procedure.
We first interpolate the photoionization model grids in nebular metallicity to construct a
 grid of strong-line ratios as a function of $U$ and $Z_*$ at fixed $Z_{\text{neb}}^{\text{in}}$ according to
 log($Z_{\text{neb}}^{\text{in}}$)=log($Z_{\text{neb}})+0.24$, where $Z_{\text{neb}}$ is the
 measured direct-method O/H given in Table~\ref{tab:props}.
We fit for $U$ and $Z_*$ by performing a $\chi^2$ minimization using the measured reddening-corrected 
strong-line ratios simultaneously.
For the fitting procedure, we do not include ratios involving lines of N because of uncertainty in the N/O vs.\ O/H relation
 at high redshifts \citep{mas14,ste14,san15,san16a,sha15,str18},
 and also exclude line ratios involving S because it has an ionization potential lower
 than that of H and may be emitted in partially-ionized regions that our photoionization models do not include.
We use line ratios of collisionally-excited metal lines relative to H$\beta$ only.  Thus, the line ratios used in the fitting
 process include [O\iii]$\lambda$5007/H$\beta$ (O3), [O\ii]$\lambda\lambda$3726/H$\beta$ (O2), and [Ne\iii]$\lambda$3869/H$\beta$ (Ne3).
Given the tight relation between O3 and Ne3 at sub-solar metallicities \citep{lev14}, a galaxy must have at minimum
 detections of O3 and O2 to fit for $U$ and $Z_*$, though Ne3 is also used when available.
Examples of this fitting process are shown in the left panels of Figures~\ref{fig9} to~\ref{fig12} for
 the four MOSDEF [O\iii]$\lambda$4363 emitters.

We estimate confidence intervals on the best-fit $U$ and $Z_*$ by perturbing the line ratios and
 direct-method O/H according to the uncertainties, and repeating the fitting procedure above on each
 realization, 5000 times in total.
The uncertainties on $U$ and $Z_*$ thus include uncertainties in the direct-method O/H that sets $Z_{\text{neb}}^{\text{in}}$.
The resulting distributions of $U$, $Z_*$, and $Z_*/Z_{\text{neb}}^{\text{in}}$ are displayed
 in the right panels of Figures~\ref{fig9} to~\ref{fig12}.  The red ``X" and lines denote the best fit
 in each panel.

Of the 18 $z>1$ auroral-line emitters in our sample, 4 have limits on the [O\ii] flux and one lacks [O\ii] coverage
 and thus cannot be fit for $U$ and $Z_*$.
For the other 13 objects and two composites, the ionization parameter is always constrained, typically
 within $\pm0.2$~dex, and has a weak covariance with $Z_*$.
The stellar metallicity is well-constrained (i.e., $1\sigma$ confidence intervals within the explored parameter space)
 for 6 $z>1$ galaxies (see, e.g., Figs.~\ref{fig9} and~\ref{fig11}), as well as the \citet{ste16} composite.
Six of the $z>1$ galaxies and the MOSDEF stack have one-sided constraints on $Z_*$ (e.g., Figs.~\ref{fig10} and~\ref{fig12}),
 which we display as $3\sigma$ upper limits.  No targets produced lower limits on $Z_*$.
The single remaining $z>1$ galaxy (C12a) has a $1\sigma$ $Z_*$ confidence interval that spans the full parameter space, thus no constraint
 could be placed on $Z_*$, although $U$ could still be constrained with larger uncertainties.
The best-fit $U$ and $Z_*$ values assuming the BPASS grids are given in Table~\ref{tab:uzstar}.

\begin{table}
 \centering
 \caption{Constraints on the ionization parameter and stellar metallicity for the $z>1$ auroral-line sample.
 }\label{tab:uzstar}
 \begin{tabular}{ l l l l }
   \hline\hline
   Object & log($U$) & $\log{(\frac{\mbox{Z}_*}{\mbox{Z}_\odot})}$ & $\log{(\frac{\mbox{Z}_{\text{neb}}^{\text{in}}}{\mbox{Z}_\odot})}$   \\
   \hline
   \hline
   \makecell{AEGIS\\11452} & -2.33$^{+0.20}_{-0.14}$ & -0.73$^{+0.51}_{-2.41}$ & -0.72$^{+0.21}_{-0.15}$ \\
   \makecell{GOODS-S\\41547} & -2.76$^{+0.16}_{-0.06}$ & $<$-0.14$^{a}$ & -0.60$^{+0.31}_{-0.22}$ \\
   \makecell{COSMOS\\1908} & -1.84$^{+0.04}_{-0.18}$ & -1.03$^{+0.08}_{-0.72}$ & -0.42$^{+0.11}_{-0.14}$ \\
   \makecell{COSMOS\\23895} & -2.67$^{+0.10}_{-0.06}$ & $<$-1.55$^{a}$ & -0.45$^{+0.25}_{-0.17}$ \\
   \makecell{S13} & -2.30$^{+0.05}_{-0.01}$ & $<$-1.95$^{a}$ & -0.49$^{+0.07}_{-0.07}$ \\
   \makecell{J14} & -2.28$^{+0.07}_{-0.14}$ & -0.85$^{+0.12}_{-2.29}$ & -0.32$^{+0.40}_{-0.22}$ \\
   \makecell{C12a} & -2.35$^{+0.22}_{-0.05}$ & ---$^{b}$ & -0.98$^{+0.21}_{-0.20}$ \\
   \makecell{C12b} & -2.80$^{+0.01}_{-0.01}$ & -2.02$^{+0.01}_{-0.06}$ & -0.30$^{+0.03}_{-0.03}$ \\
   \makecell{Ste14c} & -2.11$^{+0.04}_{-0.07}$ & -1.72$^{+0.28}_{-0.44}$ & -0.29$^{+0.04}_{-0.06}$ \\
   \makecell{Ste14a} & -2.33$^{+0.09}_{-0.06}$ & $<$-1.33$^a$ & -0.41$^{+0.06}_{-0.06}$ \\
   \makecell{Ste14b} & -2.15$^{+0.04}_{-0.07}$ & $<$-0.55$^a$ & -0.34$^{+0.02}_{-0.06}$ \\
   \makecell{C12c} & -2.27$^{+0.04}_{-0.11}$ & -1.20$^{+0.14}_{-1.94}$ & -0.68$^{+0.01}_{-0.22}$ \\
   \makecell{B14} & -2.23$^{+0.05}_{-0.01}$ & $<$-2.42$^{a}$ & -0.34$^{+0.08}_{-0.08}$ \\
   \hline
   \multicolumn{4}{c}{composites} \\
   \hline
   \makecell{MOSDEF} & -2.63$^{+0.08}_{-0.02}$ & $<$-1.00$^{a}$ & -0.57$^{+0.09}_{-0.08}$ \\
   \makecell{KBSS-LM1} & -2.85$^{+0.02}_{-0.02}$ & -2.00$^{+0.13}_{-0.21}$ & -0.30$^{+0.03}_{-0.03}$ \\
   \hline
 \end{tabular}
 \begin{flushleft}
 $^{a}$ {3$\sigma$ upper limit.}
 $^{b}$ {The 1$\sigma$ limits on $Z_*$ spanned the entire parameter space, thus no constraint or limit on $Z_*$ was obtained.}
 \end{flushleft}
\end{table}

Our methodology of fixing the nebular metallicity of the models to fit for $Z_*$ is complementary to
 the method of \citet{ste16}, who fixed $Z_*$ according to the stacked FUV continuum spectrum and fit for nebular
 metallicity.
For the \citet{ste16} composite, we obtain a best-fit log($U)=-2.85$, consistent with their best-fit value of log($U)=-2.8$ when employing
 BPASS binary models with a 100~$\msun$ IMF cutoff.
We find a best-fit stellar metallicity of $Z_*/\text{Z}_{\odot}=0.01$, somewhat lower than the best-fit value of $Z_*/\text{Z}_{\odot}=0.07$
 from \citet{ste16}.
We note that \citeauthor{ste16}\ only considered BPASS models with $Z_*/\text{Z}_{\odot}\ge0.07$ ($Z\ge0.001$), such that the best-fit stellar
 metallicity was at the lower bound of the parameter space.
In this work, we allow the stellar metallicity of the BPASS models to be as low as $Z_*/\text{Z}_{\odot}=0.0007$ ($Z=10^{-5}$).
It is possible that the FUV continuum of the KBSS-LM1 stack could favor an even lower $Z_*$ if the range of permitted metallicities was increased.
We conclude that both our method and that presented in \citet{ste16} reliably constrain $U$, but further investigation is required
 to understand whether both methods infer similar $Z_*$.
In the future, we will analyze individual $z>1$ galaxies with both auroral-line and FUV spectral continuum measurements
 to provide a more rigorous validation of these techniques and test of the stellar models.

\subsection{Stellar metallicity and O/Fe at $z>1$}\label{sec:ofe}

\begin{figure}
 \includegraphics[width=\columnwidth]{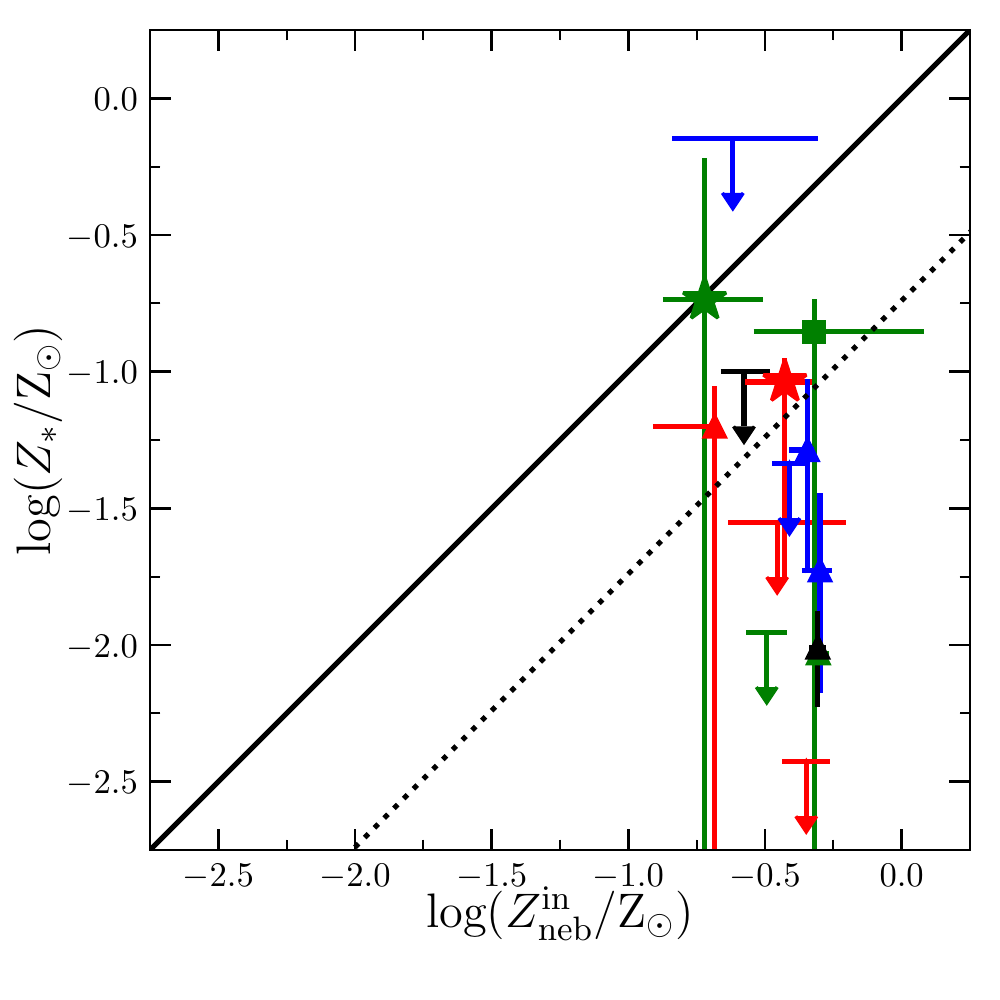}
 \centering
 \caption{
Best-fit stellar metallicity ($Z_*$) vs.\ input nebular metallicity ($Z_{\text{neb}}^{\text{in}}$) for $z>1$ galaxies.
  The input nebular metallicity is set by adding 0.24~dex to the observed direct-method 12+log(O/H) given in Table~\ref{tab:props}.
  Points and color-coding are as in Figure~\ref{fig3}.
  The solid black line shows a one-to-one relation, while the dotted black line shows $Z_*=1/5Z^{\text{in}}_{\text{neb}}$,
 the theoretical limit from pure Type II SNe enrichment \citep{nom06}.
  The $z>1$ sample strongly prefers $Z^{\text{in}}_{\text{neb}}>Z_*$ (i.e., super-solar O/Fe).
}\label{fig13}
\end{figure}

We now utilize these best-fit values of the ionization parameter and stellar metallicity to explore
 the ionization state of $z>1$ star-forming regions and the redshift evolution of these properties.
In Figure~\ref{fig13}, we compare the best-fit $Z_*$ (i.e., Fe/H) with the
 nebular metallicity $Z_{\text{neb}}^{\text{in}}$ (i.e., O/H) of each $z>1$ galaxy.
The solid black line shows a one-to-one relation where $Z_*$=$Z_{\text{neb}}^{\text{in}}$ on a solar scale or,
 equivalently, O/Fe=O/Fe$_{\odot}$.
Overall, we find a strong preference for super-solar O/Fe ($Z_*$$<$$Z_{\text{neb}}^{\text{in}}$).\footnote{
The preference for super-solar O/Fe is not driven by our assumption that log($Z_{\text{neb}}^{\text{in}}$)=log($Z_{\text{neb}})+0.24$.
If we instead assumed $Z_{\text{neb}}^{\text{in}}$=$Z_{\text{neb}}$, O/H$^{\text{in}}$ would be lower than in our fiducial case, but \textit{even lower} values of
 $Z_*$ (i.e., Fe/H) than those in Table~\ref{tab:uzstar} would be required to fit the observed line ratios.  In many cases, $Z_*<0.0007~\text{Z}_{\odot}$
 (the lowest stellar metallicity in the BPASS models) would be required (Fig.~\ref{fig8}, left column) and the inferred O/Fe would increase compared to
 our fiducial case.}
Only two galaxies have constraints or limits that are consistent with $Z_*$=$Z_{\text{neb}}^{\text{in}}$,
 and the one that is not an upper limit has a large uncertainty towards lower $Z_*$.
The remaining 11 individual galaxies and two composites favor $Z_*$$<$$Z_{\text{neb}}^{\text{in}}$.

Six galaxies and the \citet{ste16} stack have inferred $Z_*$ less then 1/5~$Z_{\text{neb}}^{\text{in}}$,
 where $\approx$5$\times$O/Fe$_{\odot}$ is the theoretical limit for $\alpha$-enhancement from pure Type II supernova
 enrichment \citep[dotted line in Fig.~\ref{fig13};][]{nom06}.
These galaxies are thus in tension with the standard picture of chemical enrichment through star formation.
It is possible that uncertainties on elemental yields in very metal-poor stars ($Z_*<0.03~\text{Z}_{\odot}$)
 could relieve this tension if metal-poor supernovae have lower Fe yields relative to O.
However, some objects fall $>1$~dex below the Type II SNe limit.
The ionizing spectra of metal-poor massive stars is highly uncertain, and few observational constraints currently
 exist \citep{sen17,sen19,sta19}.  If metal-poor massive stars produce \textit{harder} ionizing spectra than the current stellar models
 predict, then the true stellar metallicities would in fact be higher.

Alternatively, a top-heavy IMF at high redshifts may relieve tension with the theoretical Type II SNe limit.
The limit of 5$\times$O/Fe$_{\odot}$ is an IMF-integrated value assuming a canonical high-mass
 IMF slope of $\approx-2.3$.
SNe of very massive ($\gtrsim25~\msun$) stars yield $>10\times$O/Fe$_{\odot}$ \citep{nom06,kob06}, such that the IMF-integrated
 O/Fe can be larger than 5$\times$O/Fe$_{\odot}$ for top-heavy IMFs (i.e., with high-mass slope shallower than $-2.3$).
A top-heavy IMF would also imply a harder ionizing spectrum at fixed metallicity due to the increased relative
 abundance of hotter, more massive stars.
Thus, for the case of a top-heavy IMF, $>5\times$O/Fe$_{\odot}$ is allowed and a higher $Z_*$ (i.e., Fe/H)
 would best fit the observed line ratios compared to the values of $Z_*$ inferred under our fiducial model
 (assuming a \citealt{cha03} IMF), lowering the inferred O/Fe.
Recent observational and theoretical work has suggested that the IMF grows more top-heavy in low-metallicity and
 high-SFR environments common among high-redshift galaxies \citep{jer18,sch18}.
The potential impact of IMF variations on the ionizing spectrum in high-redshift star-forming regions should
 be further investigated in future work.

Despite the large uncertainty on the ionizing spectra of metal-poor stellar populations at high redshift, our results still
 strongly disfavor a scenario in which $Z_*$$=$$Z_{\text{neb}}^{\text{in}}$, which would occur at stellar metallicities
 where better observational constraints are available ($\sim1/3-1/2$~Z$_{\odot}$).
A third possibility is that the sources below the Type II SNe limit have additional ionization mechanisms beyond
 star formation that harden the galaxy-averaged ionizing spectrum, such as shocks or weak AGN.
The galaxies in the $z>1$ sample do not exhibit any obvious signatures of
 AGN, but a low-luminosity AGN in a highly star-forming galaxy may only slightly shift the observed line ratios.
However, it is unlikely that the presence of shocks or AGN are responsible for the position of the
 \citet{ste16} stack of 30 galaxies unless such a phenomenon is ubiquitous among typical star-forming
 galaxies at $z\sim2$.
Ultimately, we require more confidence in stellar models of metal-poor ($\lesssim$0.2~Z$_{\odot}$) massive stars to understand
 the behavior of galaxies falling below the Type II SNe limit.

\begin{figure}
 \includegraphics[width=\columnwidth]{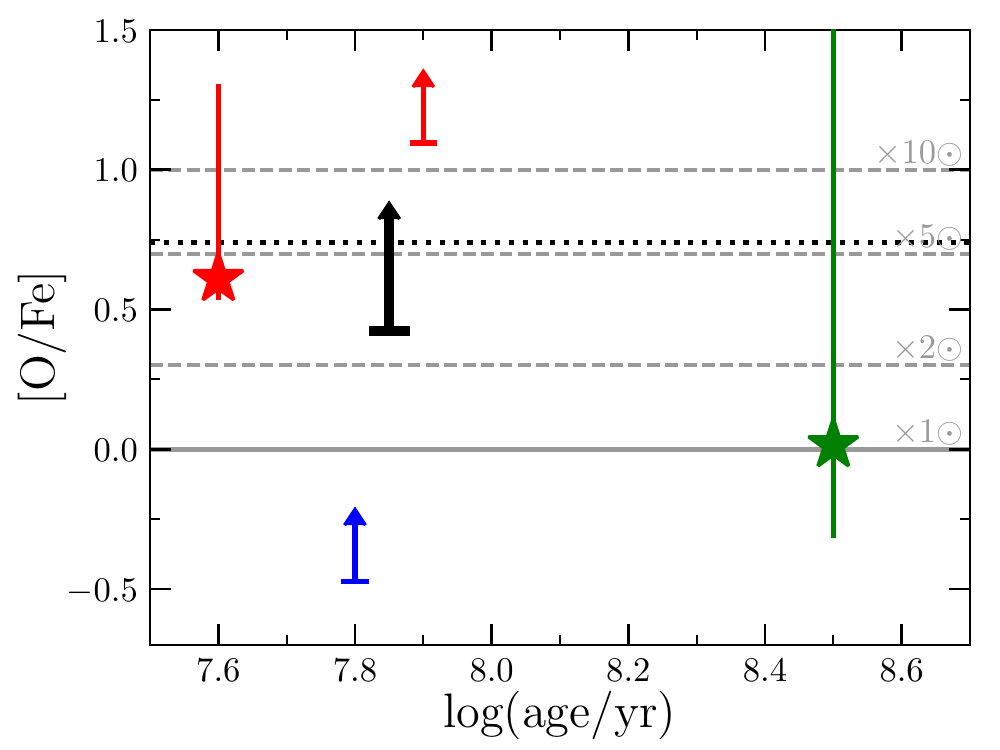}
 \centering
 \caption{
O/Fe abundance ratio relative to the solar value vs. stellar population age from SED fitting for the four MOSDEF [O\iii]$\lambda$4363 metallicity sample.
  Points are color-coded by redshift as in Figure~\ref{fig3}.  The thick black point shows the value inferred from the MOSDEF [O\iii]$\lambda$4363 composite spectrum, adopting the median age of the individual galaxies.
  The black dashed line shows the theoretical upper-limit of [O/Fe]=0.74 from pure Type II SNe enrichment at $Z_*=0.07\text{Z}_{\odot}$ \citep{nom06}.
Gray lines show 1, 2, 5, and 10 times O/Fe$_{\odot}$.
}\label{fig14}
\end{figure}

We show the O/Fe ratio as a function of stellar popuation age for the MOSDEF [O\iii]$\lambda$4363 emitters
 and stack in Figure~\ref{fig14}.
The stellar population age is derived from the best-fit SED models, and O/Fe is on a solar scale such that
 [O/Fe]=log((O/Fe)/(O/Fe$_{\odot}$)).
We find that all four galaxies have young stellar populations ($<$300~Myr), with three younger than 100~Myr.
Two of the MOSDEF galaxies are constrained to have super-solar O/Fe, while the other two do not have strong
 constraints.
The composite spectrum yields a 3$\sigma$ lower-limit of O/Fe$\approx$$3\times$O/Fe$_{\odot}$.
The ages of the MOSDEF [O\iii]$\lambda$4363 emitters are consistent with the presence of super-solar
 O/Fe from an enrichment history dominated by Type II SNe, with solar O/Fe expected to be reached only at ages $\gtrsim$1~Gyr.
The [O\iii]$\lambda$4363 emitters have ages younger than 90\% of MOSDEF star-forming galaxies matched in
 \mstar\ and redshift.
While the precise stellar population ages are significantly uncertain due to systematics associated with
 the assumed star-formation history, attenuation law, and metallicity for SED fitting, relative ages
 should be robust under our assumption of constant star formation.
Furthermore, the large sSFRs and emission-line equivalent widths of the $z>1$ sample (Table~\ref{tab:props}) suggest
 very young ages, in agreement with results from the SED fitting.
The four MOSDEF [O\iii]$\lambda$4363 emitters appear to be very young star-forming galaxies in which
 the delayed Fe enrichment from Type Ia SNe has not yet occurred for the bulk of past star formation.

\subsection{Ionization parameter at $z>1$}\label{sec:u}

The $z>1$ star-forming galaxy population has considerably higher ionization parameters than are typical of
 $z\sim0$ samples (e.g., SDSS), as evidenced by significantly higher O32 and O3 at fixed \mstar\ 
 \citep[e.g.,][]{nak14,san16a,san18,dic16,hol16,kas17,str17,kas19a}.
sSFR has been shown to strongly correlate with O32 and O3 \citep[e.g.,][]{nak14,san16a,kew15,dic16,kas19b},
 and is consequently closely tied to $U$.
However, sSFR is anticorrelated with O/H according to the \mstar-SFR-O/H relation present at low and high redshifts:
galaxies with higher SFR at fixed \mstar\ (i.e., higher sSFR) have lower O/H \citep[e.g.,][]{man10,lar10,and13,san18}.
Furthermore, $U$ is tightly anticorrelated with O/H in the local universe \citep{dop06a,dop06b,per14,sanc15,kas19b}.
It is therefore possible that the high values of $U$ inferred at high redshifts are simply a byproduct of the
 lower metallicities and higher sSFRs of high-redshift galaxies compared to the local universe.
It has been proposed that the redshift evolution of strong-line ratio sequences is primarily driven by
 enhanced $U$ at high redshifts compared to typical $z\sim0$ values at fixed O/H \citep{kew15,kew16,kas17,kas19a,kaa18}.
In other words, the $U$-O/H relation evolves with redshift such that $U$ increases at fixed O/H with increasing redshift.
We test this scenario by comparing $U$, sSFR, and direct-method O/H of our $z>1$ sample to the corresponding measurements
 of local star-forming galaxies.

\begin{figure}
 \includegraphics[width=0.97\columnwidth]{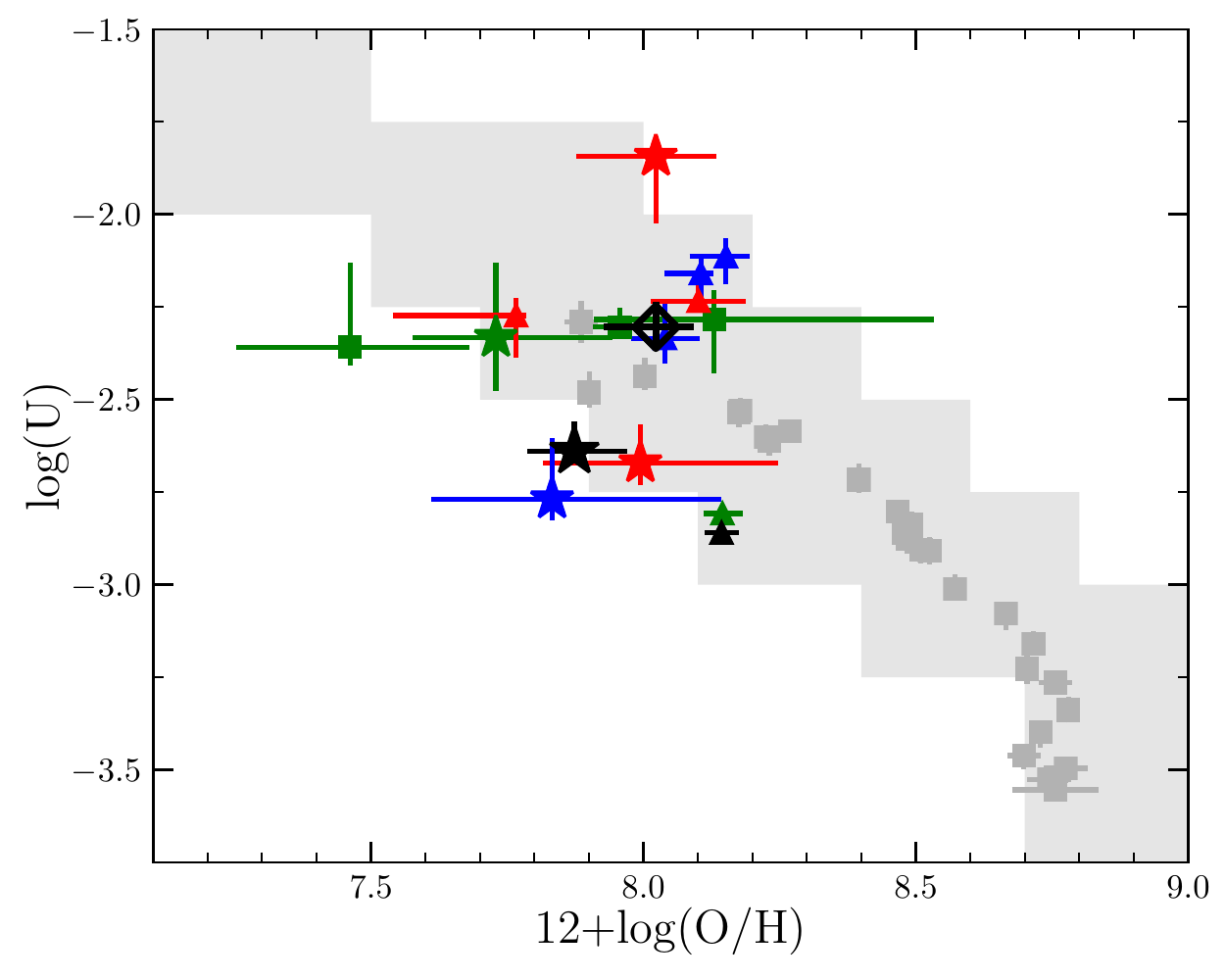}
 \includegraphics[width=0.97\columnwidth]{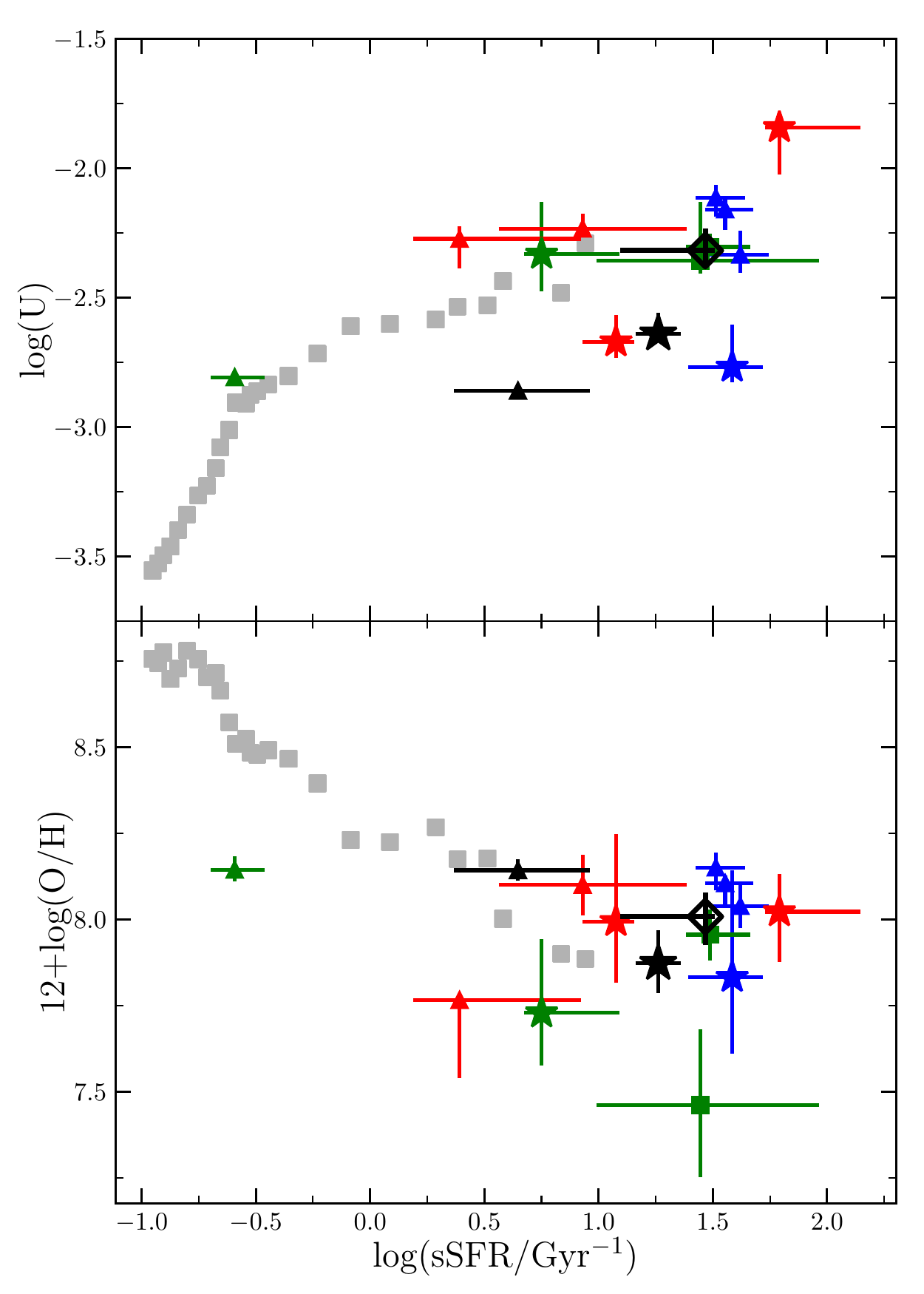}
 \centering
 \caption{
Relationships between $U$, direct-method nebular O/H, and sSFR for $z>1$
 galaxies (colored points) and stacked spectra of $z\sim0$ galaxies \citep[gray points;][]{and13}.
  Shapes and colors are as in Figure~\ref{fig3}, with the black open diamond denoting the median of the $z>1$ sample.
  The light-gray shaded region in the top panel shows the range of U and 12+log(O/H) of local \hii regions found by \citet{per14}.
  On average, galaxies at $z>1$ do not show a significant offset from the mean $z\sim0$ $U$-O/H, $U$-sSFR, and sSFR-O/H relations.
}\label{fig15}
\end{figure}

Figure~\ref{fig15} compares $U$, sSFR, and temperature-based O/H for the $z>1$ auroral-line sample
 and $z\sim0$ \mstar-binned composites of \citet{and13}.
We determine $U$ for the \citet{and13} stacks using the same models and fitting procedure described in
 Section~\ref{sec:fitting}, except that we fix $Z_*=Z_{\text{neb}}^{\text{in}}$ as expected for $z\sim0$
 galaxies with more extended star-formation histories than high-redshift galaxies.
We note that allowing $Z_*$ to vary freely does not significantly change the results.
Median values of the $z>1$ sample of individual galaxies are also displayed in each panel as a open black diamond.

As expected, we find that $U$ is anticorrelated with O/H for $z\sim0$ galaxies, as shown in the top panel of
 Figure~\ref{fig15}.  We find our mean $U$-O/H relation for $z\sim0$ SDSS stacks closely matches that
 found by \citet{per14} for $z=0$ \hii regions (shaded light gray region).
The $z>1$ auroral-line sample scatters above and below the local relation defined by the $z\sim0$ stacks.
We do not have a suitable sample size or dynamic range in O/H to look for an anticorrelation among the
 high-redshift galaxies alone.
The $z>1$ galaxies display a significant scatter in $U$ at fixed O/H, but the intrinsic scatter is similar to
 the range spanned by local \hii regions after accounting for measurement uncertainties, $\sim0.25$~dex in $U$
 at fixed O/H.
The median values of the $z>1$ sample are 12+log(O/H$)^{\text{med}}=8.02^{+0.06}_{-0.09}$
 and log($U)^{\text{med}}=-2.30^{+0.06}_{-0.06}$.
The median of the $z>1$ sample is in remarkable agreement with the $z\sim0$ $U$-O/H relation
 of both local galaxies and individual \hii regions.

We show $U$ vs.\ sSFR in the middle panel of Figure~\ref{fig15}.
We find a correlation between $U$ and sSFR among the $z\sim0$ stacks.
Once again, the $z>1$ auroral-line sample scatters both above and below the $z\sim0$ relation.
Interestingly, the lowest-sSFR galaxy in the $z>1$ sample \citep[C12b;][]{chr12a,chr12b},
 with sSFR $\sim1.5$~dex below the sample average, falls directly on the $z\sim0$ relation
 and has a particularly tightly-constrained U and O/H.
The lowest-mass \citet{and13} bins have similar sSFR to the range spanned by the $z>1$ sample (Fig.~\ref{fig5}).
The median $z>1$ values fall on the $U$-sSFR relation defined by the $z\sim0$ composites.
 
In the bottom panel of Figure~\ref{fig15}, we present O/H vs. sSFR, finding the anticorrelation
 between O/H and sSFR at $z\sim0$ as expected from the FMR.
The $z>1$ galaxies follow the same relation between O/H and sSFR displayed
 by the $z\sim0$ galaxies.  This result is expected based on the agreement between $z\sim0$, $z\sim0.8$,
 and $z\sim2.2$ samples in the FMR projection shown in the bottom panel of Figure~\ref{fig7}.
The $z>1$ median falls on the extrapolation of the relation displayed $z\sim0$ stacks with lower sSFR.

In summary, Figure~\ref{fig15} shows that, on average, the $z>1$ galaxies fall on the local relationships between
 $U$, O/H, and sSFR.
The high ionization parameters of the $z>1$ sample are therefore as expected in accordance with their
 low metallicities and high sSFRs, and are not in excess of what is expected at fixed O/H compared to
 $z\sim0$ galaxies.
This result suggests that variations in $U$ beyond local relations is not a significant driver of the evolving
 strong-line ratios of galaxies.
Combining these results with those presented in Section~\ref{sec:ofe}, we conclude that the primary cause
 of strong-line ratio evolution at fixed O/H (Fig.~\ref{fig3}) for our $z>1$ auroral-line sample is a
 harder ionizing spectrum at fixed nebular metallicity compared to what is typical in $z\sim0$ galaxy populations.
The harder spectrum at fixed O/H in the $z>1$ auroral-line sample is naturally explained by the youth of their stellar
 populations, leading to chemically-immature galaxies that have super-solar O/Fe due to the predominance of Type II SNe enrichment
 that promptly produces O and a lack of time-delayed Type Ia SNe that are the main source of Fe.
The Fe-poor massive stars produce harder ionizing spectra than their counterparts at solar O/Fe.

\section{Discussion}\label{sec:discussion}

In this work, we have used a sample of 18 individual $z>1$ galaxies with direct-method metallicities to
 test strong-line metallicity indicators, construct the mass-metallicity relationship,
 investigate the redshift invariance of the \mstar-SFR-O/H relation,
 and demonstrate that the high ionization state of this sample is driven by an ionizing spectrum that is
 harder than that of $z\sim0$ galaxies at fixed O/H.
In this section, we discuss the implications of these results for typical galaxy populations at $z\sim1-3$,
 extreme emission-line galaxy samples at $z\sim1-3$, and galaxies in the epoch of reionization at $z>6$.

\subsection{Implications for typical $z\sim1-3$ galaxies}\label{sec:typicalcomparison}

Obtaining accurate metallicity estimates from strong-line ratios for large existing spectroscopic survey
 data sets at $z\sim1-3$ (e.g., MOSDEF, \citealt{kri15}; KBSS ,\citealt{ste14}; 3D-HST, \citealt{bra12a}; FMOS-COSMOS, \citealt{kas19a})
 requires knowledge of the ionization state of the ISM in typical high-redshift galaxies.
We have shown that our $z>1$ auroral-line sample has significantly higher SFR at fixed \mstar\ than galaxies
 falling on the mean \mstar-SFR relation at the same redshifts (Fig.~\ref{fig5}).
We now put the $z>1$ auroral-line sample in the context of a representative sample of star-forming galaxies at $z\sim2.3$,
 and discuss which conclusions, if any, can be extended to the typical galaxy population at $z>1$.

\begin{figure}
 \includegraphics[width=\columnwidth]{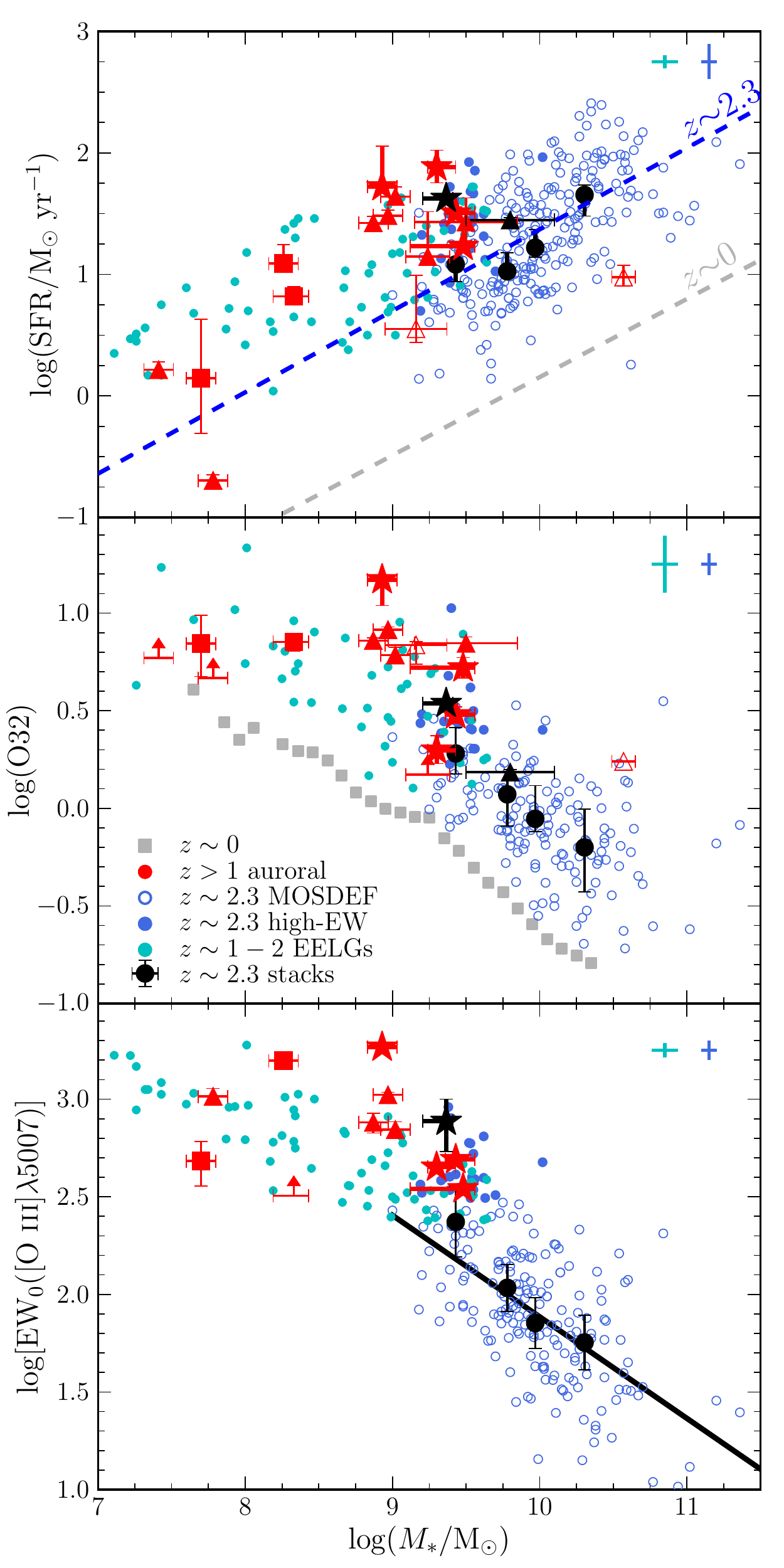}
 \centering
 \caption{SFR (top), O32 (middle), and EW$_0$([O\iii]$\lambda$5007) (bottom) vs. \mstar\ for high-redshift galaxy samples.
The $z>1$ auroral-line sample is shown as red points, with shapes as in Fig.~\ref{fig3}.
Hollow triangles denote the two probable mergers (C12b and C12c).
Open blue circles display a representative sample of $z\sim2.3$ star-forming galaxies from MOSDEF \citep{san18}, while filled blue
 circles denote the subset of that sample with EW$_0$([O\iii]$\lambda$5007)$>$300~\AA.
The $z\sim2$ extreme emission-line galaxy sample of \citet{tan18} is presented as cyan circles.
In each panel, the cyan and blue error bar denotes the mean uncertainty of the \citet{tan18} and $z\sim2.3$ MOSDEF samples, respectively.
Composite spectra are shown in black for the stack of MOSDEF [O\iii]$\lambda$4363 emitters (star), \citet[][; triangle]{ste16},
 and typical $z\sim2.3$ MOSDEF galaxies \citep[circles;][]{san18}.
The blue dashed line in the top panel shows the best-fit \mstar-SFR relation of equation~\ref{eq:sfrmstar}.
Gray points and dashed line show mean $z\sim0$ relations \citep{and13}.
The mean $z\sim2.3$ EW$_0$([O\iii]$\lambda$5007) vs.\ \mstar\ relation from \citet{red18} is shown as the solid black line in
 the bottom panel.
}\label{fig16}
\end{figure}

In Figure~\ref{fig16}, we present SFR, O32, and EW$_0$([O\iii]$\lambda$5007) as a function of stellar mass.
We compare the $z>1$ auroral-line sample (red points) to a sample of $z\sim2.3$ star-forming galaxies from MOSDEF
 representative of the typical population at log(\mstar/$\msun)=9-11$ from \citet{san18} (open blue circles).
Black circles show the mean properties of this sample obtained via spectral stacking in four \mstar\ bins,
 where we have derived the mean EW$_0$([O\iii]$\lambda$5007) in each bin according to the method described in
 Section~\ref{sec:MOSDEFo4363}.
We find that all but two galaxies in the auroral-line sample (median redshift of 2.2) fall above the mean $z\sim2.3$
 \mstar-SFR relation, with a mean offset of $\langle\Delta\text{log(SFR)}\rangle\sim0.6$~dex.
The O32 values of the typical $z\sim2.3$ sample are much lower than those of the auroral-line sample, but the typical
 sample has a significantly higher average stellar mass.  When comparing at log(\mstar/$\msun)=9.0-9.5$ where the
 two samples overlap, the auroral-line sample has O32 values $\sim0.5$ dex higher on average.
The equivalent widths of the auroral-line sample span EW$_0$([O\iii]$\lambda5007)=300-2000$~\AA, with a median value
 of $\sim500$~\AA.  This value is much larger than the mean value of the $z\sim2.3$ comparison sample ($\sim100$~\AA).
It is important to take into account the fact that EW$_0$([O\iii]$\lambda$5007) has a strong dependence on stellar mass \citep{red18}.
However, even when comparing at fixed \mstar, the auroral-line sample displays EW$_0$([O\iii]$\lambda$5007) that is larger
 by a factor of 3.

\begin{figure}
 \includegraphics[width=\columnwidth]{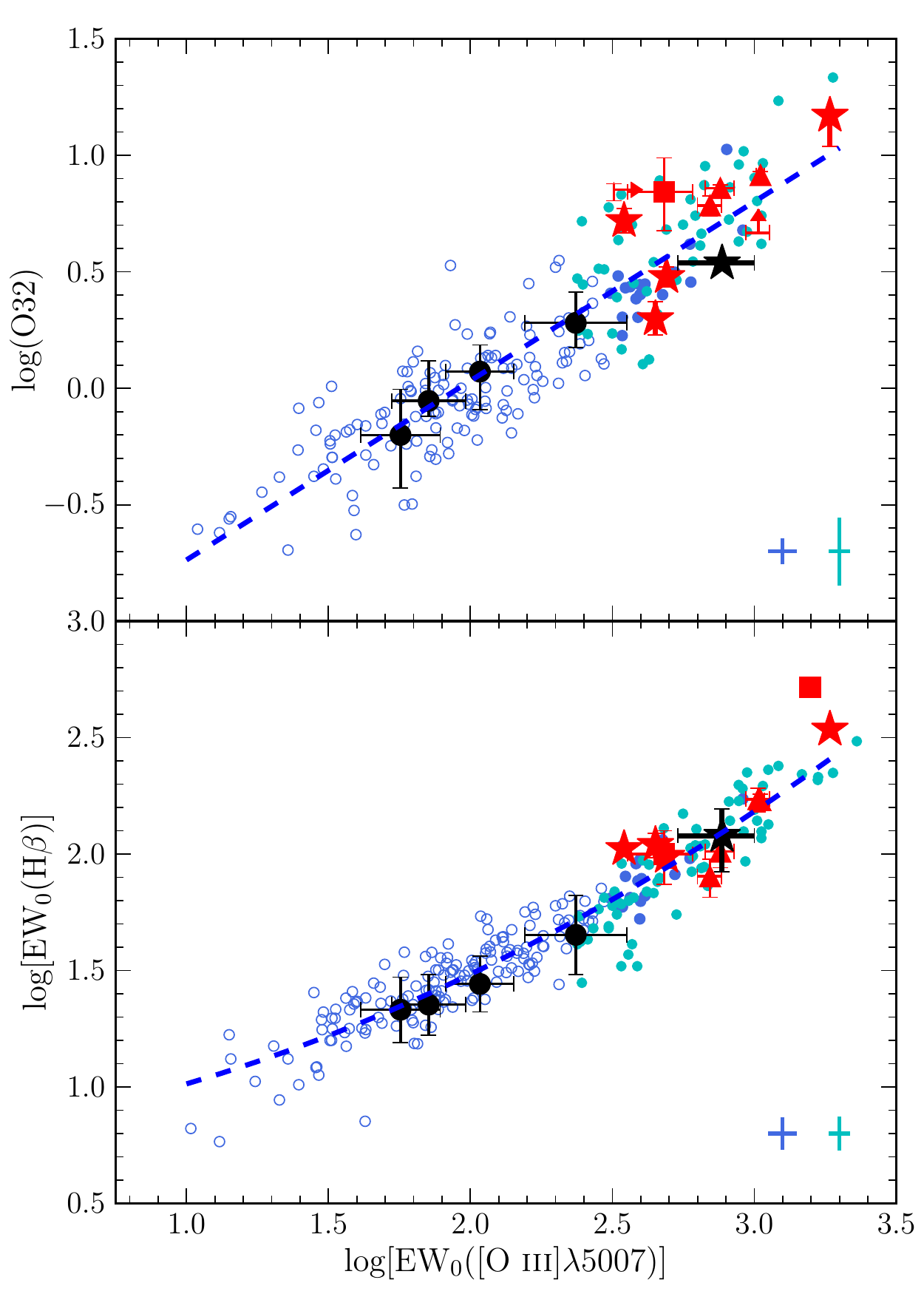}
 \centering
 \caption{
O32 (top) and EW$_0$(H$\beta$) (bottom) as a function of EW$_0$([O\iii]$\lambda$5007).
Points are as in Figure~\ref{fig16}.
Both panels display remarkably tight correlations over 2.5 orders of magnitude in EW$_0$([O\iii]$\lambda$5007).
}\label{fig17}
\end{figure}

We show O32 and EW$_0$(H$\beta$) vs.\ EW$_0$([O\iii]$\lambda$5007) in Figure~\ref{fig17}, and once again find that
 the auroral-line sample is nearly disjoint from the typical $z\sim2.3$ sample, having higher emission-line
 equivalent widths suggestive of younger ages and higher O32 values implying lower O/H and higher $U$.
We select the subset of the $z\sim2.3$ comparison sample with EW$_0$([O\iii]$\lambda5007)>300$~\AA\ (matched
 to the auroral-line sample), displayed as filled blue circles in Figures~\ref{fig16} and ~\ref{fig17}.
The EW-matched $z\sim2.3$ sample reproduces the properties of the auroral-line sample in the range of stellar mass
 overlap, selecting a subset lying above the \mstar-SFR relation at log(\mstar/$\msun)=9.0-9.5$ with similarly
 high O32 values.
However, this EW-matched subset makes up only 11\% (22/203) of the typical log(\mstar/$\msun)\gtrsim9.0$
 star-forming population at $z\sim2.3$ from MOSDEF.

Based on the high sSFRs,  O32 values, and emission-line equivalent widths, the $z>1$ auroral-line sample
 is made up of galaxies that are younger and more metal-poor than the bulk of the typical $z\sim1-3$ galaxy population
 at log(\mstar/$\msun)>9.0$, the mass range that most high-redshift spectroscopic surveys probe.
Accordingly, it is not clear that our conclusions regarding metallicity calibration evolution and O/Fe enhancement
 apply to the bulk of the $z\sim2$ star-forming population.
Based on the high-redshift analog O32 strong-line calibration of BKD18, $\sim$90\% of the $z\sim2.3$
 comparison sample has oxygen abundances higher than 12+log(O/H)=8.14, the highest direct-method O/H of the
 auroral-line sample.
Therefore, the strong-line indicators considered in Figure~\ref{fig3} have not been tested over the metallicity
 range of the bulk of the log(\mstar/$\msun)>9.0$ galaxy population at $z\sim2$.
However, given that the auroral-line sample represents the most extreme subset of the MOSDEF $z\sim2.3$ sample,
 it is expected that the strong-line ratio vs.\ O/H relations of typical $z\sim2$ galaxies will show
 \textit{at most} the same level of evolution as the auroral-line sample ($\sim$0.1~dex in O/H at fixed strong-line ratio).
The BKD18 high-redshift analog calibrations are likely good to within the same amount for the
 typical $z\sim2$ population when calculating sample average abundances.

The very young ages of the auroral-line sample implied by large emission-line equivalent widths and
 sSFRs are consistent with the super-solar O/Fe values found from the photoionization models.
However, typical $z\sim2$ galaxies at log(\mstar/$\msun)>9.0$ appear to have older stellar populations.
Star-forming galaxies with stellar populations 300~Myr to $\sim1$~Gyr old may still display some level of O/Fe enhancement
 relative to solar values, while the abundance ratios of galaxies older than $\sim1$~Gyr will lie close to O/Fe$_{\odot}$.
The age of the universe at $z=2.3$ is 2.8~Gyr, and can thus accommodate stellar populations older than 1~Gyr.
\citet{red18} find that galaxies with EW$_0$([O\iii]$\lambda5007)<100$~\AA\ (the typical value at log(\mstar/$\msun)\sim10.0$
 and $z\sim2.3$) have ages $>$1~Gyr according to SED fitting.
It is therefore possible that roughly half of the $z\sim2.3$ MOSDEF sample has O/Fe$_{\odot}$ while the other half
 has enhanced O/Fe.
However, ages inferred from SED fitting are sensitive to assumptions regarding the stellar population models, stellar metallicity,
attenuation curve, and star-formation history, such that there are large systematic uncertainties on the ages.

To confirm or deny the presence of O/Fe enhancement in the typical $z\sim2$ population requires
 measurements of either auroral-lines or high-S/N rest-FUV continuum, from which Fe/H can be inferred,
 for individual galaxies spanning the \mstar-SFR relation.
Detecting [O\iii]$\lambda$4363 is currently out of reach for galaxies on the mean \mstar-SFR relation where
 fluxes are expected to be $\sim5-200$ times weaker than for the MOSDEF [O\iii]$\lambda$4363 emitters in this work.
With current facilities, obtaining high-S/N spectra of the rest-FUV is the most viable option over a wide range of \mstar, SFR,
 and metallicity, especially for massive metal-rich galaxies.

If the $z\sim2$ star-forming population is ``mixed" as implied by the ages in \citet{red18},
 with chemically-immature low-mass galaxies having super-solar O/Fe
 and mature high-mass galaxies having solar O/Fe, it carries implications for measurements of metallicity scaling relations.
Assuming a single metallicity calibration in one of the two extremes (i.e., the high-redshift analog vs.\ $z\sim0$
 reference calibrations of BKD18) would lead to systematic biases in the slope of the MZR, such that the
 measured MZR would be artificially steepened.
If such a scenario holds, a different metallicity calibration must be applied in the low- and high-mass regimes,
 or else age-dependent calibrations must be constructed.
There is likely an age gradient above and below the \mstar-SFR relation as well, which will bias measurements of
 the \mstar-SFR-O/H relation for a mixed population.
The strong-line metallicities of the highest-sSFR local galaxies, analogous to $z\sim2$ systems, may also be biased.
To fully characterize the evolution of gas-phase metallicity with global galaxy properties, we must understand
 at what redshift O/Fe enhancement begins affecting the star-forming population, the relation of O/Fe with \mstar\ 
 and SFR, and at what redshift the full star-forming population is dominated by Type II SNe enrichment.

\citet{ste16} find super-solar O/Fe using a composite spectrum of 30 star-forming galaxies at $z\sim2.3$ falling on
 the \mstar-SFR relation, and we find consistent results when applying our methodology.
This finding may signify that many $z\sim2$ galaxies on the \mstar-SFR relation are O/Fe enhanced,
 but the sample-averaged measurement cannot address how O/Fe varies across the sample, a pressing question
 given that stellar population age (and thus O/Fe) varies with \mstar\ and sSFR.
Additionally, there are concerns that the \citet{ste16} KBSS-LM1 stack may be biased such that it is not
 representative of typical $z\sim2$ samples.
In Figure~\ref{fig15}, the KBSS-LM1 stack (black triangle) has a peculiar low ionization parameter
 for its direct-method O/H and sSFR, falling significantly below the $z\sim0$ relations and lower than
 any galaxies in our $z>1$ sample.
Expectations are that high-redshift galaxies will have \textit{at least} as high $U$ as local galaxies
 at fixed O/H and sSFR.  The offset of the KBSS-LM1 stack is much more significant than individual galaxies
 scattering low in these relations since it is an average of 30 galaxies.
It is possible that the stacking procedure has systematically biased the rest-optical lines towards
 lower-excitation ratios.
Individual determinations of O/Fe (via auroral lines or rest-UV continuum)
 at higher \mstar\ and lower sSFR than our $z>1$ auroral-line sample are
 sorely needed to validate stacking methods and explore O/Fe as a function of galaxy properties.

\citet{ste16} and \citet{str18} find that harder ionizing spectra at fixed O/H produced by super-solar O/Fe
 stellar populations are responsible for driving the well-known offset of $z\sim2$ galaxies from the local
 star-forming sequence in the [N\ii] BPT diagram.
\citet{str18} further find that elevated N/O at fixed O/H is required to explain the most highly-offset $z\sim2$ objects.
Our results are consistent with this scenario.  However, with the current sample, we cannot evaluate whether this is the
 case for the more massive galaxies for which the [N\ii] BPT offset is observed, if O/Fe enhancement alone can fully account for
 the observed offset, and what the role of N/O variation is.
Our current $z>1$ auroral-line sample lacks [N\ii] coverage and detections in most galaxies.
Future high-redshift auroral-line samples spanning a wider dynamic range in \mstar\ and metallicity will directly address
 the nature of the [N\ii] BPT diagram offset.

\subsection{Implications for $z\sim1-3$ extreme emission-line galaxies}\label{sec:eelgcomparison}

Extreme emission-line galaxies (EELGs) are a population of galaxies with spectra dominated by
 high equivalent width emission lines, including the ``green peas" in the local universe \citep{car09}.
The large equivalent widths of our $z>1$ auroral-line sample imply similarities to EELGs.
We show the EELG sample of \citet{tan18} at $1.3<z<2.4$ in Figures~\ref{fig15} and~\ref{fig16}
 (cyan points).  This sample was selected to have EW$_0$([O\iii]$\lambda5007)>300$~\AA.
We find that the $z>1$ auroral-line sample spans a similar range of SFR, O32, and EW$_0$([O\iii]$\lambda$5007)
 at fixed \mstar\ as the EELGs, which are high-excitation galaxies lying above the \mstar-SFR relation
 at log(\mstar/$\msun)\sim7-9.5$.  The high-EW $z\sim2.3$ MOSDEF subsample overlaps with the most massive EELGs.

In Figure~\ref{fig16}, the auroral-line sample again aligns with the EELG sample, with similar O32 and EW$_0$(H$\beta$)
 at fixed EW$_0$([O\iii]$\lambda$5007).
The EELGs appear to be a continuation of the relation between O32, EW$_0$(H$\beta$), and EW$_0$([O\iii]$\lambda$5007)
 defined by the $z\sim2.3$ MOSDEF sample, forming remarkably tight sequences over 2.5 orders of magnitude
 in EW$_0$([O\iii]$\lambda$5007).
We fit these relations, obtaining:
\begin{equation}\label{eq:o32ew}
\text{log(O32)} = 0.77x - 1.50
\end{equation}
\begin{equation}\label{eq:hbew}
\text{log[EW}_0(\text{H}\beta)] = 0.12x^2 + 0.12x + 0.77
\end{equation}
where $x=\text{log[EW}_0([\text{O\iii}]\lambda5007)]$, with 0.18 and 0.11~dex scatter in O32 and EW$_0$(H$\beta$), respectively.
These relations can be used to obtain rough information about the O/H and ionization parameter of galaxies with
 large equivalent widths where EW$_0$([O\iii]+H$\beta$) can be inferred from broadband photometry.

Given the common properties of the $z\sim2$ EELGs and $z>1$ auroral-line sample, we conclude that
 the EELGs are also young metal-poor galaxies with super-solar O/Fe, driving extremely hard ionizing spectra
 in their \hii regions.
By modeling the photometry and rest-optical spectra simultaneously, \citet{tan18} find ages younger than 300~Myr
 for the majority of their sample.
\citet{tan18} suggest EELGs have hard ionizing spectra and high ionization parameters based on their large O32 values.
The O32 high-redshift analog calibration of BKD18 yields metallicities spanning 12+log(O/H)=7.7$-$8.4 for the EELG
 sample, with a median of 12+log(O/H$)_{\text{med}}$=8.07, similar to the O/H distribution of the auroral-line sample.
At these metallicities, the $z=0$ $U$-O/H relation predicts high ionization parameters of log($U$)=$-2.0$ to $-2.5$ \citep{per14}.
When modeling the nebular spectra of EELGs, we suggest using BPASS models including binaries, and assuming $Z_*=1/5Z_{\text{neb}}$
 (the pure Type II SNe limit) and the $z=0$ $U$-O/H relation.

\subsection{Implications for typical $z>6$ galaxies and reionization}\label{sec:eelgcomparison}

Star-forming galaxies at $z>6$ are thought to be the sources of ionizing photons that powered
 reionization \citep[e.g.,][]{bou12,bou15,rob15,sta16}.  Through SED fitting and broadband photometric excess, these galaxies have been
 found to have very large EW$_0$([O\iii]+H$\beta$), with typical values of $\sim670$~\AA\ and extending
 up to $>1500$~\AA\ for extreme sources, and high sSFRs ($\sim10$~Gyr$^{-1}$) \citep{lab13,smi14,smi15,rob16}.
These properties are similar to those of the $z>1$ auroral-line sample and the $z\sim2$ EELGs.
Indeed, the \citet{tan18} sample was selected to be analogs of epoch of reionization galaxies.
Given the similar properties of these two samples, we suggest that typical star-forming galaxies at $z>6$ have
 super-solar O/Fe.

The age of the universe at $z=6$ is 900 Myr.  Assuming the first galaxies began forming $\sim200$~Myr after
 the big bang, the oldest possible stellar populations at $z=6$ would have an age of $\sim$700~Myr.  This maximum
 age would decrease at higher redshifts when reionization was in progress ($\sim$400 and 250~Myr at $z=8$ and 10,
 respectively).  A spectroscopically-confirmed $z\sim10$ galaxy has been found to have an age of 340~Myr, close
 to the age limit \citep{hoa18}.  \citet{lab13} find a younger age of 100~Myr for a stacked SED of $z\sim8$ galaxies.
Due to the maximum possible age of galaxies at $z>6$, \textit{every galaxy} in the epoch of reionization
 should host an O/Fe-enhanced stellar population.
The properties of our $z>1$ auroral-line sample provide observational evidence for this expectation.
Just as for $z\sim1-3$ EELGs, we suggest employing BPASS binary stellar models with $Z_*=1/5Z_{\text{neb}}$ and the local relation
 between $U$ and O/H when modeling the nebular spectra of $z>6$ galaxies.
Based on the results in Figure~\ref{fig3}, we find that the most useful strong-line ratios for determining
 gas-phase oxygen abundance from rest-optical spectroscopy of $z>6$ galaxies are O32 and Ne3O2, which will be
 accessible out to $z\sim9$ and 12, respectively, with \textit{JWST}/NIRSpec and NIRCam spectroscopy.
While it is possible that [N\ii]-based indicators may also be useful,
 these can only be observed out to $z\approx6.6$ with NIRSpec and NIRCam.

Super-solar O/Fe values carry important implications for reionization.  Massive stars with lower Fe
 abundances produce harder spectra \textit{and} a larger number of ionizing photons per unit stellar mass.
The latter makes it easier for star-forming galaxies to reionize the universe, requiring lower escape fractions.
Low-mass, faint galaxies are thought to provide the bulk of the ionizing photons due to their number \citep[e.g.,][]{bou12,bou15,fin12,sta16}.
Such galaxies should form latest and thus have the youngest stellar populations with maximal
 O/Fe$\approx$5$\times$O/Fe$_{\odot}$ based on current Type II SNe yields \citep{nom06}.
Assuming $z>6$ galaxies follow the relations in Figure~\ref{fig17}, we use equations~\ref{eq:o32ew}
 and~\ref{eq:hbew} to estimate O32 values of $z>6$ galaxies.
\citet{lab13} use stacked broadband photometry to estimate EW$_0$([O\iii]+H$\beta$)=670~\AA\ of L$^*$ galaxies
 at $z\sim8$.
This equivalent width corresponds to EW$_0$([O\iii]$\lambda$5007)=420~\AA, EW$_0$(H$\beta$)=80~\AA,
 log(O32)=0.52, and 12+log(O/H)=8.15 (0.3~Z$_{\odot}$), assuming the high-redshift analog O32 calibration
 of BKD18.  In the case of 5$\times$O/Fe$_{\odot}$, the stellar metallicity would be $Z_*$=0.06~Z$_{\odot}$.
\citet{rob16} find EW$_0$([O\iii]+H$\beta$)$\sim$1500~\AA\ for four bright galaxies at $z\sim7-9$, implying
 EW$_0$([O\iii]$\lambda$5007)=1000~\AA, EW$_0$(H$\beta$)=160~\AA, log(O32)=0.8, and 12+log(O/H)=8.0 (0.2~Z$_{\odot}$).
For the \citet{rob16} sample, maximal O/Fe enhancement implies $Z_*$=0.04~Z$_{\odot}$.
These rough estimates suggest that metal-poor massive stars with $\lesssim5\%$~Z$_{\odot}$ power the ionizing
 spectra of the star-forming galaxies responsible for reionizing the universe.

\section{Summary \& conclusions}\label{sec:summary}

We have compiled a sample of 18 individual star-forming galaxies at $z=1.4-3.7$ with detections of auroral
 [O\iii]$\lambda$4363 or O\iii]$\lambda$1663, for which temperature-based nebular oxygen abundances can
 be derived independent of strong-line ratio calibrations.
Four of these are galaxies with [O\iii]$\lambda$4363 detections from the MOSDEF survey, three of which have not
 been previously published.
We utilized this sample to investigate the chemical abundances and ionization states of star-forming
 regions at $z>1$.
For the first time, we have constructed metallicity scaling relations, including the mass-metallicity relation and fundamental
 metallicity relation, at $z>1$ based purely on direct-method metallicities.
The independently-constrained nebular metallicities enable us to unambiguously constrain the
 stellar metallicities and ionization parameters of these galaxies using photoionization models,
 providing unprecedented insight into the gas physical conditions in $z>1$ \hii regions.
We summarize our main results and conclusions below.
\begin{enumerate}
\item We investigated whether strong-line metallicity calibrations change with redshift by directly comparing
 the strong-line ratios O3, O2, R23, O32, and Ne3O2 at fixed direct-method metallicity for samples at
 $z\sim0-2.2$ (Fig.~\ref{fig3}).  Over 12+log(O/H)=7.7$-$8.1, we find that O3 and R23 are saturated and are not useful
 metallicity indicators.  The median values of the $z>1$ auroral-line sample closely match the $z\sim0$ high-redshift
 analogs of BKD18 to within $\sim0.1$ dex in O/H at fixed line ratio.  We suggest the use of the
 BKD18 high-redshift analog calibrations at $z>1$, with O32 and Ne3O2 providing the most utility
 as these ratios are monotonic with metallicity over 7.7$<$12+log(O/H)$<$8.1.
With \textit{JWST}, spectroscopic measurements of O32 and Ne3O2 can be made to $z\sim9$ and $z\sim12$, respectively.
Local \hii region calibrations should not be utilized at 12+log(O/H)$\lesssim$8.0 for this set of strong-line ratios.
Our auroral-line sample did not have sufficient [N\ii] coverage and detections to investigate N/O and nitrogen-based
 metallicity indicators.

\item We construct the stellar mass---gas-phase metallicity relation using temperature-based metallicities for the first
 time at $z>1$ (Fig.~\ref{fig4}).  We find a correlation between \mstar\ and direct-method O/H (eq.~\ref{eq:mzrtotcorrz}).
After correcting for SFR biases in the $z>1$ sample, we find that metallicity decreases by $0.4\pm0.1$~dex from
 $z\sim0$ to $z\sim2.2$ at log(\mstar/$\msun)=7.5-9.5$, consistent with results based on strong-line methods.
We find tentative evidence for a \mstar-SFR-O/H relation at $z\sim2.2$ based on direct-method metallicities (Fig.~\ref{fig6}).
Galaxies at $z>1$ have the same oxygen abundance of galaxies at $z\sim0$ at fixed \mstar\ and SFR to within 0.1~dex,
 suggesting that the FMR does not significantly evolve with redshift (Fig.~\ref{fig7}).

\item Using photoionization models, we uniquely leverage the direct-method nebular metallicity measurements to
 obtain constraints on the ionization parameter, $U$, and stellar metallicity of massive stars, $Z_*$, in star-forming regions
 in the $z>1$ auroral-line galaxies (Sec.~\ref{sec:models}).  For the majority of the $z>1$ auroral-line sample,
 we find that $Z_*$, which traces Fe/H, is lower than the nebular metallicity (i.e., O/H) on a solar abundance scale.
This result implies super-solar O/Fe values, in contrast to typical $z\sim0$ systems which have $Z_*=Z_{\text{neb}}$.
The super-solar O/Fe values can be explained by a physcally-motivated picture in which the young stellar populations of
 these galaxies have been enriched by Type II SNe that occur promptly following a star-formation event and are the
 dominant production channel for O, but have not yet been fully enriched by Type Ia SNe that are delayed
 $\sim0.3-1$~Gyr and are the dominant producer of Fe-peak elements.
The four MOSDEF [O\iii]$\lambda$4363 emitters have ages $\le300$~Myr, and are younger than 90\% of MOSDEF galaxies
 matched in redshift and \mstar.

\item On average, the $z>1$ auroral-line sample has the same $U$ as local galaxies and \hii regions
 with the same O/H, and the same $U$ and O/H as $z\sim0$ galaxies matched in sSFR (Fig.~\ref{fig15}).
The high ionzation parameters of the $z>1$ sample are driven by their high sSFRs and low oxygen abundances,
 with $U$ values as expected from local relations between $U$, sSFR, and O/H.
We conclude that evolution in $U$ at fixed O/H is not a significant driver of the evolution of strong-line
 ratios with redshift.
Harder ionizing spectra at fixed O/H compared to the local universe are a cause of strong-line ratio evolution
 for our $z>1$ auroral-line sample due to super-solar O/Fe.
Due to a lack of [N\ii] coverage and detections, we are unable to evaluate the role of N/O with the current sample
 and cannot draw conclusions about the positions of $z\sim2$ galaxies in the [N\ii] BPT diagram.

\item The $z>1$ auroral-line sample has significantly higher sSFR, O32, and emission-line equivalent widths than
 typical $z\sim2.3$ galaxies (Figs.~\ref{fig16} and~\ref{fig17}), indicating they are younger and more metal-poor
 than the bulk of the $z\sim2$ star-forming population.  It is therefore unclear if our results carry over to
 typical galaxies at these redshifts, in particular because stellar population age is expected to vary with
 \mstar\ and sSFR.  It is possible that $z\sim2$ galaxies are a mixed population,
 where low-mass and starburst galaxies are young and have super-solar O/Fe, while more massive galaxies have
 lower levels of O/Fe enhancement, potentially solar O/Fe at log(\mstar/$\msun)\gtrsim10.5$.
As a result of such variation, metallicity measurements would be biased as a function of \mstar\ and SFR when assuming
 a single calibration for the full sample, affecting the measured shape of the MZR and \mstar-SFR-O/H relation.
O/Fe constraints for galaxies falling on the \mstar-SFR relation and more massive, metal-rich galaxies are needed
 to obtain robust constraints on metallicity scaling relations in the high-redshift universe.

\item The properties of the $z>1$ auroral-line sample are closely matched to those of extreme emission-line
 galaxies at $z\sim2$ \citep{tan18}.  We conclude that EELGs have super-solar O/Fe, but similar $U$ to local
 galaxies with the same O/H.  We find that typical $z\sim2.3$ MOSDEF galaxies and EELGs form a tight sequence
 in O32 and EW$_0$(H$\beta$) as a function of EW$_0$([O\iii]$\lambda$5007) across 2.5 orders of magnitude
 (eqs.~\ref{eq:o32ew} and~\ref{eq:hbew}).

\item The emission-line equivalent widths and sSFRs of $z>6$ galaxies also match the properties of the $z>1$
 auroral-line sample.  We extend our results to typical star-forming galaxies in the epoch of reionization,
 implying more efficient production of ionizing photons from Fe-poor massive stars.
Because of a maximum stellar population age imposed by the age of the universe, all galaxies at $z>6$ are
 expected to have super-solar O/Fe.
We use the O32 and EW$_0$(H$\beta$) vs.\ EW$_0$([O\iii]$\lambda$5007) relations to estimate the nebular
 and stellar metallicities of UV-luminous $z>6$ galaxies, finding typical values of $Z_{\text{neb}}\sim0.25~Z_{\odot}$
 and $Z_*\sim0.05~Z_{\odot}$.

\end{enumerate}

Ultimately, a larger sample of $z>1$ auroral-line emitters spanning a wider dynamic range in oxygen
 abundance is needed to make a full assessment of metallicity indicators across the range of metallicities
 spanned by large spectroscopic survey datasets.
Such a sample would allow for tests of calibration shapes, not just normalization as performed in this work.
Our results suggest that finding galaxies with [O\iii]$\lambda$4363 that can be detected with current
 facilities can be achieved by identifying galaxies with EW$_0$([O\iii]$\lambda$5007)$\gtrsim$300~\AA\ and
 high SFR ($\gtrsim$10~$\msun$~yr$^{-1}$), but this selection will only yield metal-poor, low-mass galaxies.
Obtaining auroral [O\ii]$\lambda\lambda$7320,7330 measurements for massive, moderately metal-rich galaxies
 with high SFRs presents a potential avenue to extend the metallicity baseline, improving calibration tests
 and constraints on the mass-metallicity relation.

The ionizing spectrum of massive stars is so strongly tied to the production of emission lines that it
 is imperative to understand the properties of massive stars at low and high redshifts in order to
 extract robust star-formation rates, metallicities, and ionization state information from spectra.
To gain a more complete picture of massive stars at high redshifts, we must constrain O/Fe of
 individual galaxies spanning the \mstar\ and SFR range of the star-forming population.
Reaching this goal will require auroral-line measurements and high-S/N FUV continuum spectroscopy of a large number
 of high-redshift galaxies over a wide dynamic range of properties.
The increased sensitivity of \textit{JWST} and 30~m-class telescopes will be needed to achieve
 this goal.  Until that time, we must continue extrapolating from small samples with detailed
 observations to infer the chemical enrichment of the large spectroscopic datasets currently available at $z\sim1-3$.

\section*{Acknowledgements}
Based on data obtained at the W.M. Keck Observatory, which is operated as a scientific partnership
 among the California Institute of Technology, the University of California, and NASA, and was made
 possible by the generous financial support of the W.M. Keck Foundation.
We acknowledge support from NSF AAG grants AST-1312780, 1312547, 1312764, and 1313171,
 archival grant AR-13907 provided by NASA through the Space Telescope Science Institute,
 and grant NNX16AF54G from the NASA ADAP program.
We also acknowledge a NASA
contract supporting the “WFIRST Extragalactic Potential
Observations (EXPO) Science Investigation Team” (15-WFIRST15-0004), administered by GSFC.
  We additionally acknowledge the 3D-HST collaboration
 for providing spectroscopic and photometric catalogs used in the MOSDEF survey.
  We wish to extend special thanks to those of Hawaiian ancestry on
 whose sacred mountain we are privileged to be guests. Without their generous hospitality,
 the work presented herein would not have been possible.

%%%%%%%%%%%%%%%%%%%%%%%%%%%%%%%%%%%%%%%%%%%%%%%%%%

%%%%%%%%%%%%%%%%%%%% REFERENCES %%%%%%%%%%%%%%%%%%

% The best way to enter references is to use BibTeX:

\bibliographystyle{mnras}
\bibliography{tepaper}

%%%%%%%%%%%%%%%%%%%%%%%%%%%%%%%%%%%%%%%%%%%%%%%%%%

%%%%%%%%%%%%%%%%% APPENDICES %%%%%%%%%%%%%%%%%%%%%

\appendix

\section{Literature $\lowercase{z}>1$ auroral-line sample}\label{app}

In this appendix, we describe the details of the oxygen abundance and galaxy property calculations on a case-by-case basis
 for each of the [O\iii]$\lambda$4363 and O\iii]$\lambda$1663 emitters taken from the literature.
We also explain why we do not include the claimed [O\iii]$\lambda$4363 detection from \citet{yua09} in our analysis.

\subsection{Literature [O\iii]$\lambda$4363 emitters}\label{app:o4363}

\citet{bra12b} report a marginal 1.7$\sigma$ detection of [O\iii]$\lambda$4363 for SL2SJ02176-0513 (B18), a star-forming galaxy
 at $z=1.847$ magnified by a factor of 19.  
We estimate zero nebular reddening according to the observed Balmer ratio of H$\delta$/H$\beta=0.18\pm0.03$,
 assuming an intrinsic value of H$\delta$/H$\beta=0.259$.
The [O\ii]$\lambda\lambda$3726,3729 doublet is not included in the \textit{HST}/WFC3 G141 grism wavelength coverage.
Following \citet{bra12b}, we instead assume a uniform distribution of [2.5, 9.0] for the
 [O\iii]$\lambda$5007/[O\ii]$\lambda\lambda$3726,3729 ratio \citep{erb10,ate11} and the mean value of
 [O\iii]$\lambda$5007/[O\ii]$\lambda\lambda$3726,$3729=5.75$ for calculating the metallicity.
  This range of [O\iii]/[O\ii] values is typical of galaxies at the same stellar mass
 as B18 (see Fig.~\ref{fig17}).
The oxygen abundance under this set of assumptions is 12+log(O/H$)=7.56^{+0.51}_{-0.25}$.  We note that O$^+$ makes up at most
 30\% of O under the range of assumed [O\iii]/[O\ii] values, and the total oxygen abundance
 only varies by $\pm0.06$~dex.  Properties of B18 are also reported in \citet{ber18} (see~\ref{app:o1663}).

\citet{chr12a,chr12b} report a 3.4$\sigma$ detection of [O\iii]$\lambda$4363 for a lensed galaxy at $z=1.8339$,
 Abell 1689 arc ID 31.1 (C12a), magnified by a factor of 26.6.  Nebular reddening is estimated from H$\gamma$
 and H$\beta$, with H$\delta$/H$\beta=0.40\pm0.07$ corresponding to E(B-V)$_{\text{gas}}=0.31^{+0.35}_{-0.31}$,
 assuming an intrinsic ratio of H$\delta$/H$\beta=0.468$.  The direct-method oxygen abundance is
 12+log(O/H$)=7.46^{+0.23}_{-0.23}$.  We adopt the rest-frame equivalent width reported in \citet[][ID 876\_330 therein]{sta14}
 of EW$_0$([O\iii]$\lambda\lambda$4959,5007+H$\beta)=740\pm190$~\AA, and determine EW$_0$([O\iii]$\lambda$5007)
 and EW$_0$(H$\beta$) based on the measured O3 ratio.

\citet{sta13} report a detection of [O\iii]$\lambda$4363 for CSWA~141 (S13) at $z=1.425$ with a significance
 of 7.3$\sigma$.  S13 is magnified by a factor of 5.5.  Nebular reddening is estimated from H$\beta$,
 H$\gamma$, and H$\delta$, and is found to be small (E(B-V)$_{\text{gas}}=0.03$).
The total oxygen abundance is 12+log(O/H$)=7.96^{+0.07}_{-0.07}$.
The stellar mass of S13 is log(\mstar/$\msun)=8.33^{+0.10}_{-0.14}$ (private communication, D. Stark \& R. Mainali).

A marginal 1.7$\sigma$ detection of [O\iii]$\lambda$4363 is reported by \citet{jam14} for CSWA~20 (J14),
 a star-forming galaxy at $z=1.433$ magnified by a factor of 11.5.  H$\alpha$, H$\beta$, and H$\gamma$
 are used to determine nebular reddening.  The total oxygen abundance of J14 based on [O\iii]$\lambda$4363 is
 12+log(O/H$)=8.13^{+0.35}_{-0.21}$.
\citet{pat18} report a stellar mass of log(\mstar/$\msun)=10.3\pm0.3$ for J14, estimated by fitting the galaxy SED obtained
 from SDSS broadband photometry, with $z$-band being the reddest filter at $\lambda_{\text{rest}}\approx3700$~\AA.
\citet{jam14} measure a line width of $34.9\pm0.7$~km~s$^{-1}$ for the narrow systemic component of strong nebular emission
 features in the spectrum of J14, which also display a broad blueshifted component.
The high stellar mass of \citet{pat18} is inconsistent with such narrow line widths for standard disk and spheroid geometries,
 requiring unphysically large virial coefficients to reconcile the two in dynamical modeling.
Additionally considering systematic uncertainties in lens modeling and the lack of broadband photometry blueward of the Balmer break,
 we conclude that the reported stellar mass of J14 from \citet{pat18} is not reliable and likely overestimates the true stellar mass.
We do not report \mstar\ or sSFR for J14 in Table~\ref{tab:props}.

\subsection{Literature O\iii]$\lambda\lambda$1661,1666 emitters}\label{app:o1663}

\citet{chr12a,chr12b} report O\iii]$\lambda$1663 detections for three strongly lensed galaxies at $z>1$.
Abell 1689 arc ID 31.1 (C12a) has detections of both O\iii]$\lambda$1663 and O\iii]$\lambda$4363.
C12a has a total oxygen abundance of 12+log(O/H$)=7.34^{+0.44}_{-0.70}$ based on O\iii]$\lambda$1663.
Despite a remarkable 21.2$\sigma$ detection of O\iii]$\lambda$1663,
 the uncertainty on the metallicity is very large due to the large uncertainty of the dust correction and the
 extreme sensitivity of O\iii]$\lambda$1663/[O\iii]$\lambda$5007 to reddening.  The oxygen abundance derived
 using O\iii]$\lambda$1663 is consistent with that based on [O\iii]$\lambda$4363 (see~\ref{app:o4363}), although
 the uncertainty is significantly lower when using [O\iii]$\lambda$4363.  We adopt the total oxygen abundance
 based on [O\iii]$\lambda$4363 for this target.

SMACS J0304 (C12b) at $z=1.9634$ has a 7.5$\sigma$ detection of O\iii]$\lambda$1663 \citep{chr12a,chr12b}.  Nebular reddening is estimated
 using H$\alpha$, H$\beta$, H$\gamma$, and H$\delta$.  The direct-method oxygen abundance is 12+log(O/H$)=8.14^{+0.04}_{-0.04}$.
\citet{chr12a} note C12b as a merging system based on its complex morphology, with 5 distinct components at close
 projected separation ($\lesssim20$~kpc, uncorrected for magnification) in \textit{HST} imaging.

A 12.7$\sigma$ O\iii]$\lambda$1663 detection is reported for M2031 \citep[C12c;][]{chr12a,chr12b}.
H$\beta$ and H$\epsilon$ suggest little nebular reddening assuming an intrinsic ratio
 of H$\epsilon$/H$\beta=0.159$ \citep{ost06}.  The total oxygen abundance is 12+log(O/H$)=7.77^{+0.02}_{-0.30}$.
C12c is an interacting system, with a companion located at the same redshift ($\Delta v<50$~km~s$^{-1}$)
 and 11.2~kpc projected separation in the source plane \citep{pat16}.

\citet{jam14} report a 2.4$\sigma$ detection of O\iii]$\lambda$1663 for the gravitationally-lensed galaxy CSWA~20 (J14),
 which also has a marginal detection of [O\iii]$\lambda$4363 (see~\ref{app:o4363}).
The direct-method oxygen abundance derived from O\iii]$\lambda$1663 is 12+log(O/H$)=7.86^{+0.15}_{-0.14}$.
The metallicity based on [O\iii]$\lambda$4363 is $\sim1\sigma$ consistent with this value.
Given that both [O\iii]$\lambda$4363 and O\iii]$\lambda$1663 are marginally detected (1.7$\sigma$ and 2.4$\sigma$, respectively),
 the derived O/H uncertainties are comparable, and the large sensitivity to reddening correction for O\iii]$\lambda$1663,
 we utilize the metallicity based on [O\iii]$\lambda$4363 for J14.
Our results do not change significantly when instead utilizing the O\iii]$\lambda$1663-based metallicity.

\citet{bay14} present a 6.9$\sigma$ O\iii]$\lambda$1663 detection for a lensed star-forming galaxy at $z=3.6252$ (B14).
H$\beta$ and H$\gamma$ are both detected, but H$\gamma$ is contaminated by skylines.  The measured
 H$\gamma$/H$\beta=0.52\pm0.02$ is 2.5$\sigma$ inconsistent with the maximum value of 0.468 in the
 case of no nebular reddening \citep{ost06}, again suggesting H$\gamma$ is unreliable.
Following \citet{bay14}, we adopt A$_{\text{V}}=1.0\pm0.2$ from their SED fitting and assume that the nebular and
 stellar reddening is equal.  This assumption yields 12+log(O/H$)=8.10^{+0.09}_{-0.08}$.

\citet{sta14} present 3.5$\sigma$ detections of O\iii]$\lambda$1663 for two lensed galaxies: Abell 860\_359 at $z=1.7024$ (S14a)
 and MACS 0451 ID 1.1b at $z=2.0596$ (S14b).  Nebular reddening is estimated from H$\alpha$ and H$\beta$ for each source.
[O\ii]$\lambda\lambda$3726,3729 is not detected for either galaxy,
 but 2$\sigma$ upper limits are reported.  We adopt the 2$\sigma$ [O\ii]
 upper limit values when calculating the metallicity, and a uniform distribution between zero and 3$\sigma$ for [O\ii]
 when creating realizations for uncertainty estimation.  Under this assumption, both objects are still dominated by
 doubly-ionized oxygen, with $\sim20$\% of O in O$^+$.  Varying the assumed [O\ii] values between zero and 3$\sigma$
 changes the total oxygen abundances by $<0.07$~dex.  The derived direct-method oxygen abundances are
 12+log(O/H$)=8.05^{+0.11}_{-0.08}$ for S14a and 12+log(O/H$)=7.31^{+0.09}_{-0.14}$ for S14b.
The EW$_0$([O\iii]$\lambda$5007) and EW$_0$(H$\beta$) of 860\_359 is inferred from the reported
 EW$_0$([O\iii]$\lambda\lambda$4959,5007+H$\beta)=1550\pm150$~\AA\ based on the measured O3 ratio.

\citet{ste14} present three unlensed galaxies from the KBSS survey with detections of O\iii]$\lambda$1663.
These galaxies are BX74 (Ste14a), BX418 (Ste14b), and BX660 (Ste14c) at $z=2.1889$, 2.3054, and 2.1742, respectively,
 with 7.5$\sigma$, 7.7$\sigma$, and 5.5$\sigma$ detections of O\iii]$\lambda$1663.
Line ratios for these three galaxies are presented in both \citet{ste14} and \citet{erb16}, where individual line fluxes
 were derived from the latter in \citet{pat18}.
The rest-optical line ratios disagree between \citet{ste14} and \citet{erb16}, with the latter yielding significantly
 lower O/H for all three galaxies.
We utilize the measurements in \citet{ste14} because this work included a more robust treatment of differential
 slit losses between filters, resulting in more accurate line ratios.
For all three galaxies, we use the combination of line ratios from \citet{ste14} and H$\alpha$ fluxes from \citet{erb16}
 to derive observed line fluxes and uncertainties.
We also take stellar masses from \citet{ste14}, where the photometry used for SED fitting was corrected for
 rest-optical emission-line contamination.
We obtain SFRs, temperatures, densities, and metallicities that are consistent with the values reported in \citet{ste14}.
The total oxygen abundances are 12+log(O/H$)=8.03^{+0.06}_{-0.06}$ for Ste14a,
 12+log(O/H$)=8.10^{+0.02}_{-0.06}$ for Ste14b, and 12+log(O/H$)=8.15^{+0.04}_{-0.06}$ for Ste14c. 

\citet{koj17} report a 6.5$\sigma$ detection of O\iii]$\lambda$1663 for COSMOS 12805 (K17),
 an unlensed galaxy at $z=2.159$.
A 2$\sigma$ upper limit is reported for [O\ii]$\lambda\lambda$3726,3729.
We adopt the same strategy as for the two galaxies from \citet{sta14}, finding a total
 oxygen abundance of 12+log(O/H$)=8.26^{+0.03}_{-0.30}$.
Assuming the 2$\sigma$ [O\ii] upper limit for the abundance calculation gives 34\% of total O
 in the singly-ionized state.  The significant uncertainty of the [O\ii] flux and, consequently,
 the O$^+$ abundance is the cause of the large lower error on O/H.

\citet{ber18} present a very high S/N rest-UV spectrum of SL2SJ02176-0513 (B18) with S/N$>$50 in each
 component of the O\iii]$\lambda\lambda$1661,1666 doublet.  Rest-optical \textit{HST}/grism spectroscopy
 has also revealed a marginal [O\iii]$\lambda$4363 detection \citep[][see~\ref{app:o4363}]{bra12b}.
We adopt the updated magnification factor and stellar mass of B18 from \citet{ber18}.
The rest-UV spectrum reveals an extremely high level of ionization based on strong nebular C\iv
 and He\ii, and models presented in \citet{ber18} suggest $<$2\% of O is in O$^+$ with a non-negligible
 fraction in O$^{3+}$.  We employ their oxygen ionization correction factor of O/H=1.055$\times$O$^{2+}$/H
 and find the total oxygen abundance to be 12+log(O/H$)=7.53^{+0.09}_{-0.05}$ using O\iii]$\lambda$1663.
We adopt this metallicity value for B18, which is consistent with but better constrained than
 the abundance based on the marginal [O\iii]$\lambda$4363 detection from \citet{bra12b}.

The O\iii]$\lambda$1663 doublet is detected at 14.3$\sigma$ significance for the highly-magnified
 Lynx arc at $z=3.357$ \citep[V04;][]{fos03,vil04}.  Reddening cannot be estimated from Balmer lines since only
 H$\beta$ is available.  Following \citet{pat18}, we assume zero nebular reddening based on SED
 modeling of the rest-optical and rest-UV spectrum \citep{vil04}.  In the absence of a detection of [O\ii],
 we follow the same procedure as for the two galaxies from \citet{sta14}.  We note that the Lynx arc is
 highly ionized, with O$^+$ making up at most 4\% of O (3$\sigma$ upper limit), such that the [O\ii] non-detection
 has a negligible impact on the nebular metallicity determination.  The total oxygen abundance is
 12+log(O/H$)=7.76^{+0.04}_{-0.04}$.  While a stellar mass is given in \citet{vil04}, it is important to note that
 they only observe one bright knot of the highly-magnified arc with a particularly low mass-to-light ratio (possibly
 a single lensed \hii region) and that the stellar mass is that of an instantaneous burst simple stellar population
 able to power the H$\beta$ emission.  The stellar mass from \citet{vil04} thus only represents the mass of a star
 cluster ionizing the \hii region, not the global stellar mass of the galaxy, and is
 not suitable for studying metallicity scaling relations with global galaxy properties.

\subsection{The case of the Yuan \& Kewley (2009) [O\iii]$\lambda$4363 detection}\label{app:y09}

A detection of [O\iii]$\lambda$4363 for Abell 1689 Lens 22.3 at $z=1.706$ was presented in \citet{yua09},
 with a magnification of 15.5.  The reported [O\iii]$\lambda$4363 flux is 0.27$\pm$0.1 on a normalized
 scale where the H$\beta$ flux is 1.0.
\citet{yua09} measure a redshift from [O\iii]$\lambda$4363 of $z=1.696$ that is $\Delta z=0.009$ lower than
 the redshift measured from strong rest-optical lines, $z=1.705$.
This redshift offset corresponds to a velocity offset of $\sim1000$~km~s$^{-1}$ ($\Delta\lambda=15$~\AA\ in the rest-frame),
 and calls into question the legitimacy of the [O\iii]$\lambda$4363 detection given that [O\iii]$\lambda$4363 emission
 should be at rest with respect to other rest-optical nebular emission lines.
Based on the redshift from the strong lines ($z=1.705$), the rest-frame centroid of the line in question is 4348~\AA, close
 to the wavelength of H$\gamma$ (4342~\AA).
Given that the spectrum in \citet{yua09} has a low spectral resolution of $R\sim500$ (9~\AA\ resolution in the rest-frame at
 $\lambda=4350$~\AA), we argue that H$\gamma$ was misidentified as [O\iii]$\lambda$4363.
Indeed, H$\gamma$ is almost always significantly stronger than [O\iii]$\lambda$4363 (the case for every other source in our $z>1$ auroral-line sample),
 and should be easily detected if [O\iii]$\lambda$4363 is also detected unless H$\gamma$ is contaminated by a sky line.
\citet{yua09} report a detector response-corrected ratio of H$\alpha$/H$\beta=5.03$.  This Balmer ratio implies a nebular reddening
 of E(B-V)$_{\text{gas}}=0.57$.
Based on this E(B-V)$_{\text{gas}}$ value, the expected H$\gamma$/H$\beta$ ratio is 0.34, consistent with the
 reported [O\iii]$\lambda$4363/H$\beta=0.27\pm0.1$ from \citet{yua09}.
We conclude that the claimed [O\iii]$\lambda$4363 detection in \citet{yua09} is in fact a misidentification of H$\gamma$,
 and exclude this source from our literature $z>1$ auroral-line sample.
\citet{gbu19} have recently obtained a deep $R\sim3000$ spectrum of Lens~22.3 and confirmed that the line reported in
 \citet{yua09} is H$\gamma$ and no significant [O\iii]$\lambda$4363 is present.

\section{Calibration sample biases of Curti et al. (2017)}\label{app:c17}

\begin{figure}
 \includegraphics[width=\columnwidth]{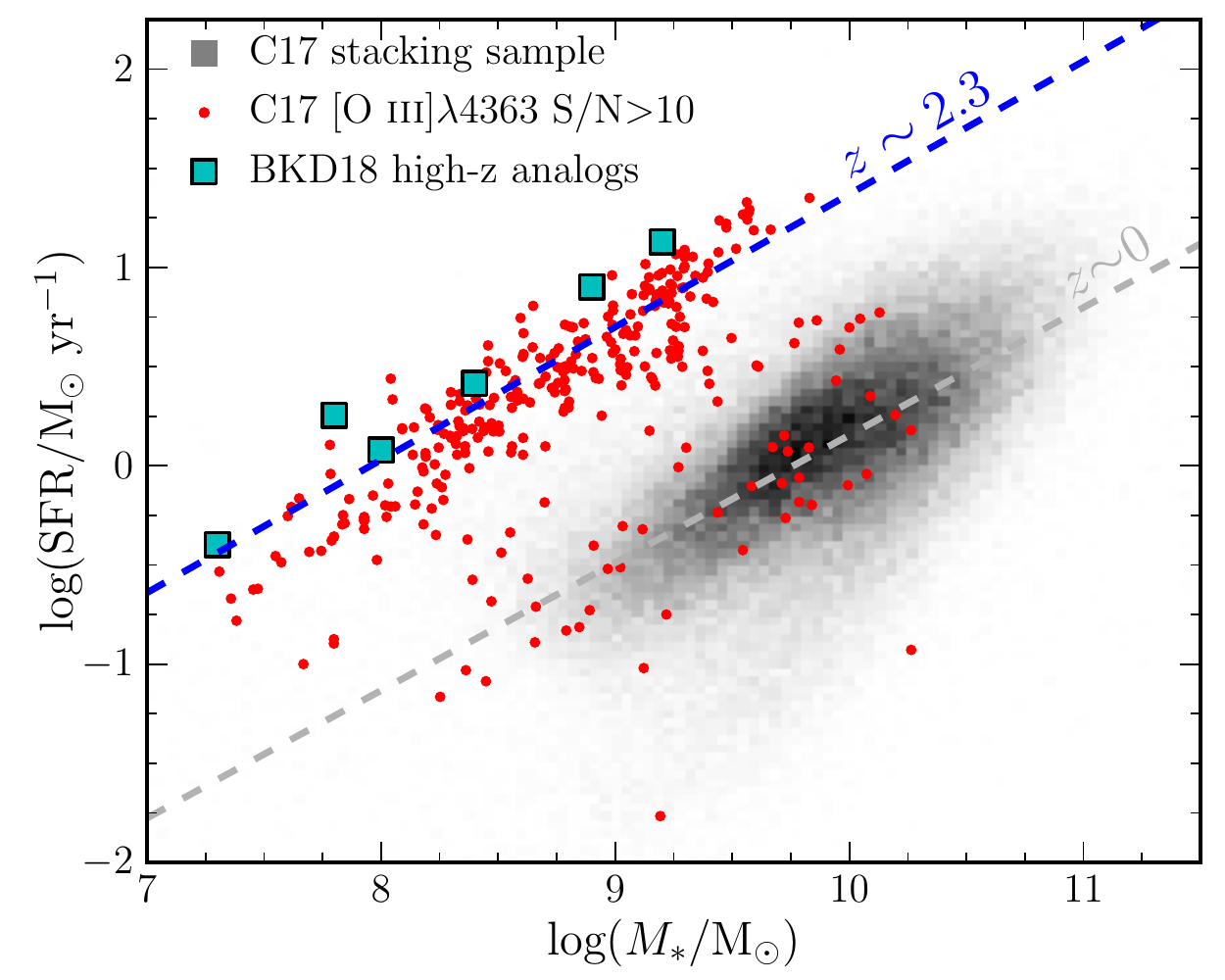}
 \centering
 \caption{
SFR vs.\ \mstar\ for the \citet{cur17} $z\sim0$ SDSS stacking sample (gray histogram) and the subset with
 [O\iii]$\lambda$4363 at $>$10$\sigma$ (red points).  The gray and blue dashed lines show the mean \mstar-SFR relation at
 $z\sim0$ and $z\sim2.3$, respectively.
The cyan squares show the median \mstar\ and SFR for each bin of the \citet{bia18} high-redshift analog composites.
}\label{figb1}
\end{figure}

\citet{cur17} constructed $z\sim0$ metallicity calibrations using stacked spectra in bins of line ratio in the
 O3 vs.\ O2 diagram.
The stacked spectra only reached down to 12+log(O/H$)\approx8.4$, and C17 supplemented the calibration sample
 with individual galaxies meeting the original selection criteria that additionally had [O\iii]$\lambda$4363
 detected with $>$10$\sigma$ significance.
The C17 calibration fits are thus dominated by this [O\iii]$\lambda$4363-detected subsample at 12+log(O/H$)\lesssim8.4$,
 and will be affected by biases associated with the cut on [O\iii]$\lambda$4363 S/N.
To investigate biases in the C17 calibration sample, we have used their selection criteria to select the same
 $z\sim0$ stacking sample from the MPA-JHU DR7 catalogs \citep{kau03,bri04,tre04,sal07}.
We select the individual [O\iii]$\lambda$4363-detected subsample by further requiring S/N$>$10 for [O\iii]$\lambda$4363.
We plot the C17 stacking sample (gray histogram) and [O\iii]$\lambda$4363-detected subsample (red points) in
 the SFR vs.\ \mstar\ diagram in Figure~\ref{figb1}.
For comparison, we also show the mean $z\sim0$ and $z\sim2.3$ \mstar-SFR relations from Fig.~\ref{fig5}, and the
 median \mstar\ and SFRs of the BKD18 high-redshift analog stacks (cyan squares).

We find that the vast majority of the C17 [O\iii]$\lambda$4363-detected subsample has low mass (log(\mstar/$\msun)\lesssim9.5$)
 and lies significantly above the mean $z\sim0$ \mstar-SFR relation.
The C17 [O\iii]$\lambda$4363-detected subsample is in fact coincident with the $z\sim2.3$ \mstar-SFR relation (blue dashed line)
 and the BKD18 high-redshift analogs, selected to have a similar offset in the BPT diagram as $z\sim2$ galaxies.
The C17 subset with S/N$>$10 for [O\iii]$\lambda$4363 thus has high-sSFR typical of high-redshift galaxies that has
 been found to correlate with extreme ISM conditions \citep{kew15,kaa17,kaa18,kas19b}.
Consequently, the C17 calibrations trace extreme ISM conditions typical at $z>1$ at 12+log(O/H$)\lesssim8.4$ where
 the individual [O\iii]$\lambda$4363-detected subsample dominates, and normal $z\sim0$ ISM conditions only at
 12+log(O/H$)\gtrsim8.4$ where the representative stacking sample dominates.
The agreement between the $z>1$ auroral-line sample and C17 calibrations in Fig.~\ref{fig3} does not indicate that
 $z\sim0$ calibrations are applicable at high redshifts, but instead is a result of the biased nature of the C17
 calibrations at low metallicities.

%%%%%%%%%%%%%%%%%%%%%%%%%%%%%%%%%%%%%%%%%%%%%%%%%%

% Don't change these lines
\bsp    % typesetting comment
\label{lastpage}
\end{document}